\documentclass[11pt,a4paper]{article}
\usepackage{jheppub}
\usepackage{tikz}
\usepackage{subcaption}

\usepackage{footmisc}

\allowdisplaybreaks

\title{\boldmath M5-branes Probing Flux Backgrounds}

\author[a,b]{Ibrahima Bah,}
\author[c]{Federico Bonetti,}
\author[a]{Enoch Leung}
\author[a]{and Peter Weck} 
\affiliation[a]{Department of Physics and Astronomy, Johns Hopkins University,\\
3400 North Charles Street, Baltimore, MD 21218, USA}
\affiliation[b]{Institute for Advanced Study, \\
Olden Lane, Princeton, New Jersey 08540, USA}
\affiliation[c]{Mathematical Institute, University of Oxford,\\
Woodstock Road, Oxford, OX2 6GG, UK}

\emailAdd{iboubah@jhu.edu}
\emailAdd{federico.bonetti@maths.ox.ac.uk}
\emailAdd{yleung5@jhu.edu}
\emailAdd{pweck1@jhu.edu}

\abstract{We analyze the global symmetries and anomalies of 4d $\cN = 1$ field theories that
arise from a stack of $N$ M5-branes probing a class of flux backgrounds. These backgrounds consist of a resolved $\mathbb{C}^2 / \mathbb{Z}_k$ singularity fibered over a smooth Riemann surface of genus $g \ge 2$, supported by a non-trivial $G_4$-flux configuration labeled by a collection of $2(k-1)$ flux quanta, $\{N_i\}$. 
For $k=2$, this setup defines a non-trivial superconformal field theory (SCFT) in the IR, which is holographically dual to an explicit $AdS_5$ solution first described by Gauntlett, Martelli, Sparks, and Waldram. The generalization to $k \ge 3$ is hard to tackle directly within holography. Instead, in this paper we lay the groundwork for a systematic analysis of such a generalization by adopting anomaly inflow methods to identify continuous and discrete global symmetries of the 4d field theories. 
We also compute the 't Hooft anomalies for continuous symmetries at leading order in the limit of large $N$, $N_i$.}
 

\newcommand{\cN}{\mathcal{N}}

\begin{document}


\maketitle
\flushbottom

\addtocontents{toc}{\protect\setcounter{tocdepth}{2}}


 
\section{Introduction and summary}

Geometric and brane engineering provide a  powerful framework for the construction
of non-trivial quantum field theories (QFTs) and the analysis of their strongly coupled regimes.
A prominent example is furnished by 4d superconformal field theories
(SCFTs) realized as the low-energy limit of M5-brane configurations in M-theory.
Large classes of strongly coupled 4d SCFTs can be realized by wrapping
M5-branes on a Riemann surface with defects, preserving $\cN = 2$ \cite{Gaiotto:2009we,Gaiotto:2009hg}
or $\cN =1$ \cite{Bah:2011vv,Bah:2012dg,Maruyoshi:2009uk,Benini:2009mz,Bah:2011je} supersymmetry. 


In this work we aim to investigate a class of constructions
that remains largely unexplored: M5-branes probing flux backgrounds.
As a case study, we consider M5-branes wrapped on a Riemann surface and
probing a resolved $\mathbb C^2/\mathbb Z_k$ 
singularity. 
Such setups should be contrasted with M5-branes
probing an unresolved $\mathbb C^2/\mathbb Z_k$ singularity,
which yields well studied 6d (1,0) SCFTs \cite{Brunner:1997gk,Blum:1997fw,Blum:1997mm,Intriligator:1997dh,Brunner:1997gf,Hanany:1997gh}. The latter may
be further compactified to four dimensions on a Riemann surface, see e.g.~\cite{Gaiotto:2015usa,Franco:2015jna,DelZotto:2015rca,Hanany:2015pfa,Morrison:2016nrt,Razamat:2016dpl,Bah:2017gph,Apruzzi:2018oge,Kim:2018lfo}.

Our analysis is motivated by a class of M-theory solutions, first discussed by Gauntlett, Martelli, Sparks, and Waldram (GMSW) \cite{Gauntlett:2004zh}, of the form $AdS_5 \times_w M_6$, a warped product of $AdS_5$ and an internal space $M_6$. These solutions can be interpreted as the near-horizon geometry of a stack of M5-branes probing a resolved $\mathbb C^2 /\mathbb Z_2$ singularity, fibered over a smooth genus-$g$ Riemann surface $\Sigma_g$, and stabilized by a $G_4$-flux configuration threading
four-cycles constructed with $\Sigma_g$ and the resolution two-cycle
of $\mathbb C^2/\mathbb Z_2$ \cite{Bah:2019vmq}, the latter being a $\mathbb C \mathbb P^1$ fibration over $\Sigma_g \times S^2$. Such an interpretation is supported by the features of an equivalent description of the internal space $M_6$ as a fibration of a 4d space $M_4$ over $\Sigma_g$, where $M_4$ is the resolution of the orbifold $S^4/\mathbb{Z}_2$, obtained by the blow-up of the $\mathbb{Z}_2$ fixed points at the north and south poles of $S^4$.

The setups described above admit a natural generalization, 
in which the group $\mathbb Z_2$ is replaced by $\mathbb Z_k$ with $k \ge 3$.
More precisely, we consider a different topology for $M_6$:
we still take $M_6$ to be 
a fibration of a 4d space $M_4$ over $\Sigma_g$,
but now $M_4$ is the resolution of $S^4/\mathbb Z_k$, obtained by blowing up the fixed points of the $\mathbb Z_k$ action at the poles of $S^4$. 
The blow-up procedure generates a collection of
$k-1$ two-cycles at each pole. A non-trivial $G_4$-flux threads $M_4$ as well as the four-cycles
obtained by combining these resolution cycles with $\Sigma_g$.
In total, we have $2k-1$ flux parameters:
one flux quantum $N$ is interpreted as the number of M5-branes in the stack,
while two independent sets of $k-1$ flux quanta 
$N_{{\rm N}_i}$, $N_{{\rm S}_i}$
($i=1,\dots,k-1$)  describe the resolution
of the orbifold singularities at the north and south poles of $S^4$.

The above discussion gives a concrete characterization of the
topology of $M_6$
and the flux configuration threading it. The key physical question
is whether this putative topology and its flux data
can correspond
to actual well-defined M-theory  setups that yield non-trivial 4d field theories.
A possible inroad into this problem is to 
search for explicit $AdS_5$ 
 solutions in 11d supergravity in which the internal space
has the topology of $M_6$ and is supported by   
a $G_4$-flux configuration with the prescribed flux quanta.
In other words, these solutions would be the $k \ge 3$
generalization of the $k=2$ GMSW solutions.
The BPS systems governing $AdS_5$ solutions
preserving $\cN = 1$ superconformal symmetry
are well-understood \cite{Gauntlett:2004zh,Bah:2015fwa}. 
A direct search for $AdS_5$ solutions of the desired kind for $k\ge 3$,
however, turns out to be prohibitively hard. 
Another possible strategy could be to try
to construct a supergravity solution that
describes the flux background   probed by the M5-brane stack.
Even in the case of $k=2$, however, such a solution is not available,
suggesting that it might be particularly challenging to push
forward this approach for general $k$.

Faced with the difficulties outlined above,
we turn to a different methodology for investigating if
the topology and flux configuration under examination
can yield interesting physics. Our approach
is based on anomaly inflow. More precisely, 
the central working assumption of this paper is that
the topology of $M_6$ and the associated flux configuration
can be regarded as admissible boundary conditions
for the 11d supergravity fields 
in the vicinity of a codimension-7 object, 
extended along four
non-compact spacetime dimensions
and furnishing a low-energy description of the 
wrapped M5-brane stack probing the flux background.
The boundary conditions specified by $M_6$, and the flux threading it,
induce an anomalous gauge variation of the low-energy M-theory
effective action. Systematic methods have been developed
\cite{Bah:2019rgq}---building on  \cite{Freed:1998tg,Harvey:1998bx}---which 
take the topology and flux data on $M_6$ as input,
and yield an inflow anomaly polynomial $I_6^\text{inflow}$
that encodes the anomalous gauge variation induced by the boundary.
According to the inflow paradigm,
this variation cancels exactly against the 't Hooft anomalies
of the 4d degrees of freedom
that capture the IR dynamics of the wrapped M5-branes probing the flux background.



The main goal of this paper is
the explicit  implementation of this circle
of ideas to the $M_6$ input data described before. This proves   to be a non-trivial task.
We uncover a rich pattern of symmetries and associated anomaly theories,
demonstrating the power and flexibility of inflow methods.

Our analysis involves two classes of global symmetries in 4d.
The first class consists of   ordinary continuous symmetries associated with the isometries
of $M_6$.
The second class is associated with  cohomology classes in $M_6$
and consists both of higher-form symmetries \cite{Gaiotto:2014kfa}
and ordinary symmetries (i.e.~0-form symmetries).
Expansion of the M-theory 3-form
onto  non-trivial
cohomology classes of $M_6$  yields a collection of
external $\mathrm U(1)$ $p$-form gauge fields (with $p$ ranging from 0 to 3)
that enters the 5d
low-energy effective action of M-theory reduced on $M_6$. 
Following the general recipe of \cite{Bah:2020uev}, we analyze the topological mass terms
for these fields in the 5d low-energy effective action,
thereby identifying which $\mathrm U(1)$ gauge symmetries in 5d are spontaneously broken
to discrete, cyclic subgroups. Our findings 
are summarized in table \ref{field_content}.
Depending on the choice of boundary conditions,
these discrete 5d gauge symmetries correspond to different 
discrete global
symmetries in four dimensions.\footnote{For
a related discussion for 3d ABJM-type models, see \cite{Bergman:2020ifi}.}

The main focus of this work is the study of continuous
symmetries. As a result, we integrate out
the $\mathrm U(1)$ $p$-form gauge fields that are topologically massive
in 5d---and thus correspond to broken $\rm U(1)$'s.
The residual, unbroken symmetries include ordinary symmetries,
as well as ``$(-1)$-form'' symmetries associated with the axionic
fields in table \ref{field_content}, which are best understood in terms of
anomalies in the space of coupling constants \cite{Cordova:2019jnf}.
Notice that a careful bookkeeping of  
 broken symmetries is essential
in order to compute correctly 
the 't Hooft anomalies  for    unbroken   symmetries,
as discussed in \cite{Bah:2019vmq} for the case of $k=2$.

The key ingredient in the computation of the
't Hooft anomalies for continuous  symmetries
is the construction of the class $E_4$, which is the closed, equivariant
completion of the cohomology class describing the background
$G_4$-flux configuration \cite{Bah:2019rgq, Hosseini:2020vgl}. Here, equivariance refers to the
action of the continuous isometries of $M_6$.\footnote{In the process of
carrying out this construction, we verify the absence of cohomological
obstructions~\cite{WU1993381}. The latter would signal the spontaneous breaking 
of some of the continuous symmetries associated with isometries of $M_6$.
See  \cite{Bah:2021mzw,Bah:2021hei} for a realization of this mechanism
in the context of wrapped M5-brane setups.}
We are mainly interested in determining those terms
in the inflow anomaly polynomial that are leading in the limit
of large flux quanta  $N$, $N_{\mathrm N_i}$, $N_{\mathrm S_i}$.
To this end, it is sufficient to 
 integrate $E_4^3$ along the internal $M_6$
directions. This procedure accounts for the effect of the two-derivative
$C_3 \wedge G_4 \wedge G_4$ coupling in the low-energy M-theory effective action. In general,
$I_6^{\rm inflow}$ also receives contributions from the higher-derivative
  coupling $C_3 \wedge X_8$ in the   action---where $X_8$ is a certain combination of Pontryagin classes constructed from the 11d metric, reported in \eqref{X8_def}---but these terms are subleading
  in the limit of large $N$, $N_{\mathrm N_i}$, $N_{\mathrm S_i}$, and fall
beyond the scope of this work.   
(They are studied in the special case~$k=2$ in~\cite{Bah:2019vmq}.)


 The full result of the calculation outlined above (including the contribution of the axionic fields)
is recorded in appendix \ref{full_anomaly_polynomial_appendix}.
Section \ref{subsec_results} contains more compact expressions 
valid in some special cases of interest  (we do not report the axionic terms).
Equation \eqref{full_k_3} gives the full answer
for $k=3$, for arbitrary  values of the flux parameters 
$N$, $N_{\mathrm N_i}$, $N_{\mathrm S_i}$.
In \eqref{all_eq_I6} we record the result for generic $k$,
in the special case in which all the resolution flux quanta 
$N_{\mathrm N_i}$, $N_{\mathrm S_i}$ are equal.

The rest of this paper is organized as follows. In section \ref{geometric_setup_section}
we describe the geometry and flux configuration of the internal space $M_6$
for generic $k \ge 2$, determining useful bases for the relevant (co)homology groups of $M_6$.
Section \ref{sec:symmetries} is devoted to the analysis of the topological mass terms
originating from reduction of 11d supergravity on $M_6$. In section \ref{sec:inflow}
we present the computation of the inflow anomaly polynomial for   continuous
symmetries, and record the results of this calculation.
We conclude with a brief discussion in section \ref{sec_conclusion}. Several appendices collect useful technical material.

\section{Geometric setup}\label{geometric_setup_section}

We discuss in this section the salient topological features of the 6d internal space that is described by the fiber bundle
\begin{equation}
	M_4 \hookrightarrow M_6 \rightarrow \Sigma_g \, ,
\end{equation}
with the fiber $M_4$ being the manifold obtained by resolving the fixed points of the orbifold $S^4/\mathbb{Z}_k$, i.e.
\begin{equation}
	M_4 = [S^4/\mathbb{Z}_k]_\mathrm{resolved}
\end{equation}
for $k \geq 2$, and $\Sigma_g$ is a higher-genus Riemann surface with $g \geq 2$.

\subsection{Orbifold action and resolution of singularities}

Under the orbifold action of $\mathbb{Z}_k$, the isometry group $\mathrm{SO}(5) \supset \mathrm{SO}(4) \cong \mathrm{SU}(2)_L \times \mathrm{SU}(2)_R$ of the four-sphere $S^4$ is reduced to a $\mathrm{U}(1)_L \times \mathrm{SU}(2)_R$ subgroup.\footnote{Note that there is an enhanced symmetry for $k=2$ where we still have $\mathrm{SU}(2)_L \times \mathrm{SU}(2)_R$ as the isometry group \cite{Bah:2019vmq}.} With a slight abuse of notation, $\mathrm{SU}(2)_R$ is identified as the R-symmetry of the theory, and $\mathrm{U}(1)_L$ is identified as a flavor symmetry. For the purpose of illustrating the topology of the orbifold, we can write the metric of $S^4/\mathbb{Z}_k$ as
\begin{gather}
	ds^2(S^4/\mathbb{Z}_k) = d\eta^2 + \sin^2 \eta \bigg[\frac{1}{k^2} \, D\varphi^2 + \frac{1}{4} \, ds^2(S^2_\psi)\bigg] \, ,\label{S4_Zk_metric}\\
		\ D\varphi = d\varphi + \frac{k}{2} \cos{\theta} \, d\psi \, , \quad ds^2(S^2_\psi) = d\theta^2 + \sin^2 \theta \, d\psi^2 \, ,\label{trivial_connection_forms}
\end{gather}
where $\eta,\theta \in [0,\pi]$, while the angular coordinates $\varphi$ and $\psi$ both have periodicities of $2\pi$, thus allowing us to identify $\mathrm{U}(1)_L = \mathrm{U}(1)_\varphi$ and $\mathrm{SU}(2)_R = \mathrm{SU}(2)_\psi$. We make a few remarks regarding the metric shown above. Firstly, the angle $\psi$ is the usual azimuthal angle of $S^2_\psi$ where the circle $S^1_\psi$ vanishes at $\theta = 0,\pi$. Secondly, the latter term of the metric \eqref{S4_Zk_metric} is written, up to the factor of $1/k^2$ due to the orbifold action, as a Hopf fibration of $S^3$ over $S^2_\psi$ with fiber $S^1_\varphi$. It should also be noted that there exist two orbifold fixed points at $\eta = 0,\pi$ which are locally (charge-$k$) single-center Taub-NUT spaces.

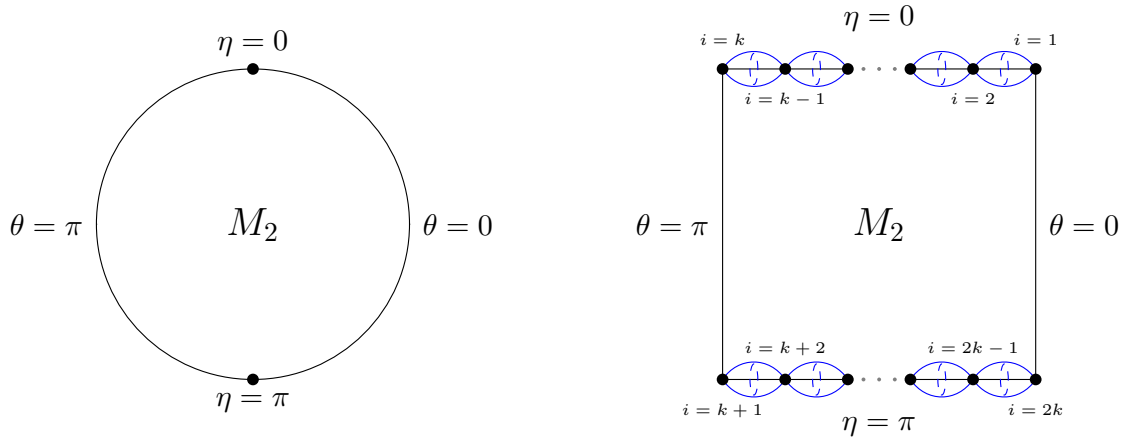
\begin{figure}[t!]
	\centering
		\resizebox{\textwidth}{!}{
		\begin{tikzpicture}
			\draw (2,2) circle (2);
			\fill (2,0) circle (0.075) node[below=1] {$\eta = \pi$};
			\fill (2,4) circle (0.075) node[above=1] {$\eta = 0$};
			\fill (4,2) circle (0) node[right=1] {$\theta = 0$};
			\fill (0,2) circle (0) node[left=1] {$\theta = \pi$};
			\node (S4Zk) at (2,2) {\Large $M_2$};
			\draw (8,0) -- (9.6,0);
			\fill[black!50] (9.8,0) circle (0.025);
			\fill[black!50] (10,0) circle (0.025) node[color=black,below=10] {$\eta = \pi$};
			\fill[black!50] (10.2,0) circle (0.025);
			\draw (10.4,0) -- (12,0);
			\draw (8,4) -- (9.6,4);
			\fill[black!50] (9.8,4) circle (0.025);
			\fill[black!50] (10,4) circle (0.025) node[color=black,above=10] {$\eta = 0$};
			\fill[black!50] (10.2,4) circle (0.025);
			\draw (10.4,4) -- (12,4);
			\draw (8,0) -- node[left=1] {$\theta = \pi$} (8,4);
			\draw (12,0) -- node[right=1] {$\theta = 0$} (12,4);
			\node (M4) at (10,2) {\Large $M_2$};
			\draw[color=blue] (8,0) .. controls (8.2,0.3) and (8.6,0.3) .. (8.8,0);
			\draw[color=blue] (8,0) .. controls (8.2,-0.3) and (8.6,-0.3) .. (8.8,0);
			\draw[dashed,color=blue] (8.4,0) ellipse (0.05 and 0.2);
			\draw[color=blue] (8.8,0) .. controls (9,0.3) and (9.4,0.3) .. (9.6,0);
			\draw[color=blue] (8.8,0) .. controls (9,-0.3) and (9.4,-0.3) .. (9.6,0);
			\draw[dashed,color=blue] (9.2,0) ellipse (0.05 and 0.2);
			\draw[color=blue] (10.4,0) .. controls (10.6,0.3) and (11,0.3) .. (11.2,0);
			\draw[color=blue] (10.4,0) .. controls (10.6,-0.3) and (11,-0.3) .. (11.2,0);
			\draw[dashed,color=blue] (10.8,0) ellipse (0.05 and 0.2);
			\draw[color=blue] (11.2,0) .. controls (11.4,0.3) and (11.8,0.3) .. (12,0);
			\draw[color=blue] (11.2,0) .. controls (11.4,-0.3) and (11.8,-0.3) .. (12,0);
			\draw[dashed,color=blue] (11.6,0) ellipse (0.05 and 0.2);
			\draw[color=blue] (8,4) .. controls (8.2,4.3) and (8.6,4.3) .. (8.8,4);
			\draw[color=blue] (8,4) .. controls (8.2,3.7) and (8.6,3.7) .. (8.8,4);
			\draw[dashed,color=blue] (8.4,4) ellipse (0.05 and 0.2);
			\draw[color=blue] (8.8,4) .. controls (9,4.3) and (9.4,4.3) .. (9.6,4);
			\draw[color=blue] (8.8,4) .. controls (9,3.7) and (9.4,3.7) .. (9.6,4);
			\draw[dashed,color=blue] (9.2,4) ellipse (0.05 and 0.2);
			\draw[color=blue] (10.4,4) .. controls (10.6,4.3) and (11,4.3) .. (11.2,4);
			\draw[color=blue] (10.4,4) .. controls (10.6,3.7) and (11,3.7) .. (11.2,4);
			\draw[dashed,color=blue] (10.8,4) ellipse (0.05 and 0.2);
			\draw[color=blue] (11.2,4) .. controls (11.4,4.3) and (11.8,4.3) .. (12,4);
			\draw[color=blue] (11.2,4) .. controls (11.4,3.7) and (11.8,3.7) .. (12,4);
			\draw[dashed,color=blue] (11.6,4) ellipse (0.05 and 0.2);
			\fill (8,0) circle (0.075) node[below=5] {\tiny $i=k+1$};
			\fill (8.8,0) circle (0.075) node[above=5] {\tiny $i=k+2$};
			\fill (9.6,0) circle (0.075);
			\fill (10.4,0) circle (0.075);
			\fill (11.2,0) circle (0.075) node[above=5] {\tiny $i=2k-1$};
			\fill (12,0) circle (0.075) node[below=5] {\tiny $i=2k$};
			\fill (8,4) circle (0.075) node[above=5] {\tiny $i=k$};
			\fill (8.8,4) circle (0.075) node[below=5] {\tiny $i=k-1$};
			\fill (9.6,4) circle (0.075);
			\fill (10.4,4) circle (0.075);
			\fill (11.2,4) circle (0.075) node[below=5] {\tiny $i=2$};
			\fill (12,4) circle (0.075) node[above=5] {\tiny $i=1$};
		\end{tikzpicture}
		}
	\caption{Illustration of the topology of $M_4$, with the $\psi$ and $\varphi$ angles suppressed, before ({\it left}) and after ({\it right}) the resolution of the orbifold singularities at $\eta = 0,\pi$. We may roughly think of the coordinates $\eta$ and $\theta$ as the 2d ``latitude'' and ``longitude'' respectively. The circle $S^1_\psi$ vanishes along the entire boundary $\partial M_2$, whereas $S^1_\varphi$ vanishes only at the monopoles, which are labeled by the index $i = 1,\dots,2k$. The blue bubbles overlaid on $\partial M_2$ depict the resolution two-cycles connected with unit-charge Kaluza-Klein monopoles after the orbifold singularities are blown up.}
	\label{base_space_illustration}
\end{figure}

We analyze in this work the scenario where the two orbifold singularities are resolved through blow-up. More details of this procedure are recorded in appendix \ref{orbifold_singularity_resolution_appendix}. The resultant resolved manifold, $M_4$, is locally a multi-center Gibbons-Hawking space with $k-1$ aligned two-cycles separated by $k$ (unit-charge) Kaluza-Klein monopoles at $\eta = 0,\pi$ respectively, hence reducing the $\mathrm{SU}(2)_\psi$ isometry into its $\mathrm{U}(1)_\psi$ subgroup. Accordingly, the fiber $M_4$ stands on its own as the fiber bundle
\begin{equation}
	S^1_\varphi \hookrightarrow M_4 \rightarrow S^1_\psi \times M_2 \, ,
\end{equation}
where $M_2$ is the compact 2d space spanned by $\eta$ and $\theta$, whose boundary $\partial M_2$ is described by the four intervals with $\eta,\theta = 0,\pi$. An illustration of the topology of $M_4$ before and after the resolution of the orbifold singularities is provided in figure \ref{base_space_illustration}. Moreover, a metric on $M_4$ can be cast into the schematic form,
\begin{gather}
	ds^2(M_4) = ds^2(M_2) + R_\psi^2(\eta,\theta) d\psi^2 + R_\varphi^2(\eta,\theta) D\varphi^2 \, ,\label{M4_metric}\\
	D\varphi = d\varphi - L(\eta,\theta) d\psi \, .
\end{gather}
The function $L(\eta,\theta)$ encodes information about the fibration in $M_4$ whose significance will be discussed shortly. The functions $R_\psi(\eta,\theta)$ and $R_\varphi(\eta,\theta)$ parameterize respectively the radii of the circles $S^1_\psi$ and $S^1_\varphi$ with respect to the position on $M_2$. In particular, $R_\psi(\eta,\theta)$ vanishes everywhere on the boundary of $M_2$, whereas $R_\varphi(\eta,\theta)$ is nonvanishing everywhere on $M_2$ except at the positions of the $2k$ Kaluza-Klein monopoles on $\partial M_2$.

It is instructive to use a single periodic parameter, $t \sim t + 1$, to parameterize the boundary $\partial M_2$. The locations of the monopoles, $t_i$ for $i=1,\dots,2k$, further divides $\partial M_2$ into distinct intervals, so that
\begin{equation}
	\partial M_2 = \bigcup_{i=1}^{2k} \, [t_i,t_{i+1}] \, .
\end{equation}
Without loss of generality, we pick a convention that $t_1$ corresponds to $(\eta,\theta)=(0,0)$ and $t_{2k}$ corresponds to $(\eta,\theta)=(\pi,0)$. As derived in appendix \ref{orbifold_singularity_resolution_appendix}, $L(\eta,\theta)$ is a piecewise constant function on $\partial M_2$, described by
\begin{equation}
	\ell_i \equiv L(t_i < t < t_{i+1}) = \begin{cases} \displaystyle i - \frac{k}{2} & \mathrm{if} \ 1 \leq i \leq k \, ,\\[2ex] \displaystyle \frac{3k}{2} - i & \mathrm{if} \ k + 1 \leq i \leq 2k \, .\end{cases}\label{l_i_expression}
\end{equation}
The difference $n_i = \ell_i - \ell_{i-1}$ measures the Kaluza-Klein charge of the $i$-th monopole. In particular, the charge of each monopole for $1 \leq i \leq k$ is $+1$, while for $k+1 \leq i \leq 2k$ the charge is $-1$, with the relative sign accounting for the opposite orientations relative to $M_2$ at $\eta = 0,\pi$. The sum of the charges along $\eta=0$ and $\eta=\pi$ is equal to $k$ and $-k$ respectively as expected. With each resolution two-cycle, there is an associated $\mathrm{U}(1)$ gauge symmetry, so there is an overall $\mathrm{U}(1)^{k-1}$ symmetry at $\eta = 0$ and similarly at $\eta = \pi$.

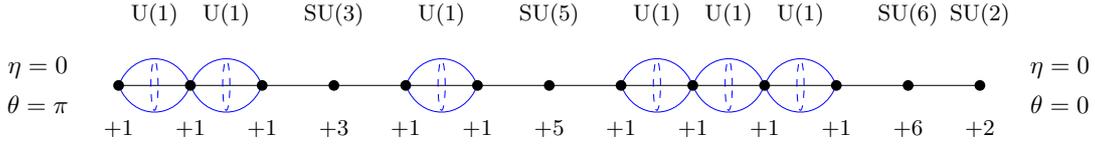
\begin{figure}[t!]
	\centering
		\resizebox{\textwidth}{!}{
		\begin{tikzpicture}
			\draw node[left=10] {\begin{tabular}{c} \small $\eta = 0$ \\[0.25ex] \small $\theta = \pi$ \end{tabular}} (0,0) -- (12,0) node[right=10] {\begin{tabular}{c} \small $\eta = 0$ \\[0.25ex] \small $\theta = 0$ \end{tabular}};
			\draw[color=blue] (0,0) .. controls (0.2,0.5) and (0.8,0.5) .. (1,0);
			\draw[color=blue] (0,0) .. controls (0.2,-0.5) and (0.8,-0.5) .. (1,0);
			\draw[dashed,color=blue] (0.5,0) ellipse (0.05 and 0.35) node[above=20,black] {\footnotesize $\mathrm{U}(1)$};
			\fill (0,0) circle (0.075) node[below=10] {\footnotesize $+1$};
			\draw[color=blue] (1,0) .. controls (1.2,0.5) and (1.8,0.5) .. (2,0);
			\draw[color=blue] (1,0) .. controls (1.2,-0.5) and (1.8,-0.5) .. (2,0);
			\draw[dashed,color=blue] (1.5,0) ellipse (0.05 and 0.35) node[above=20,black] {\footnotesize $\mathrm{U}(1)$};
			\fill (1,0) circle (0.075) node[below=10] {\footnotesize $+1$};
			\fill (2,0) circle (0.075) node[below=10] {\footnotesize $+1$};
			\fill (3,0) circle (0.075) node[below=10] {\footnotesize $+3$} node[above=20] {\footnotesize $\mathrm{SU}(3)$};
			\draw[color=blue] (4,0) .. controls (4.2,0.5) and (4.8,0.5) .. (5,0);
			\draw[color=blue] (4,0) .. controls (4.2,-0.5) and (4.8,-0.5) .. (5,0);
			\draw[dashed,color=blue] (4.5,0) ellipse (0.05 and 0.35) node[above=20,black] {\footnotesize $\mathrm{U}(1)$};
			\fill (4,0) circle (0.075) node[below=10] {\footnotesize $+1$};
			\fill (5,0) circle (0.075) node[below=10] {\footnotesize $+1$};
			\fill (6,0) circle (0.075) node[below=10] {\footnotesize $+5$} node[above=20] {\footnotesize $\mathrm{SU}(5)$};
			\draw[color=blue] (7,0) .. controls (7.2,0.5) and (7.8,0.5) .. (8,0);
			\draw[color=blue] (7,0) .. controls (7.2,-0.5) and (7.8,-0.5) .. (8,0);
			\draw[dashed,color=blue] (7.5,0) ellipse (0.05 and 0.35) node[above=20,black] {\footnotesize $\mathrm{U}(1)$};
			\fill (7,0) circle (0.075) node[below=10] {\footnotesize $+1$};
			\draw[color=blue] (8,0) .. controls (8.2,0.5) and (8.8,0.5) .. (9,0);
			\draw[color=blue] (8,0) .. controls (8.2,-0.5) and (8.8,-0.5) .. (9,0);
			\draw[dashed,color=blue] (8.5,0) ellipse (0.05 and 0.35) node[above=20,black] {\footnotesize $\mathrm{U}(1)$};
			\fill (8,0) circle (0.075) node[below=10] {\footnotesize $+1$};
			\draw[color=blue] (9,0) .. controls (9.2,0.5) and (9.8,0.5) .. (10,0);
			\draw[color=blue] (9,0) .. controls (9.2,-0.5) and (9.8,-0.5) .. (10,0);
			\draw[dashed,color=blue] (9.5,0) ellipse (0.05 and 0.35) node[above=20,black] {\footnotesize $\mathrm{U}(1)$};
			\fill (9,0) circle (0.075) node[below=10] {\footnotesize $+1$};
			\fill (10,0) circle (0.075) node[below=10] {\footnotesize $+1$};
			\fill (11,0) circle (0.075) node[below=10] {\footnotesize $+6$} node[above=20] {\footnotesize $\mathrm{SU}(6)$};
			\fill (12,0) circle (0.075) node[below=10] {\footnotesize $+2$} node[above=20] {\footnotesize $\mathrm{SU}(2)$};
		\end{tikzpicture}
		}
	\caption{An example of a partially resolved orbifold fixed point for $k = 25$. Here we only show the interval corresponding to $\eta = 0$. (There is an analogous situation at $\eta = \pi$ with a generally different charge configuration.) Each number denotes the charge of the monopole above it, while each blue bubble represents a resolution two-cycle sandwiched between two unit-charge monopoles. Note that there is no two-cycle adjacent to any monopole whose charge is greater than one. There is a $\mathrm{U}(1)$ gauge symmetry associated with each two-cycle, and an enhanced $\mathrm{SU}(|n_i|)$ gauge symmetry associated with each monopole with charge $1 \leq |n_i| \leq k$.}
	\label{partially_resolved_setups_example}
\end{figure}

Following the same line of argument, it would be intriguing to generalize our results to partially resolved orbifold fixed points where there is a reduced number of monopoles but with generally non-unit charges, provided that (with the appropriate orientations) they sum up to $k$ along each of the two boundary intervals at $\eta = 0,\pi$, i.e.
\begin{equation}
	\sum_{i} n_i = \begin{cases} +k & \mathrm{at} \ \eta = 0 \ \mathrm{with} \ 1 \leq n_i \leq k \, ,\\ -k & \mathrm{at} \ \eta = \pi \ \mathrm{with} \ -k \leq n_i \leq -1 \, .\end{cases}
\end{equation}
For each $|n_i| \geq 2$, a $\mathrm{U}(1)^{|n_i|+1}$ subgroup in the fully resolved setup enhances to an $\mathrm{SU}(|n_i|)$ gauge symmetry \cite{Sen:1997kz}. Geometrically, this corresponds to collapsing a pair of adjacent resolution two-cycles into a charge-$|n_i|$ monopole. We show an example of such a partially resolved setup in figure \ref{partially_resolved_setups_example} to illustrate the enhancement of gauge symmetries. A further analysis of these general scenarios is left to future work.

\subsection{Topological twists}

The twisting of $M_4$ over the Riemann surface introduces $\mathrm{U}(1)$ connections over $\Sigma_g$ to the global angular forms $d\psi$ and $D\varphi$ originally defined on $M_4$. Specifically, we preserve 4d $\mathcal{N} = 1$ supersymmetry by performing the following topological twist \cite{Bah:2011vv,Bah:2012dg},
\begin{equation}\label{top_twist}
	d\psi \to D\psi = d\psi - 2 \pi \chi A_1^\Sigma \, ,
\end{equation}
where $\chi = 2 - 2g$ is the Euler characteristic of $\Sigma_g$, and $A_1^\Sigma$ is the local antiderivative of the normalized volume form $V_2^\Sigma$ as defined by
\begin{equation}
	\int_{\Sigma_g} V_2^\Sigma = 1 \, .
\end{equation}
Similarly, we may also promote
\begin{equation}\label{flav_twist}
	d\varphi \to D\varphi = d\varphi - L D\psi - 2 \pi \zeta A_1^\Sigma
\end{equation}
for some integer flavor twist parameter $\zeta$. As we will soon see, the topological twist of the $\mathrm{U}(1)_\psi \times \mathrm{U}(1)_\varphi$ isometry group over the Riemann surface leads to non-trivial relations in the homology of $M_6$.

\subsection{Homology of $M_6$ and sum rules}

The physics of the 5d supergravity theory obtained by reducing M-theory on the 6d internal space, which we will explore in more depth in the next section, depends crucially on the topology of $M_6$. Therefore, it is important for us to understand the homology of $M_6$.

\paragraph{One-cycles}

Based on the earlier discussion, we observe that there is no one-cycle in $M_4$, so the only one-cycles present in $M_6$ are those associated with the Riemann surface $\Sigma_g$. There are in total $2g$ one-cycles in $M_6$ that are formally the pullbacks (via the dual cohomology) of the standard $\mathcal{A}$ and $\mathcal{B}$ cycles of the Riemann surface, which we denote as
\begin{equation}
	\mathcal{C}_1^{\Sigma,u} \in \big\{\mathcal{C}_1^{\mathcal{A},p}, \mathcal{C}_1^{\mathcal{B},p}\big\} \ ,
\end{equation}
where $u = 1,\dots,2g$ and $p = 1,\dots,g$. The intersection pairing between the $\mathcal{A}$ and $\mathcal{B}$ cycles is as usual given by
\begin{equation}
	\big\langle\mathcal{C}_1^{\mathcal{A},p},\mathcal{C}_1^{\mathcal{A},q}\big\rangle = \big\langle\mathcal{C}_1^{\mathcal{B},p},\mathcal{C}_1^{\mathcal{B},q}\big\rangle = 0 \, , \quad \big\langle\mathcal{C}_1^{\mathcal{A},p},\mathcal{C}_1^{\mathcal{B},q}\big\rangle = -\big\langle\mathcal{C}_1^{\mathcal{B},q},\mathcal{C}_1^{\mathcal{A},p}\big\rangle = \delta^{pq} \, ,
\end{equation}
which can be compactly expressed by the following intersection matrices \cite{Bah:2020uev},
\begin{equation}\label{one-cycle_intersection}
	\mathcal{K}^{uv} = \begin{pmatrix} 0 & \delta^{pq} \\ -\delta^{pq} & 0 \end{pmatrix} \, , \quad \mathcal{K}_{uv} = \begin{pmatrix} 0 & \delta_{pq} \\ -\delta_{pq} & 0 \end{pmatrix} \, ,
\end{equation}
the latter being constructed such that $\mathcal{K}^{uv} \mathcal{K}_{vw} = -\delta^u_w$\,. To complete the discussion of one-cycles, we note that there exist dual harmonic one-forms $\lambda_{1,v}$ in de Rham cohomology that can be chosen to be orthonormal to the one-cycles $\mathcal{C}_1^{\Sigma,u}$, i.e.\footnote{In the remainder of this paper, any discussion invoking the use of cohomology groups is implicitly understood to be within the context of de Rham cohomology unless otherwise specified.}
\begin{equation}
	\int_{\mathcal{C}_1^{\Sigma,u}} \lambda_{1,v} = \delta^u_v \, .
\end{equation}

\paragraph{Two-cycles}

The Riemann surface $\Sigma_g$ is a two-cycle on its own, but when pulled back to $M_6$ it remains to be a bona fide two-cycle only at the positions of the monopoles where $S^1_\psi$ and $S^1_\varphi$ vanish simultaneously, thus yielding a set of $2k$ two-cycles in $M_6$ which we denote schematically as
\begin{equation}
	\mathcal{C}_2^{\Sigma,i} = \Sigma_g|_{t=t_i}\label{Sigma_two_cycles}
\end{equation}
for $i=1,\dots,2k$. We also have the resolution two-cycles resulting from the blow up of the orbifold singularities of $S^4/\mathbb{Z}_k$ at $\eta=0,\pi$. Each of these resolution two-cycles extends between adjacent monopoles at $t=t_i,t_{i+1}$ on $\partial M_2$, where $S^1_\psi$ vanishes but $S^1_\varphi$ is nonvanishing. By the same token, there exist two other two-cycles stretching between $\eta=0$ and $\eta=\pi$, or in other words, on the intervals $t_k < t < t_{k+1}$ and $t_{2k} < t < t_1$ respectively. In terms of figure \ref{base_space_illustration}, these two-cycles can be visualized as bubbles sitting on the $\theta=0,\pi$ intervals. Overall, this gives us another set of $2k$ two-cycles,\footnote{For the sake of comprehension, we are ignoring the generally non-trivial bundle structures when writing such schematic expressions of cycles as direct products of subspaces.}
\begin{equation}
	\mathcal{C}_2^i = [t_i,t_{i+1}] \times S^1_\varphi \, .\label{resolution_two_cycles}
\end{equation}
We hereafter refer to these collectively as the ``resolution two-cycles.''

The na\"{i}ve counting of the two-cycles listed above tells us that the total number of two-cycles in $M_6$ is $4k$. However, it turns out that the twisting of the $\mathrm{U}(1)_\psi \times \mathrm{U}(1)_\varphi$ bundle trivializes certain linear combinations of these two-cycles. As explained in detail in appendix \ref{cohomology_class_representatives_appendix}, this can be most easily seen by working with the dual de Rham cohomology group, $H^2(M_6) \cong H_2(M_6)$, and we find that the following homological relation is satisfied,
\begin{equation}
	\mathcal{C}_2^{\Sigma,i+1} - \mathcal{C}_2^{\Sigma,i} = (\chi \ell_i - \zeta) \, \mathcal{C}_2^i \, .\label{two_cycles_recurrence_relation}
\end{equation}
We see from \eqref{two_cycles_recurrence_relation} that only one of the $\mathcal{C}_2^{\Sigma,i}$ is independent of the $\mathcal{C}_2^i$. We end up with two overall homological sum rules relating the $2k$ cycles $\mathcal{C}_2^i$,
\begin{equation}
	\sum_{i=1}^{2k} \mathcal{C}_2^i = 0 \, , \qquad \sum_{i=1}^{2k} \ell_i \, \mathcal{C}_2^i = 0 \, .\label{two_cycles_sum_rules}
\end{equation}
Note that the first sum rule can be heuristically understood as the fact that the sum of all the two-cycles $\mathcal{C}_2^i$ forms the boundary of the 3d space composed of $S^1_\varphi$ and $M_2$. The second Betti number can then be computed as $b_2(M_6) = 4k - (2k - 1) - 2 = 2k - 1$, where the former factor of $2k-1$ comes from the recurrence relation \eqref{two_cycles_recurrence_relation}, while the factor of $2$ comes from the two sum rules \eqref{two_cycles_sum_rules}.\footnote{Note that if we were to restrict to $M_4$ only, then we would have $b_2(M_4) = 2k-2$ instead.} For practical purposes, it is convenient to pick some complete basis of $2k-1$ two-cycles, $\mathcal{C}_2^\alpha$, from among those listed in \eqref{Sigma_two_cycles} and \eqref{resolution_two_cycles}, and construct the corresponding dual cohomology class representatives, $\omega_{2,\beta}$, by orthonormalizing their inner products, i.e.
\begin{equation}
	\int_{\mathcal{C}_2^\alpha} \omega_{2,\beta} = \delta^\alpha_\beta \, .
\end{equation}

\paragraph{Three-cycles}

We can pair up one of the resolution two-cycles $\mathcal{C}_2^i$ and one of the Riemann surface one-cycles $\mathcal{C}_1^{\Sigma,u}$ to form a three-cycle in $M_6$,
\begin{equation}
	\mathcal{C}_3^{i,u} = \mathcal{C}_2^i \times \mathcal{C}_1^{\Sigma,u} = [t_i,t_{i+1}] \times S^1_\varphi \times \mathcal{C}_1^{\Sigma,u} \, .\label{three_cycles}
\end{equation}
It is straightforward to check that these are all the non-trivial three-cycles residing in $M_6$. Note in particular that there is no one-cycle in $M_4$ that can be paired with the Riemann surface. Given the structure of \eqref{three_cycles}, it is unsurprising that the two sum rules for the two-cycles are directly inherited, i.e.
\begin{equation}
	\sum_{i=1}^{2k} \mathcal{C}_3^{i,u} = 0 \, , \qquad \sum_{i=1}^{2k} \ell_i \, \mathcal{C}_3^{i,u} = 0\label{three_cycles_sum_rules}
\end{equation}
for each $u = 1,\,\cdots,2g$. These sum rules can be worked out through an analysis resembling that for the two-cycles. The third Betti number is simply given by $b_3(M_6) = (2k)(2g) - 2g(2) = 4g(k - 1)$. As in the previous cases, we can always choose a complete basis of three-cycles $\mathcal{C}_3^x$, with $x = 1,\dots,b_3(M_6)$ being a collective index for $(i,u)$, whose inner product with the dual cohomology class representatives, $\Lambda_{3,y}$, is orthonormal.

\paragraph{Four-cycles}

The obvious candidate for a four-cycle in $M_6$ is just
\begin{equation}
	\mathcal{C}_{4,\mathrm{C}} = M_4 \, .\label{bulk_four_cycle}
\end{equation}
We can also pair up a resolution two-cycle with the Riemann surface, i.e.
\begin{equation}
	\mathcal{C}_{4,i} = \mathcal{C}_2^i \times \Sigma_g = [t_i,t_{i+1}] \times S^1_\varphi \times \Sigma_g \, .\label{resolution_four_cycles}
\end{equation}
Once again, we seek out relations between these four-cycles by working with the dual cohomology group, $H^4(M_6)$. We can repeat essentially the same exercise as before to arrive at the homological relations,
\begin{equation}
	\sum_{i=1}^{2k} \mathcal{C}_{4,i} = \chi \mathcal{C}_{4,\mathrm{C}} \, , \qquad \sum_{i=1}^{2k} \ell_i \, \mathcal{C}_{4,i} = \zeta \mathcal{C}_{4,\mathrm{C}} \, .\label{four_cycles_sum_rules}
\end{equation}
Unlike the cases we saw earlier, the right-hand sides of the relations above do not vanish; they are equal to $M_4$ up to multiplicative factors of the topological twists. These two sum rules are a manifestation of the non-trivial structure of the bundle $M_4 \hookrightarrow M_6 \rightarrow \Sigma_g$. For example, because of the twist associated with the $\mathrm{U}(1)_\psi$ bundle, the sum of all $\mathcal{C}_{4,i}$ does not form the boundary of any manifold as one might intuitively expect, reflecting the fact that $M_6$ is not simply a product manifold. In fact, the two topological twists trivialize different linear combinations of four-cycles (or four-forms, in cohomology). The vanishing of the RHS of \eqref{two_cycles_sum_rules} and \eqref{three_cycles_sum_rules}, on the other hand, can be attributed to the absence of any two-cycle or three-cycle in $M_4$ playing the role of $\mathcal{C}_{4,\mathrm{C}}$ in \eqref{four_cycles_sum_rules}.
As expected from Poincar\'{e} duality, the fourth Betti number is $b_4(M_6) = 2k - 2 + 1 = 2k - 1 = b_2(M_6)$. We again note that a set of basis four-cycles $\mathcal{C}_4^\alpha$ and dual cohomology class representatives $\Omega_4^\beta$ can be suitably chosen such that they are orthonormal to one another.

Here we would like to make a detour and mention that one can define the flux quanta,
\begin{equation}
	N = \int_{\mathcal{C}_{4,\mathrm{C}}} \frac{G_4}{2\pi} \, , \qquad N_i = \int_{\mathcal{C}_{4,i}} \frac{G_4}{2\pi} \, ,
\end{equation}
where $G_4$ is the M-theory four-form flux. They obey sum rules analogous to \eqref{four_cycles_sum_rules}, i.e.
\begin{equation}
	\sum_{i=1}^{2k} N_i = \chi N \, , \qquad \sum_{i=1}^{2k} \ell_i N_i = \zeta N \, ,\label{flux_quanta_sum_rules}
\end{equation}
which implies that there are only $2k-1$ independent flux quanta characterizing $G_4$. 

\paragraph{Five-cycles}

Similarly to the discussion of three-cycles, the only five-cycles in $M_6$ are
\begin{equation}
	\mathcal{C}_5^u = \mathcal{C}_{4,\mathrm{C}} \times \mathcal{C}_1^{\Sigma,u} \, ,
\end{equation}
where $u = 1,\dots,b_5(M_6)$ with $b_5(M_6) = b_1(M_6) = 2g$ by virtue of Poincar\'{e} duality.

\begin{table}[t!]
	\centering
	\begin{tabular}{|c||cccccccc|} 
		\hline
		& $b_0$ & $b_1$ & $b_2$ & $b_3$ & $b_4$ & $b_5$ & $b_6$ & $\chi$\\
		\hline
		\hline
		$\Sigma_g$ & $1$ & $2g$ & $1$ & - & - & - & - & $2(1-g)$\\
		$M_4$ & $1$ & $0$ & $2k-2$ & $0$ & $1$ & - & - & $2k$\\
		$M_6$ & $1$ & $2g$ & $2k-1$ & $4g(k-1)$ & $2k-1$ & $2g$ & $1$ & $4k(1-g)$\\
		\hline
	\end{tabular}
	\caption{Betti numbers and Euler characteristics of $\Sigma_g$, $M_4$, $M_6$.}
	\label{Betti_numbers_Euler_characteristics_table}
\end{table}

The Betti numbers and the Euler characteristics of $\Sigma_g$, $M_4$, $M_6$ are tabulated in table \ref{Betti_numbers_Euler_characteristics_table}.\footnote{Recall that the Euler characteristic of a manifold $M$ can be computed using the alternating sum, $\chi(M) = \sum_{i=0}^{\mathrm{dim} M} (-1)^i b_i(M)$.} As a sanity check, all the Betti numbers are consistent with Poincar\'{e} duality. It is also explicitly verified that $\chi(M_6) = \chi(\Sigma_g) \, \chi(M_4)$ as expected from the Serre spectral sequence.\footnote{Everywhere else we use the symbol $\chi$ to refer specifically to the Euler characteristic of the Riemann surface $\Sigma_g$.}

\paragraph{Natural basis of (co)homology classes}

We will frequently employ a natural basis of homology classes, and hence cohomology classes, in the rest of this paper. In addition to the standard $\mathcal{A}$ and $\mathcal{B}$ cycles of the Riemann surface as the obvious choice of basis one-cycles $\mathcal{C}_1^{\Sigma,u}$, we choose the basis four-cycles $\mathcal{C}_{4,\alpha}$ to be
\begin{equation}
	\mathcal{C}_{4,1 \leq \alpha \leq k-1} = \mathcal{C}_{4,1 \leq i \leq k-1} \, , \quad \mathcal{C}_{4,\alpha=k} = \mathcal{C}_{4,\mathrm{C}} \, , \quad \mathcal{C}_{4,k+1 \leq \alpha \leq 2k-1} = \mathcal{C}_{4,k+1 \leq i \leq 2k-1} \, ,\label{natural_four_cycles}
\end{equation}
which are respectively the $k-1$ four-cycles $\mathcal{C}_{4,i}$, as in \eqref{resolution_four_cycles}, in the ``north'' where $\eta = 0$, the four-cycle $\mathcal{C}_{4,\mathrm{C}} = M_4$, and the $k-1$ four-cycles $\mathcal{C}_{4,i}$ in the ``south'' where $\eta = \pi$. As an aside, it is convenient to adopt an intuitive naming convention,
\begin{equation}
	\mathcal{C}_{4,1 \leq i \leq k-1} \equiv \mathcal{C}_{4,\mathrm{N}_i} \, , \qquad \mathcal{C}_{4,k+1 \leq i \leq 2k-1} \equiv \mathcal{C}_{4,\mathrm{S}_{2k-i}} \, .\label{cohomology_class_basis_convention}
\end{equation}
We select the basis two-cycles $\mathcal{C}_2^\alpha$ to be those which are Poincar\'{e}-dual to the basis four-cycles described above, so that their dual cohomology class representatives obey
\begin{equation}
	\int_{M_6} \Omega_4^\alpha \wedge \omega_{2,\beta} = \delta^\alpha_\beta \, .
\end{equation}
Last but not least, we pick the basis three-cycles $\mathcal{C}_3^x$ to be combinations of the basis two-cycles $\mathcal{C}_2^{\alpha \neq k}$ and the basis one-cycles $\mathcal{C}_1^{\Sigma,u}$. 



\section{Continuous and discrete flavor symmetries}
\label{sec:symmetries}
In this section we analyze continuous zero-, one-, and two-form global symmetries for the 4d field theories
of interest, and study their breaking to discrete subgroups. 
More precisely, we identify the global symmetries of the 4d QFTs from the gauge symmetries of 11d supergravity reduced on $M_6$, taking care to account for spontaneous symmetry breaking from topological mass terms in the 5d low-energy effective action. Expansion of the three-form potential $C_3$ on the cohomology classes of $M_6$ provides part of the 5d spectrum of massless gauge fields. Additional gauge fields arise from the isometries of $M_6$. In particular, we can couple the $\mathrm{U}(1)_\psi$, $\mathrm{U}(1)_\varphi$ isometries to a pair of massless, abelian gauge fields $A_1^\psi$, $A_1^\varphi$, respectively. However, these abelian gauge fields will not participate in the St\"uckelberg mechanism of primary interest in this section, so we will turn them off until section \ref{sec:inflow}, and focus for the time being on the cohomology sector of the spectrum. The independence of gauge fields associated with isometries from those in the cohomology sector is given explicitly by \eqref{isometry_cohomology_decoupled_convention}.



The three-form potential $C_3$ gives rise to massless abelian $p$-form gauge fields in the reduction to 5d via the Kaluza-Klein mechanism. These are the fields in one-to-one correspondence with the cohomology classes of $M_6$. 
We can identify them by expanding the variation of $C_3$ away from its background value as
\begin{equation}
	 \frac{\delta C_3}{2\pi} =  \frac{a_0^x}{2\pi} \, \Lambda_{3,x} + \frac{A_1^\alpha}{2\pi} \wedge \omega_{2,\alpha} + \frac{B_2^u}{2\pi} \wedge \lambda_{1,u} + \frac{c_3}{2\pi} \, . \label{C3_Expansion}
\end{equation}
Recall that $\Omega_4^\alpha$, $\Lambda_{3,x}$, $\omega_{2,\alpha}$, and $\lambda_{1,u}$ are the de Rham cohomology class representatives of $M_6$ introduced in section \ref{geometric_setup_section}.  The zero-, one-, two-, and three-forms $a_0^x$, $A_1^\alpha$, $B_2^u$, $c_3$ are dynamical, abelian 5d gauge potentials, with field strengths quantized in units of $2\pi$. 
In terms of these field strengths  $f_1^x = da_0^x$, $F_2^\alpha = dA_1^\alpha$, $H_3^u = dB_2^u$, and $\gamma_4 = dc_3$, we can express the four-form flux as
\begin{equation}\label{G4_Expansion}
	\frac{G_4}{2\pi} = N_\alpha \Omega_4^\alpha + \frac{f_1^x}{2\pi} \wedge \Lambda_{3,x} + \frac{F_2^\alpha}{2\pi} \wedge \omega_{2,\alpha} +  \frac{H_3^u}{2\pi} \wedge \lambda_{1,u} + \frac{\gamma_4}{2\pi} \, .
\end{equation}
The fluxes $N_\alpha$ are also quantized,
\begin{equation}
	\int_{\mathcal{C}_{4,\alpha}} \frac{G_4}{2\pi} = N_\alpha \in \mathbb{Z} \, ,\label{N_alpha_quantization}
\end{equation}
in virtue of $G_4$-flux quantization in M-theory \cite{Witten:1996md}.\footnote{Following the argument in \cite{Bah:2020uev,Hsieh:2020jpj}, we assume that there is no half-integral contribution to the flux quantization condition as far as the setup in this paper in concerned.}

At low energies, some of the continuous, abelian $p$-form gauge symmetries are spontaneously broken to discrete subgroups. In order to see this, we must consider the effects of topological terms in the 5d low-energy effective action. 
The relevant topological term in the 11d low-energy effective action is the Chern-Simons coupling,
\begin{equation}
	S_\mathrm{CS} = -\frac{2\pi}{6} \int_{M_{11}}  \frac{C_3}{2\pi} \frac{G_4}{2\pi}  \frac{G_4}{2\pi} \, .
\end{equation}
Note that here and in the equations to follow we have suppressed wedge products.
Kinetic terms for the 5d gauge fields descend from 11d kinetic terms for $G_4$ via a standard Kaluza-Klein reduction, and likewise topological terms in the 5d low-energy effective action are obtained from reduction of $S_\mathrm{CS}$. These topological terms in the 5d effective action can be expressed in terms of a six-form as
\begin{equation}
S_\mathrm{CS} = 2\pi \int_{\mathcal{M}_5} I_5^{(0)} \, , \qquad dI_5^{(0)} = I_6 \, .
\end{equation}
The six-form $I_6$ can be compactly expressed in terms of intersection numbers,
\begin{equation}
	\begin{gathered}
		\mathcal{K}_{\alpha \beta \gamma} \equiv \int_{M_6} \omega_{2,\alpha} \, \omega_{2,\beta} \, \omega_{2,\gamma} \, , \quad \mathcal{K}_\alpha^\beta \equiv \int_{M_6} \omega_{2,\alpha} \, \Omega_4^\beta \, , \quad \mathcal{K}_{xy} \equiv \int_{M_6} \Lambda_{3,x} \, \Lambda_{3,y} \, ,\\ 
		\mathcal{K}_{uv}^\alpha \equiv \int_{M_6} \lambda_{1,u} \, \lambda_{1,v} \, \Omega_4^\alpha \, , \quad \mathcal{K}_{u \alpha x} \equiv \int_{M_6} \lambda_{1,u} \, \omega_{2,\alpha} \, \Lambda_{3,x} \, .
	\end{gathered}
\end{equation}
With the field strengths $F_2^I$ turned off, $I_6$ can be written as
\begin{align}
	I_6= & -\frac{1}{6}  \, \mathcal{K}_{\alpha \beta \gamma} \, \frac{F_2^\alpha}{2\pi} \frac{F_2^\beta}{2\pi} \frac{F_2^\gamma}{2\pi} 
	-  \mathcal{K}_{u \alpha x}  \, \frac{f_1^x}{2\pi} \frac{F_2^\alpha}{2\pi} \frac{H_3^u}{2\pi} 
 + \frac{1}{2} \, \mathcal{K}_{xy} \frac{f_1^x}{2\pi} \frac{f_1^y}{2\pi} \frac{\gamma_4}{2\pi}
 \nonumber\\
& + \frac{1}{2}  N_\alpha \, \mathcal{K}_{uv}^\alpha \, \frac{H_3^u}{2\pi} \frac{H_3^v}{2\pi} 
 - N_\beta \, \mathcal{K}_\alpha^\beta \, \frac{F_2^\alpha}{2\pi} \frac{\gamma_4}{2\pi}  .\label{E4_cubed}
\end{align}
Note that the last two terms are quadratic in the external field strengths. These are the topological mass terms which spontaneously break one of the $b_2(M_6)$ zero-form symmetries and all $b_1(M_6)$ one-form symmetries to discrete subgroups. To see this more directly, we choose a basis of one-forms with intersection numbers as in \eqref{one-cycle_intersection}, and a Poincar\'{e}-dual set of two- and four-forms, i.e. $\mathcal{K}_\alpha^\beta=\delta_\alpha^\beta$, so that 
\begin{equation}
I_6 \, \supset \, -N \, \frac{\tilde{H}_{3,p}}{2\pi}\frac{H_3^p}{2\pi}- N_\alpha \, \frac{F_2^\alpha}{2\pi} \frac{\gamma_4}{2\pi} \, ,
\end{equation}
where we have split the index $u=1,\dots,2g$ as
\begin{equation}
H_3^u=(H_{3}^p,\tilde{H}_{3,p}), \qquad p=1,\dots,g \, .
\end{equation}
With an appropriate basis rotation one can pick out the single one-form gauge field $\mathcal{A}_1$ which couples to $\gamma_4$ \cite{Bah:2020uev},
\begin{equation}
-N_\alpha F_2^\alpha \gamma_4=-n \, d\mathcal{A}_1 \gamma_4 \, , \qquad n=\text{gcd}(N_\alpha) \, .\label{basis_rotation_discrete_gauge_field}
\end{equation}
We denote the remaining $b_2(M_6)-1$ gauge fields which do not appear in any topological mass term by $\mathcal{A}_1^{\hat{\alpha}}$. Under this choice of basis, the topologically massive contributions to the 5d low-energy effective action can be written as
\begin{equation}
\frac{1}{2 \pi} \int_{\mathcal{M}_5} \left(-n c_3 \, d\mathcal{A}_1-N \tilde{B}_2^i \, dB_{2,i}\right) \, .
\end{equation}
As discussed in \cite{Bah:2020uev}, the gauge fields $\mathcal{A}_1$, $c_3$, and $B_2^i$, $\tilde{B}_{2,i}$ are thus effectively continuum descriptions of discrete gauge fields, with gauge groups $\mathbb{Z}_n$, $\mathbb{Z}_n$, and $\mathbb{Z}_N \times \mathbb{Z}_N$, respectively. The spontaneous symmetry breaking of the original continuous gauge symmetries to these subgroups is governed by a St\"uckelberg mechanism. The resulting 5d gauge symmetry groups are summarized in table \ref{field_content}.

Note that in the limit where the $2k-2$ northern and southern flux quanta $\{N_{\mathrm{N},i},N_{\mathrm{S},i}\}$ (see our convention as defined around \eqref{cohomology_class_basis_convention}) are taken to be zero in \eqref{basis_rotation_discrete_gauge_field}, the integer $n$ simply becomes $N$, the only non-trivial flux parameter. In this case, we see that the self-dual $\mathbb{Z}_N$ two-form symmetry of the 6d SCFT would yield a $\mathbb{Z}_N$ zero-form symmetry and a $\mathbb{Z}_N$ two-form symmetry upon reduction to 4d. By turning on additional flux parameters, these discrete symmetries of the 4d theory are broken to discrete $\mathbb{Z}_n$ zero- and two-form symmetries \cite{Bah:2020uev}. 

The full holographic interpretation of these symmetry groups as global symmetries of a dual 4d SCFT would require a choice of boundary conditions. Different boundary conditions are generally associated with different boundary SCFTs. We refer the reader to \cite{Bah:2020uev} for a detailed discussion of possible scenarios.

\begin{table}[t!]
	\centering
	\def\arraystretch{1.5}
	\begin{tabular}{|c|c|c|}
		\hline
	\textbf{Gauge fields} & \textbf{Multiplicity} & \textbf{5d gauge symmetries} \\
		\hline
		\hline
		$c_3$ & $b_0(M_6)=1$ & $\mathbb{Z}_n$ two-form symmetry
		\\
		\hline
		$B_2^u=(B_2^i,\tilde{B}_{2,i})$ & $b_1(M_6)=2g$ & $(\mathbb{Z}_N \times \mathbb{Z}_N)^g$ one-form symmetry
		\\
		\hline
		$A_1^\alpha=(\mathcal{A}_1,\mathcal{A}_1^{\hat{\alpha}})$ & $b_2(M_6)=2k-1$ & $\mathbb{Z}_n$ and $\mathrm{U}(1)^{2k-2}$ zero-form symmetries
		\\
		\hline
		$a_0^x$ & $b_3(M_6)=4g(k-1)$ & axionic
		\\
		\hline
	\end{tabular}
	\caption{Summary of the 5d $p$-form gauge fields and symmetry groups arising from cohomology classes of the internal space $M_6$. Recall that $n \equiv \text{gcd}(N_\alpha)$.}
	\label{field_content}
\end{table}




\section{Inflow anomaly polynomial}
\label{sec:inflow}


This section is devoted to the computation of the inflow anomaly polynomial for the continuous symmetries of the internal space $M_6$ we have described. The computation will require us to first integrate out
the topologically massive  fields corresponding to discrete symmetries.
For the sake of simplicity, we restrict attention to cases in which the flavor twist parameter $\zeta$ in \eqref{flav_twist} is fixed to zero.

%
%
%
%

\subsection{Anomaly inflow methods for wrapped M5-branes}
\label{sec:inflow_background}
Consider a stack of $N$ M5-branes with worldvolume $W_6$ in a background spacetime $M_{11}$. We are interested in setups in which four dimensions $W_4$ are left external and while the rest of the brane worldvolume directions are wrapped on a smooth, compact Riemann surface. The global symmetries of the field theory defined on the extended spacetime $W_4$ can admit 't Hooft anomalies. These anomalies are fully determined by the fibration
\begin{equation} 
M_4 \hookrightarrow M_{6} \rightarrow \Sigma_g \, ,
\end{equation}
and can be encoded in a six-form anomaly polynomial $I_{6}^\mathrm{QFT}$ using the descent formalism \cite{Freed:1998tg,Harvey:1998bx}. As described in \cite{Bah:2019rgq}, this 't Hooft anomaly polynomial can be computed via anomaly inflow. Since the full M-theory is anomaly-free, the anomaly polynomial $I_6^\mathrm{4d QFT}$ associated with the 4d theory must be exactly canceled by a combination of contributions from the classical anomalous variation of the effective 11d supergravity action and from decoupled modes,
\begin{equation}\label{anomaly_cancellation}
I_6^\mathrm{inflow} + I_6^\mathrm{4d QFT} + I_6^\mathrm{decoupled} = 0 \, .
\end{equation}
This mechanism allows us to access $I_6^\mathrm{4d QFT}$ directly via $-I_6^\mathrm{inflow}$ in the large-$N$ limit, where the contributions from decoupled modes are expected to be subleading.

To identify the inflow contribution $I_6^\mathrm{inflow}$ to the anomaly in 4d, we compute the fiber integral
\begin{equation}\label{I6_integral}
I_6^\mathrm{inflow}=\int_{M_6} \mathcal{I}_{12} \, ,
\end{equation}
where $\mathcal{I}_{12}$ is a characteristic class formally defined on a fiducial 12d space $M_{12}$ such that $\partial M_{12} = M_{11}$, and given explicitly by
\begin{equation}
  \mathcal{I}_{12} =-\frac{1}{6} \, E_4^3-E_4 X_8 \, .\label{I12}
\end{equation}
The $N$ M5-branes act as a magnetic source for the four-form flux, resulting in classical anomalous variation of the 11d effective action related to $\mathcal{I}_{12}$ via descent. This anomalous variation is encoded by the form $E_4$, defined to be $G_4/2\pi$ evaluated near the brane stack under suitable boundary conditions  \cite{Bah:2019vmq,Bah:2019rgq} with the normalization
\begin{equation}
  \int_{M_4} E_4 = N \, .
\end{equation}
$E_4$ is closed, globally defined, and invariant under the symmetries of the 4d theory. The eight-form $X_8$ is given in terms of the first and second Pontryagin classes of the tangent bundle $TM_{11}$ by
\begin{equation} \label{X8_def}
X_8=\frac{1}{192} \left[ p_1^2(TM_{11})-4p_2(TM_{11}) \right] \, .
\end{equation}
Note that the first term in \eqref{I12} scales as $N^3$, while the second term is linear in $N$. Since we are interested here in large-$N$ perturbative anomalies, we will restrict attention to the $E_4^3$ term for the rest of this paper. We now turn to a more detailed discussion of the construction of the form $E_4$.

\subsection{Construction of $E_4$}
In order to compute the $d=4$ inflow anomaly polynomial \eqref{I6_integral}, we must first obtain the globally defined and closed four-form $E_4$. On top of the $p$-form gauge fields introduced in the previous section, $E_4$ also depends on the abelian gauge fields $A_1^\psi$, $A_1^\varphi$ associated with the isometries of $M_6$, with field strengths given by
\begin{equation}
F_2^I = dA_1^I, \qquad I,J \in \{\psi,\varphi\} \, .
\end{equation} 
The construction of $E_4$ follows from the expansion of $G_4/2\pi$ in \eqref{G4_Expansion},
\begin{equation}
	E_4 = N_\alpha \big(\Omega_4^\alpha\big)^\mathrm{eq} + N \, \frac{f_1^x}{2\pi} \wedge (\Lambda_{3,x})^\mathrm{eq} + N \, \frac{F_2^\alpha}{2\pi} \wedge (\omega_{2,\alpha})^\mathrm{eq} + N \, \frac{H_3^u}{2\pi} \wedge (\lambda_{1,u})^\mathrm{eq} + N \, \frac{\gamma_4}{2\pi} \, ,\label{E4_local_expression}
\end{equation}
where all forms shown are defined on the fiducial space $M_{12}$.
Here $f_1^x$, $F_2^\alpha$, $H_3^u$, and $\gamma_4$ represent background field strengths, which we have re-scaled by factors of $N$ so as to make the large-$N$ scaling of the anomaly polynomial explicit. The forms $(\Omega_4^\alpha)^\mathrm{eq}$, $(\Lambda_{3,x})^\mathrm{eq}$, $(\omega_{2,\alpha})^\mathrm{eq}$, and $(\lambda_{1,u})^\mathrm{eq}$ are extensions of the de Rham cohomology class representatives $\Omega_4^\alpha$, $\Lambda_{3,x}$, $\omega_{2,\alpha}$, and $\lambda_{1,u}$ on $M_6$ to the full $M_{12}$. These forms can be constructed by first gauging the representative of a given de Rham cohomology class with respect to the isometries, and then constructing a closed and globally defined completion. Suppressing indices labeling individual cohomology classes, these completions are of the forms
\begin{equation}
  \begin{aligned}
	\lambda_1^\mathrm{eq} & = \lambda_1^\mathrm{g} \, ,\\
	\omega_2^\mathrm{eq} & = \omega_2^\mathrm{g} + \frac{F_2^I}{2\pi} \, \omega_{0,I} \, ,\\
	\Lambda_3^\mathrm{eq} & = \Lambda_3^\mathrm{g} + \frac{F_2^I}{2\pi} \, \Lambda_{1,I}^\mathrm{g} \, ,\\
	\Omega_4^\mathrm{eq} & = \Omega_4^\mathrm{g} + \frac{F_2^I}{2\pi} \, \Omega_{2,I}^\mathrm{g} + \frac{F_2^I}{2\pi} \frac{F_2^J}{2\pi} \, \Omega_{0,IJ} \, .
  \end{aligned}\label{Completions}
\end{equation}
where the label ``$\mathrm{eq}$'' stands for ``equivariant,'' while ``$\mathrm{g}$'' stands for ``gauged.'' The details about the parameterization of the auxiliary forms $\omega_{0,I}$, $\Lambda_{1,I}$, $\Omega_{2,I}$, and $\Omega_{0,IJ}$ are left to appendix \ref{cohomology_class_representatives_appendix}. Several ambiguities in the choice of such forms, and the corresponding field redefinitions that keep $I_6^\mathrm{inflow}$ invariant, are discussed in appendix \ref{basis_independence_appendix}.

\subsection{Integrating out massive fields}
With \eqref{E4_local_expression} in hand, the $E_4^3$ term in \eqref{I12} can be expanded as in \eqref{E4_cubed_expansion}. However, in order to study the perturbative anomalies for continuous global symmetries in 4d, we must first integrate out the topologically massive 5d fields. In particular, as discussed in section \ref{sec:symmetries}, the fields $\mathcal{A}_1$, $c_3$, and $B_2^i$, $\tilde{B}_{2,i}$ correspond to topologically massive gauge fields in 5d, and thus cannot be interpreted as background fields for continuous symmetries in the 4d theory. All topological mass terms quadratic in external field strengths can be eliminated from the polynomial \eqref{E4_cubed_expansion} using the equations of motion for the 5d fields $c_3$ and $B_2^u$, respectively,
\begin{gather}
	N_\alpha N_\beta \, \mathcal{J}_I^{\alpha\beta} \, \frac{F_2^I}{2\pi} + N N_\beta \, \mathcal{K}_\alpha^\beta \, \frac{F_2^\alpha}{2\pi} - \frac{1}{2} \, N^2 \, \mathcal{K}_{xy} \, \frac{f_1^x}{2\pi} \frac{f_1^y}{2\pi} = 0 \, ,\label{c3_equation_of_motion}\\
	N_\alpha \, \mathcal{J}_{Iux}^\alpha \, \frac{f_1^x}{2\pi} \frac{F_2^I}{2\pi}  + \, \frac{1}{2} \, N_\alpha \, \mathcal{K}_{uv}^\alpha \, \frac{H_3^v}{2\pi} \,  - N \, \mathcal{K}_{u \alpha x} \, \frac{f_1^x}{2\pi} \frac{F_2^\alpha}{2\pi} = 0 \, ,\label{B2_equation_of_motion}
\end{gather} 
where we have defined the integrals
\begin{gather}
	\mathcal{J}_I^{\alpha\beta} \equiv \frac{1}{2} \int_{M_6} \! \Big(\Omega_{2,I}^\alpha \, \Omega_4^\beta + \Omega_{2,I}^\beta \, \Omega_4^\alpha\Big) \, , \quad \mathcal{J}_{Iux}^\alpha \equiv \int_{M_6} \! \Big(\Lambda_{1,xI} \, \lambda_{1,u} \, \Omega_4^\alpha - \lambda_{1,u} \, \Omega_{2,I}^\alpha \, \Lambda_{3,x}\Big).
\end{gather}
The forms $\Omega_{2,I}^\alpha$ and $\Lambda_{1,xI}$ introduced in constructing the closed completions $(\Omega_4^\alpha)^\mathrm{eq}$ and $(\Lambda_{3,x})^\mathrm{eq}$ are defined only up to the addition of harmonic forms.  As a result, the  mixing between field strengths associated with isometries and those associated with the cohomology classes of $M_6$ implied by \eqref{c3_equation_of_motion} and \eqref{B2_equation_of_motion} can be removed by an appropriate choice of $\Omega_{2,I}^\alpha$ and $\Lambda_{1,xI}$ that fixes
\begin{equation}
	N_\alpha N_\beta \, \mathcal{J}_I^{\alpha\beta}  =0 \, ,\qquad N_\alpha \, \mathcal{J}_{Iux}^\alpha = 0  \label{isometry_cohomology_decoupled_convention}
\end{equation}
for all $I$ and $x$.
A detailed demonstration is provided in appendix \ref{cohomology_class_representatives_appendix}.
Under this condition, all dependence on the field strengths $F_2^I$ associated with the isometries of $M_6$ drops out of the equations of motion \eqref{c3_equation_of_motion} and \eqref{B2_equation_of_motion}.
Therefore we can safely restrict to the cohomology sector when integrating out the topologically massive fields. The St\"uckelberg mechanism governing the spontaneous symmetry breaking to the discrete subgroups described in section \ref{sec:symmetries} is thus unaffected by the presence of gauge fields coupled to the isometries of $M_6$.

\subsection{Results for continuous symmetries} \label{subsec_results}
After integrating out topologically massive fields from \eqref{E4_cubed_expansion}, we can compute the inflow anomaly polynomial for continuous symmetries.
In appendix \ref{full_anomaly_polynomial_appendix}, we record the full anomaly polynomial $I_6^\textrm{inflow}$ in the large-$N$ limit for $\chi <0$ and $\zeta=0$, including all continuous higher-form symmetries. We stress that $I_6^\mathrm{inflow}$ encodes the perturbative anomalies associated with continuous, but not discrete symmetries of the total worldvolume theory. A formal treatment of the latter will require an application of the technology developed in \cite{Bah:2020uev} using differential cohomology, which we defer to future work. Despite the appearance of multiple auxiliary functions in \eqref{inflow_anomaly_polynomial}, the inflow anomaly polynomial is solely a function of the field strengths $f_1^x$, $F_2^I$, $F_2^\alpha$, and the parameters $k$, $\chi$, $N_\alpha$. The precise functional dependence of $I_6^\textrm{inflow}$ on these parameters, however, is contingent on the specific choice of the various forms appearing in \eqref{Completions}. Nonetheless, the difference in $I_6^\mathrm{inflow}$ resulting from distinct choices can be compensated by redefinitions of $F_2^\alpha$. We refer the reader to appendix \ref{basis_independence_appendix} for a detailed discussion. Under appropriate field redefinitions, we verified that \eqref{inflow_anomaly_polynomial} reproduces the $k=2$ inflow anomaly polynomial in \cite{Bah:2019vmq}, which describes the 4d field theory dual of the GMSW solution.

To illustrate the generalized construction developed here, consider first the case when $k=3$, and let us restrict attention to the portion of $I_6^\mathrm{inflow}$ corresponding to zero-form symmetries by setting all $f_1^x = 0$. In addition to the decoupling convention \eqref{isometry_cohomology_decoupled_convention}, we adopt the convention that the bases of two- and four-cohomology classes are Poincar\'{e}-dual to one another, 
\begin{equation}
	\mathcal{K}^\alpha_\beta = \int_{M_6} \Omega_4^\alpha \wedge \omega_{2,\beta} = \delta^\alpha_\beta \, .
\end{equation}
As a result, with the axionic field strengths $f_1^x$ turned off, integrating out $\mathcal{A}_1$ using \eqref{c3_equation_of_motion} is equivalent to imposing 
\begin{equation}
\sum_{\alpha=1}^{2k-1} N_\alpha F_2^\alpha =0
\end{equation}
as in \cite{Bah:2019vmq}.  In the four-cycle basis \eqref{natural_four_cycles}, with $2k-1$ independent associated flux quanta $N$, $N_{\mathrm{N}_i}$, $N_{\mathrm{S}_i}$, the $k=3$ inflow anomaly polynomial can then be expressed as
\begin{equation} \label{full_k_3}
I_6^\mathrm{inflow} = \frac{I_6^\mathrm{N}}{(2\pi)^3} +\left( N_{\mathrm{N}_i} \leftrightarrow N_{\mathrm{S}_i} \, , F_2^{\mathrm{N}_i} \leftrightarrow -F_2^{\mathrm{S}_i}\right) \, ,
\end{equation}
where we have defined
\begin{align}
 I_6^\mathrm{N} = \, &
  \frac{3 \chi N^3}{64} (F_2^\psi )^3 -\frac{N}{2\chi}(N_{\mathrm{N}_1}^2+N_{\mathrm{N}_1} N_{\mathrm{N}_2}+N_{\mathrm{N}_2}^2) (F_2^\psi )^3 \nonumber
\\
&
 +\frac{1}{3\chi^2}(N_{\mathrm{N}_1}+N_{\mathrm{N}_2}) (2 N_{\mathrm{N}_1}^2+N_{\mathrm{N}_1} N_{\mathrm{N}_2}+2 N_{\mathrm{N}_2}^2)
  (F_2^\psi )^3
 \nonumber \\
  &
 -\frac{1}{9\chi^3 N}(N_{\mathrm{N}_1}^2+N_{\mathrm{N}_1}
   N_{\mathrm{N}_2}+N_{\mathrm{N}_2}^2-N_{\mathrm{S}_2}^2-N_{\mathrm{S}_2} N_{\mathrm{S}_1}-N_{\mathrm{S}_1}^2)^2 (F_2^\psi )^3
 \nonumber  \\
   &
 +  \frac{1}{18 \chi ^2} (2 N_{\mathrm{N}_1}^3+3 N_{\mathrm{N}_1}^2 N_{\mathrm{N}_2}-3 N_{\mathrm{N}_1} N_{\mathrm{N}_2}^2-2 N_{\mathrm{N}_2}^3 ) (F_2^\psi )^2F_2^\varphi  -\frac{\chi N^3}{48} (F_2^\varphi )^2F_2^\psi 
 \nonumber   \\
   &
   + \frac{N}{9 \chi}( N_{\mathrm{N}_1}^2+ N_{\mathrm{N}_1} N_{\mathrm{N}_2} + N_{\mathrm{N}_2}^2 ) (F_2^\varphi )^2F_2^\psi
    + \frac{9 N^2}{8} ( N_{\mathrm{N}_1}F_2^{\mathrm{N}_1}+N_{\mathrm{N}_2}F_2^{\mathrm{N},2} ) (F_2^\psi )^2
 \nonumber    \\
   &
 -\frac{3N }{2\chi} N_{\mathrm{N}_1}N_{\mathrm{N}_2}(F_2^{\mathrm{N}_1}+F_2^{\mathrm{N}_2})(F_2^\psi )^2
    -\frac{9N }{4\chi} (N_{\mathrm{N}_1}^2 F_2^{\mathrm{N}_1}+ N_{\mathrm{N}_2}^2 F_2^{\mathrm{N}_2})(F_2^\psi )^2
  \nonumber  \\
 & 
+ \frac{1}{\chi^2}(N_{\mathrm{N}_1}^2+N_{\mathrm{N}_1}
   N_{\mathrm{N}_2}+N_{\mathrm{N}_2}^2)(N_{\mathrm{N}_1}F_2^{\mathrm{N}_1}
  + N_{\mathrm{N}_2}
   F_2^{\mathrm{N}_2})
 (F_2^\psi )^2
\nonumber \\
 &
     -\frac{N^2}{6}( N_{\mathrm{N}_1}F_2^{\mathrm{N}_1}+N_{\mathrm{N}_2}F_2^{\mathrm{N}_2} ) (F_2^\varphi )^2
   \nonumber    \\
   & -\frac{N}{3 \chi } \left[N_{\mathrm{N}_1} (N_{\mathrm{N}_1}+2 N_{\mathrm{N}_2})F_2^{\mathrm{N}_1} - N_{\mathrm{N}_2} (N_{\mathrm{N}_2}+2
   N_{\mathrm{N}_1})F_2^{\mathrm{N}_2}\right]F_2^\psi F_2^\varphi
  \nonumber \\
   &
 -\frac{3\chi N^3}{4} \left[(F_2^{\mathrm{N}_1})^2-F_2^{\mathrm{N}_1} F_2^{\mathrm{N}_2}+(F_2^{\mathrm{N}_2})^2\right]F_2^\psi  +\frac{3N^2}{2}\left[N_{\mathrm{N}_1}(F_2^{\mathrm{N}_1})^2+N_{\mathrm{N}_2} (F_2^{\mathrm{N}_2})^2\right]F_2^\psi  
   \nonumber \\
 &  
 +N^2 (N_{\mathrm{N}_1}+N_{\mathrm{N}_2}) \left[(F_2^{\mathrm{N}_1})^2-F_2^{\mathrm{N}_1} F_2^{\mathrm{N}_2}+(F_2^{\mathrm{N}_2})^2\right]F_2^\psi  
 -\frac{3N}{2\chi}  \left[N_{\mathrm{N}_1}F_2^{\mathrm{N}_1}+N_{\mathrm{N}_2} F_2^{\mathrm{N}_2}\right]^2 F_2^\psi  
   \nonumber \\
 &  
   -\frac{2N}{3\chi}(N_{\mathrm{N}_1}^2+N_{\mathrm{N}_1}N_{\mathrm{N}_2}+N_{\mathrm{N}_2}^2-N_{\mathrm{S}_1}^2-N_{\mathrm{S}_1}N_{\mathrm{S}_2}-N_{\mathrm{S}_2}^2)\left[(F_2^{\mathrm{N}_1})^2-F_2^{\mathrm{N}_1}F_2^{\mathrm{N}_2}+(F_2^{\mathrm{N}_2})^2\right]
    F_2^\psi  
 \nonumber  \\
& +\frac{N^2}{6}  \left[3N_{\mathrm{N}_2}(F_2^{\mathrm{N}_1})^2-3N_{\mathrm{N}_1}(F_2^{\mathrm{N}_2})^2+(N_{\mathrm{N}_1}-N_{\mathrm{N}_2})(F_2^{\mathrm{N}_1} +F_2^{\mathrm{N}_2})^2 \right] F_2^\varphi
  \nonumber \\
   &  -\frac{\chi N^3}{12}
   (F_2^{\mathrm{N}_1}+F_2^{\mathrm{N}_2}) \left[8 (F_2^{\mathrm{N}_1})^2-11 F_2^{\mathrm{N}_1} F_2^{\mathrm{N}_2}+8 (F_2^{\mathrm{N}_2})^2\right]
  \nonumber \\
   &  
   + N^2(N_{\mathrm{N}_1}F_2^{\mathrm{N}_1}+N_{\mathrm{N}_2} F_2^{\mathrm{N}_2})  \left[(F_2^{\mathrm{N}_1})^2-F_2^{\mathrm{N}_1}
   F_2^{\mathrm{N}_2}+(F_2^{\mathrm{N}_2})^2\right]\nonumber\\
   & + \mathcal{O}(N,N_{\mathrm{N}_1},N_{\mathrm{N}_2},N_{\mathrm{S}_1},N_{\mathrm{S}_2}) \, . \label{k3_I6}
\end{align}
Note that in addition to the symmetry of $I_6^\text{inflow}$ under
\begin{equation}
	N_{\mathrm{N}_i} \leftrightarrow N_{\mathrm{S}_i} \, , \qquad F_2^{\mathrm{N}_i} \leftrightarrow -F_2^{\mathrm{S}_i} \, ,\label{invariance_NS}
\end{equation}
of which we have made explicit use above, the inflow anomaly polynomial is invariant under the simultaneous exchanges
\begin{equation}\label{invariance_EW}
N_{\mathrm{N}_i} \leftrightarrow N_{\mathrm{N}_{k-i}} \, , \quad F_2^{\mathrm{N}_i} \leftrightarrow F_2^{\mathrm{N}_{k-i}} \, , \quad N_{\mathrm{S}_i} \leftrightarrow N_{\mathrm{S}_{k-i}} \, , \quad F_2^{\mathrm{S}_i} \leftrightarrow F_2^{\mathrm{S}_{k-i}} \, , \quad F_\varphi \rightarrow -F_\varphi,
\end{equation}
Both symmetries are in fact present for any $k$. The symmetry under \eqref{invariance_NS} can be exhibited for general $k$ most easily in the case in which all resolution flux quanta are equal,
\begin{equation}
N_{\mathrm{N}_i}, N_{\mathrm{S}_i} =N_\mathrm{N}
\end{equation}
for all $i = 1,2,\dots,k-1$. For these uniform flux configurations, the large-$N$ inflow anomaly polynomial for continuous zero-form symmetries is given by
\begin{align}\label{all_eq_I6}
	(2\pi)^3 I_6^\mathrm{inflow} &= \frac{15 k^2 \chi^3 N^3 - 60 k^2 (k^2-1) \chi N N_\mathrm{N}^2 + 16k(k^2 - 1)(3 k^2 - 2) N_\mathrm{N}^3}{1440\chi^2} (F_2^\psi )^3 \nonumber \\
	& \phantom{=\ } - \frac{\chi^2 N^3 - 2 (k^2 - 1) N N_\mathrm{N}^2}{24\chi} F_2^\psi (F_2^\varphi)^2 \nonumber\\
	& \phantom{=\ } + \sum_{i=1}^{k-1} \frac{3k \chi N^2 N_\mathrm{N} - 6(k^2 - 2ki +2i^2) N N_\mathrm{N}^2}{8\chi}  \left(F_2^{\mathrm{N}_i}   - F_2^{\mathrm{S}_i}  \right)(F_2^\psi )^2 \nonumber\\
	& \phantom{=\ } - \sum_{i=1}^{k-1} \frac{(k-2i) N N_\mathrm{N}^2}{\chi}   \left(F_2^{\mathrm{N}_i}   - F_2^{\mathrm{S}_i}  \right)F_2^\psi F_2^\varphi - \sum_{i=1}^{k-1} \frac{N^2 N_\mathrm{N}}{2k}  \left(F_2^{\mathrm{N}_i}   - F_2^{\mathrm{S}_i}  \right)(F_2^\varphi)^2 \nonumber\\
	& \phantom{=\ } - \sum_{i=1}^{k-1} \frac{k \chi N^3 - 2(k^2 - 2ki + 2i^2 + 2) N^2 N_\mathrm{N}}{4}  \left[(F_2^{\mathrm{N}_i})^2  + (F_2^{\mathrm{S}_i})^2 \right]F_2^\psi\nonumber\\
	& \phantom{=\ }
	 + \sum_{i=1}^{k-2} \frac{k \chi N^3 - 2(k^2 - 2ki + 2i^2 - k + 2i) N^2 N_\mathrm{N}}{4}  \left(F_2^{\mathrm{N}_i}   F_2^{\mathrm{N}_{i+1}}   + F_2^{\mathrm{S}_i}   F_2^{\mathrm{S}_{i+1}} \right)F_2^\psi \nonumber\\
	& \phantom{=\ } - \sum_{i=1}^{k-1} \sum_{j=1}^{k-1} \frac{N N_\mathrm{N}^2}{\chi}  \left(F_2^{\mathrm{N}_i}   + F_2^{\mathrm{S}_i}  \right) \left(F_2^{\mathrm{N}_j}  + F_2^{\mathrm{S}_j} \right)F_2^\psi\nonumber\\
	& \phantom{=\ } + \sum_{i=1}^{k-1} \frac{(k-2i) N^2 N_\mathrm{N}}{2}  \left[(F_2^{\mathrm{N}_i})^2  + (F_2^{\mathrm{S}_i})^2 \right] F_2^\varphi \nonumber\\
	& \phantom{=\ } - \sum_{i=1}^{k-2} \frac{(k-2i-1) N^2 N_\mathrm{N}}{2}   \left(F_2^{\mathrm{N}_i}   F_2^{\mathrm{N}_{i+1}}   + F_2^{\mathrm{S}_i}   F_2^{\mathrm{S}_{i+1}} \right)F_2^\varphi\nonumber\\
	& \phantom{=\ } - \sum_{i=1}^{k-1} \frac{2 \chi N^3}{3} \left[(F_2^{\mathrm{N}_i})^3  - (F_2^{\mathrm{S}_i})^3 \right]\nonumber\\
	& \phantom{=\ } - \sum_{i=1}^{k-2} \frac{(k-2i-2) \chi N^3}{4} \left[(F_2^{\mathrm{N}_i})^2  F_2^{\mathrm{N}_{i+1}}   - (F_2^{\mathrm{S}_i})^2  F_2^{\mathrm{S}_{i+1}} \right]\nonumber\\
	& \phantom{=\ } + \sum_{i=1}^{k-2} \frac{(k-2i) \chi N^3}{4} \left[F_2^{\mathrm{N}_i}   (F_2^{\mathrm{N}_{i+1}})^2  - F_2^{\mathrm{S}_i}   (F_2^{\mathrm{S}_{i+1}} )^2\right]\nonumber\\
	& \phantom{=\ } + \sum_{i=1}^{k-1} \sum_{j=1}^{k-1} N^2 N_\mathrm{N} \left[(F_2^{\mathrm{N}_i})^2  - (F_2^{\mathrm{S}_i})^2 \right] \left(F_2^{\mathrm{N}_j}  + F_2^{\mathrm{S}_j} \right)\nonumber\\
	& \phantom{=\ } - \sum_{i=1}^{k-2} \sum_{j=1}^{k-1} N^2 N_\mathrm{N} \left(F_2^{\mathrm{N}_i}   F_2^{\mathrm{N}_{i+1}}   - F_2^{\mathrm{S}_i}   F_2^{\mathrm{S}_{i+1}} \right) \left(F_2^{\mathrm{N}_j}  + F_2^{\mathrm{S}_j} \right)\nonumber\\
  & \phantom{=\ } + \mathcal{O}(N,N_\mathrm{N}) \, .
\end{align}
We observe the invariance of the expression under the exchange, $F_2^{\mathrm{N}_i} \leftrightarrow -F_2^{\mathrm{S}_i}$, as expected.

\subsection{Parity symmetries}\label{parity_symmetries_subsection}

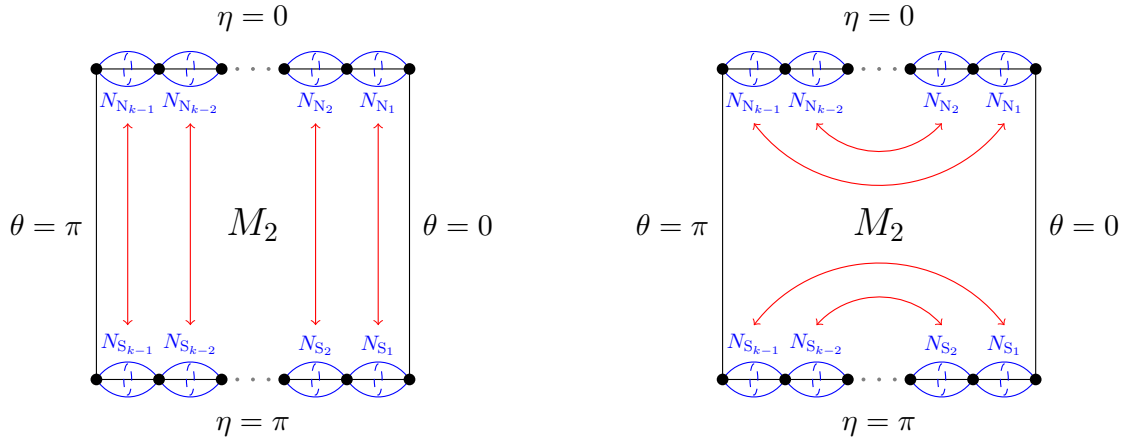
\begin{figure}[t!]
	\centering
    \resizebox{\textwidth}{!}{
		\begin{tikzpicture}
			\draw (0,0) -- (1.6,0);
			\fill[black!50] (1.8,0) circle (0.025);
			\fill[black!50] (2,0) circle (0.025) node[color=black,below=10] {$\eta = \pi$};
			\fill[black!50] (2.2,0) circle (0.025);
			\draw (2.4,0) -- (4,0);
			\draw (0,4) -- (1.6,4);
			\fill[black!50] (1.8,4) circle (0.025);
			\fill[black!50] (2,4) circle (0.025) node[color=black,above=10] {$\eta = 0$};
			\fill[black!50] (2.2,4) circle (0.025);
			\draw (2.4,4) -- (4,4);
			\draw (0,0) -- node[left=1] {$\theta = \pi$} (0,4);
			\draw (4,0) -- node[right=1] {$\theta = 0$} (4,4);
			\node (M4) at (2,2) {\Large $M_2$};
			\draw[color=blue] (0,0) .. controls (0.2,0.3) and (0.6,0.3) .. (0.8,0);
			\draw[color=blue] (0,0) .. controls (0.2,-0.3) and (0.6,-0.3) .. (0.8,0);
			\draw[dashed,color=blue] (0.4,0) ellipse (0.05 and 0.2) node[color=blue,above=5] {\scalebox{0.7}{$N_{\mathrm{S}_{k-1}}$}};
			\draw[color=blue] (0.8,0) .. controls (1,0.3) and (1.4,0.3) .. (1.6,0);
			\draw[color=blue] (0.8,0) .. controls (1,-0.3) and (1.4,-0.3) .. (1.6,0);
			\draw[dashed,color=blue] (1.2,0) ellipse (0.05 and 0.2) node[color=blue,above=5] {\scalebox{0.7}{$N_{\mathrm{S}_{k-2}}$}};
			\draw[color=blue] (2.4,0) .. controls (2.6,0.3) and (3,0.3) .. (3.2,0);
			\draw[color=blue] (2.4,0) .. controls (2.6,-0.3) and (3,-0.3) .. (3.2,0);
			\draw[dashed,color=blue] (2.8,0) ellipse (0.05 and 0.2) node[color=blue,above=5] {\scalebox{0.7}{$N_{\mathrm{S}_2}$}};
			\draw[color=blue] (3.2,0) .. controls (3.4,0.3) and (3.8,0.3) .. (4,0);
			\draw[color=blue] (3.2,0) .. controls (3.4,-0.3) and (3.8,-0.3) .. (4,0);
			\draw[dashed,color=blue] (3.6,0) ellipse (0.05 and 0.2) node[color=blue,above=5] {\scalebox{0.7}{$N_{\mathrm{S}_1}$}};
			\draw[color=blue] (0,4) .. controls (0.2,4.3) and (0.6,4.3) .. (0.8,4);
			\draw[color=blue] (0,4) .. controls (0.2,3.7) and (0.6,3.7) .. (0.8,4);
			\draw[dashed,color=blue] (0.4,4) ellipse (0.05 and 0.2) node[color=blue,below=5] {\scalebox{0.7}{$N_{\mathrm{N}_{k-1}}$}};
			\draw[color=blue] (0.8,4) .. controls (1,4.3) and (1.4,4.3) .. (1.6,4);
			\draw[color=blue] (0.8,4) .. controls (1,3.7) and (1.4,3.7) .. (1.6,4);
			\draw[dashed,color=blue] (1.2,4) ellipse (0.05 and 0.2) node[color=blue,below=5] {\scalebox{0.7}{$N_{\mathrm{N}_{k-2}}$}};
			\draw[color=blue] (2.4,4) .. controls (2.6,4.3) and (3,4.3) .. (3.2,4);
			\draw[color=blue] (2.4,4) .. controls (2.6,3.7) and (3,3.7) .. (3.2,4);
			\draw[dashed,color=blue] (2.8,4) ellipse (0.05 and 0.2) node[color=blue,below=5] {\scalebox{0.7}{$N_{\mathrm{N}_2}$}};
			\draw[color=blue] (3.2,4) .. controls (3.4,4.3) and (3.8,4.3) .. (4,4);
			\draw[color=blue] (3.2,4) .. controls (3.4,3.7) and (3.8,3.7) .. (4,4);
			\draw[dashed,color=blue] (3.6,4) ellipse (0.05 and 0.2) node[color=blue,below=5] {\scalebox{0.7}{$N_{\mathrm{N}_1}$}};
			\fill (0,0) circle (0.075);
			\fill (0.8,0) circle (0.075);
			\fill (1.6,0) circle (0.075);
			\fill (2.4,0) circle (0.075);
			\fill (3.2,0) circle (0.075);
			\fill (4,0) circle (0.075);
			\fill (0,4) circle (0.075);
			\fill (0.8,4) circle (0.075);
			\fill (1.6,4) circle (0.075);
			\fill (2.4,4) circle (0.075);
			\fill (3.2,4) circle (0.075);
			\fill (4,4) circle (0.075);
			\draw[color=red,<->] (0.4,0.7) -- (0.4,3.3);
			\draw[color=red,<->] (1.2,0.7) -- (1.2,3.3);
			\draw[color=red,<->] (2.8,0.7) -- (2.8,3.3);
			\draw[color=red,<->] (3.6,0.7) -- (3.6,3.3);
			\draw (8,0) -- (9.6,0);
			\fill[black!50] (9.8,0) circle (0.025);
			\fill[black!50] (10,0) circle (0.025) node[color=black,below=10] {$\eta = \pi$};
			\fill[black!50] (10.2,0) circle (0.025);
			\draw (10.4,0) -- (12,0);
			\draw (8,4) -- (9.6,4);
			\fill[black!50] (9.8,4) circle (0.025);
			\fill[black!50] (10,4) circle (0.025) node[color=black,above=10] {$\eta = 0$};
			\fill[black!50] (10.2,4) circle (0.025);
			\draw (10.4,4) -- (12,4);
			\draw (8,0) -- node[left=1] {$\theta = \pi$} (8,4);
			\draw (12,0) -- node[right=1] {$\theta = 0$} (12,4);
			\node (M4) at (10,2) {\Large $M_2$};
			\draw[color=blue] (8,0) .. controls (8.2,0.3) and (8.6,0.3) .. (8.8,0);
			\draw[color=blue] (8,0) .. controls (8.2,-0.3) and (8.6,-0.3) .. (8.8,0);
			\draw[dashed,color=blue] (8.4,0) ellipse (0.05 and 0.2) node[color=blue,above=5] {\scalebox{0.7}{$N_{\mathrm{S}_{k-1}}$}};
			\draw[color=blue] (8.8,0) .. controls (9,0.3) and (9.4,0.3) .. (9.6,0);
			\draw[color=blue] (8.8,0) .. controls (9,-0.3) and (9.4,-0.3) .. (9.6,0);
			\draw[dashed,color=blue] (9.2,0) ellipse (0.05 and 0.2) node[color=blue,above=5] {\scalebox{0.7}{$N_{\mathrm{S}_{k-2}}$}};
			\draw[color=blue] (10.4,0) .. controls (10.6,0.3) and (11,0.3) .. (11.2,0);
			\draw[color=blue] (10.4,0) .. controls (10.6,-0.3) and (11,-0.3) .. (11.2,0);
			\draw[dashed,color=blue] (10.8,0) ellipse (0.05 and 0.2) node[color=blue,above=5] {\scalebox{0.7}{$N_{\mathrm{S}_2}$}};
			\draw[color=blue] (11.2,0) .. controls (11.4,0.3) and (11.8,0.3) .. (12,0);
			\draw[color=blue] (11.2,0) .. controls (11.4,-0.3) and (11.8,-0.3) .. (12,0);
			\draw[dashed,color=blue] (11.6,0) ellipse (0.05 and 0.2) node[color=blue,above=5] {\scalebox{0.7}{$N_{\mathrm{S}_1}$}};
			\draw[color=blue] (8,4) .. controls (8.2,4.3) and (8.6,4.3) .. (8.8,4);
			\draw[color=blue] (8,4) .. controls (8.2,3.7) and (8.6,3.7) .. (8.8,4);
			\draw[dashed,color=blue] (8.4,4) ellipse (0.05 and 0.2) node[color=blue,below=5] {\scalebox{0.7}{$N_{\mathrm{N}_{k-1}}$}};
			\draw[color=blue] (8.8,4) .. controls (9,4.3) and (9.4,4.3) .. (9.6,4);
			\draw[color=blue] (8.8,4) .. controls (9,3.7) and (9.4,3.7) .. (9.6,4);
			\draw[dashed,color=blue] (9.2,4) ellipse (0.05 and 0.2) node[color=blue,below=5] {\scalebox{0.7}{$N_{\mathrm{N}_{k-2}}$}};
			\draw[color=blue] (10.4,4) .. controls (10.6,4.3) and (11,4.3) .. (11.2,4);
			\draw[color=blue] (10.4,4) .. controls (10.6,3.7) and (11,3.7) .. (11.2,4);
			\draw[dashed,color=blue] (10.8,4) ellipse (0.05 and 0.2) node[color=blue,below=5] {\scalebox{0.7}{$N_{\mathrm{N}_2}$}};
			\draw[color=blue] (11.2,4) .. controls (11.4,4.3) and (11.8,4.3) .. (12,4);
			\draw[color=blue] (11.2,4) .. controls (11.4,3.7) and (11.8,3.7) .. (12,4);
			\draw[dashed,color=blue] (11.6,4) ellipse (0.05 and 0.2) node[color=blue,below=5] {\scalebox{0.7}{$N_{\mathrm{N}_1}$}};
			\fill (8,0) circle (0.075);
			\fill (8.8,0) circle (0.075);
			\fill (9.6,0) circle (0.075);
			\fill (10.4,0) circle (0.075);
			\fill (11.2,0) circle (0.075);
			\fill (12,0) circle (0.075);
			\fill (8,4) circle (0.075);
			\fill (8.8,4) circle (0.075);
			\fill (9.6,4) circle (0.075);
			\fill (10.4,4) circle (0.075);
			\fill (11.2,4) circle (0.075);
			\fill (12,4) circle (0.075);
			\draw[color=red,<->] (11.6,0.7) arc (atan(1.2/1.6):180-atan(1.2/1.6):{sqrt(4)});
			\draw[color=red,<->] (8.4,3.3) arc (180+atan(1.2/1.6):360-atan(1.2/1.6):{sqrt(4)});
			\draw[color=red,<->] (10.8,0.7) arc (atan(0.7/0.8):180-atan(0.7/0.8):{sqrt(1.13)});
			\draw[color=red,<->] (9.2,3.3) arc (180+atan(0.7/0.8):360-atan(0.7/0.8):{sqrt(1.13)});
		\end{tikzpicture}
    }
	\caption{Illustration of the effects of the involutions $\mathcal{P}_{\mathrm{NS}}$ ({\it left}), and $\mathcal{P}_{\mathrm{EW}_\varphi}$ or $\mathcal{P}_{\mathrm{EW}_\psi}$ ({\it right}) on the flux quanta $N_{\mathrm{N}_i}$, $N_{\mathrm{S}_i}$. For example, under the action of $\mathcal{P}_{\mathrm{NS}}$, the roles of, say, $N_{\mathrm{N}_1}$ and $N_{\mathrm{S}_1}$, are interchanged, as indicated by one of the red double arrows in the diagram on the left. The two-form field strengths $F_2^{\mathrm{N}_i}$, $F_2^{\mathrm{S}_i}$ follow a similar exchange pattern.}
	\label{I6_parity_illustration}
\end{figure}

As noted above, the inflow anomaly polynomial for continuous, zero-form symmetries with general $k$ is invariant under the same exchanges \eqref{invariance_NS} and \eqref{invariance_EW} as the $k=3$ polynomial. In other words, $I_6^\mathrm{inflow}$ is even under the involutions,
\begin{gather}
	\mathcal{P}_\mathrm{NS}: \qquad N_{\mathrm{N}_i} \leftrightarrow N_{\mathrm{S}_i} \, , \qquad F_2^{\mathrm{N}_i} \leftrightarrow -F_2^{\mathrm{S}_i} \, , \label{P_NS}\\
	\mathcal{P}_{\mathrm{EW}_\varphi}: \quad N_{\mathrm{N}_i},N_{\mathrm{S}_i} \leftrightarrow N_{\mathrm{N}_{k-i}},N_{\mathrm{S}_{k-i}} \, , \quad F_2^{\mathrm{N}_i},F_2^{\mathrm{S}_i} \leftrightarrow F_2^{\mathrm{N}_{k-i}},F_2^{\mathrm{S}_{k-i}} \, , \quad F_2^\varphi \rightarrow -F_2^\varphi \, ,\label{P_EW_phi}
\end{gather}
and odd under the involution,
\begin{gather}
\mathcal{P}_{\mathrm{EW}_\psi}: \quad N_{\mathrm{N}_i},N_{\mathrm{S}_i} \leftrightarrow N_{\mathrm{N}_{k-i}},N_{\mathrm{S}_{k-i}} \, , \quad F_2^{\mathrm{N}_i},F_2^{\mathrm{S}_i} \leftrightarrow -F_2^{\mathrm{N}_{k-i}},-F_2^{\mathrm{S}_{k-i}} \, , \quad F_2^\psi \rightarrow -F_2^\psi \, .\label{P_EW_psi}
\end{gather}
We illustrate the effects of these involutions on the configuration of the ``resolution fluxes,'' $N_{\mathrm{N}_i}$, $N_{\mathrm{S}_i}$, in figure \ref{I6_parity_illustration}. In addition, one can apply the actions of $\mathcal{P}_\mathrm{NS}$ and $\mathcal{P}_{\mathrm{EW}_\varphi}$ (or $\mathcal{P}_{\mathrm{EW}_\psi}$) successively to find that $I_6^\mathrm{inflow}$ is also even (or odd, in the case of $\mathcal{P}_{\mathrm{EW}_\psi}$) under
\begin{gather}
  \mathcal{P}_\mathrm{NS} \mathcal{P}_{\mathrm{EW}_\varphi}: \quad N_{\mathrm{N}_i} \leftrightarrow N_{\mathrm{S}_{k-i}} \, , \quad F_2^{\mathrm{N}_i} \leftrightarrow -F_2^{\mathrm{S}_{k-i}} \, , \quad F_2^\varphi \to -F_2^\varphi \, ,\\
  \mathcal{P}_\mathrm{NS} \mathcal{P}_{\mathrm{EW}_\psi}: \quad N_{\mathrm{N}_i} \leftrightarrow N_{\mathrm{S}_{k-i}} \, , \quad F_2^{\mathrm{N}_i} \leftrightarrow -F_2^{\mathrm{S}_{k-i}} \, , \quad F_2^\psi \to -F_2^\psi \, .
\end{gather}
Note that we can identify $\mathcal{P}_\mathrm{NS}$ and $\mathcal{P}_{\mathrm{EW}_\psi}$ (or $\mathcal{P}_{\mathrm{EW}_\varphi}$) with generators of the dihedral group $D_2$ acting on $N_{\mathrm{N}_i}$, $N_{\mathrm{S}_i}$, $F_2^{\mathrm{N}_i}$, $F_2^{\mathrm{S}_i}$.

Here we offer a geometric interpretation of the aforementioned symmetries in the anomaly polynomial in terms of discrete isometries of $M_4$. With reference to \eqref{M4_metric}, or \eqref{M4_metric_explicit} for a more explicit version, the metric of $M_4$ is invariant under each of the following discrete transformations,
\begin{gather}
  \mathcal{P}_\eta: \quad \eta \to \pi - \eta \, ,\\
	\mathcal{P}_{\theta_\varphi}: \quad \theta \to \pi - \theta \, , \quad \varphi \to -\varphi \, ,\\
	\mathcal{P}_{\theta_\psi}: \quad \theta \to \pi - \theta \, , \quad \psi \to -\psi \, .
\end{gather}
We interpret these as distinct parity symmetries of $M_4$. As can be seen in figure \ref{I6_parity_illustration}, the exchange of flux quanta induced by $\mathcal{P}_\eta$ is identical to that by $\mathcal{P}_{\mathrm{NS}}$, leading us to identify the former as the geometric origin of the invariance of $I_6^\mathrm{inflow}$ under the latter. Similarly, we are led to the conclusion that $\mathcal{P}_{\theta_\varphi}$ and $\mathcal{P}_{\theta_\psi}$ are respectively the geometric origins of the even and odd parities of $I_6^\mathrm{inflow}$ under $\mathcal{P}_{\mathrm{EW}_\varphi}$ and $\mathcal{P}_{\mathrm{EW}_\psi}$. Note further that the global angular form $(D\varphi)^\mathrm{g} = D\varphi + A_1^\varphi$ has a definite (odd) parity only if $A_1^\varphi \to -A_1^\varphi$ under the action of $\mathcal{P}_{\theta_\varphi}$. Likewise, the global angular form $(D\psi)^\mathrm{g} = D\psi + A_1^\psi$ has a definite (odd) parity only if $A_1^\psi \to -A_1^\psi$ under the action of $\mathcal{P}_{\theta_\psi}$. These explain the sign changes of $F_2^\varphi$ and $F_2^\psi$ in \eqref{P_EW_phi} and \eqref{P_EW_psi} respectively.


We would like to remind the reader that the invariance of $I_6^\mathrm{inflow}$ under \eqref{P_NS} and \eqref{P_EW_phi} is made manifest as a consequence of choosing the natural basis of (co)homology classes introduced in section \ref{geometric_setup_section}. In another basis the transformation properties of $I_6^\mathrm{inflow}$ are generally obscured by nontrivial redefinitions of the flux quanta and field strengths.


%
%



\section{Conclusion} \label{sec_conclusion}

In this work, we have started a systematic exploration of
setups featuring wrapped M5-branes probing a family of flux backgrounds,
consisting of a resolved $S^4/\mathbb{Z}_k$ fibered over a higher-genus Riemann surface $\Sigma_g$.
This class of setups for $k \ge 3$ is challenging to analyze directly
in holography or in the probe picture. 
Anomaly inflow techniques, however, provide a powerful inroad
into the investigation of these setups.
Indeed, these methods hinge on robust topological and flux
data, which can be inferred without   solving explicitly 
the supersymmetry conditions and equations of motion.

Our findings exhibit a rich pattern of global symmetries and 't Hooft anomalies.
This is already visible if we restrict our attention to continuous
symmetries only, and focus on terms in the anomaly polynomial
that are leading in the limit of large $N$, $N_{\mathrm N_i}$, $N_{\mathrm S_i}$.
Our computations can be extended in two  natural directions.
Firstly, we could retain the topologically massive fields
in our analysis in order to investigate 't Hooft anomalies involving
discrete symmetries. This task was addressed
for $k=2$ in \cite{Bah:2020uev}  (see in particular appendix E), and it would be interesting to consider the general case $k \ge 3$
using the results of this paper as a starting point.
Secondly, we could access the subleading terms in 
$N$, $N_{\mathrm N_i}$, $N_{\mathrm S_i}$ by taking into account the effect
of the higher-derivative topological term
$C_3 \wedge X_8$ in the M-theory effective action.
This requires a determination of the class $X_8$, computed 
in the presence of non-zero backgrounds for the connections
$A_1^\psi$, $A_1^\varphi$ associated with the
isometries $\mathrm U(1)_\psi$,
$\mathrm U(1)_\varphi$.
A possible strategy to achieve this goal is to consider
local expressions, given in terms of differential forms,
to compute the class $X_8$. This approach has been successful 
in the case $k=2$ \cite{Bah:2019vmq}. The approach
of that paper, however, relies on a special feature of the case $k=2$:
the resolved $\mathbb C^2/\mathbb Z_2$ singularity
admits a description both in terms of a Taub-NUT geometry
and in terms of an Eguchi-Hanson geometry. The latter description
greatly facilitates the computation of $X_8$ based on local
expressions, but is unfortunately unavailable for $k\ge 3$.
It would therefore be interesting to
revisit the problem of the determination of the class $X_8$
for $k \ge 3$, possibly exploiting more refined computational techniques such as spectral sequences in equivariant cohomology.

%
%

According to the inflow paradigm, the output $I_6^{\rm inflow}$ of the 
inflow computation is to be identified with minus the anomaly polynomial
of the 4d field theory realized by the wrapped M5-branes.
The results of this work give us access to the cubic terms in
$N$, $N_{\mathrm N_i}$, $N_{\mathrm S_i}$ inside the quantity
$I_6^\text{4d QFT} = -I_6^{\rm inflow}$.
Armed with this knowledge, one can perform manipulations on 
$I_6^\text{4d QFT}$ to learn about the 4d dynamics.
In an upcoming  paper \cite{Bah:2021iaa}, we adopt $a$-maximization \cite{Intriligator:2003jj}
on $I_6^\text{4d QFT}$ to explore whether the 4d IR dynamics can define
a non-trivial SCFT, and to compare this putative SCFT to other
theories originating from wrapped M5-branes.

\acknowledgments

We are grateful to 
Zohar Komargodski,
Shlomo Razamat, Alessandro Tomasiello, 
Thomas Waddleton, and 
Gabi Zafrir
 for interesting
conversations and correspondence. 
The work of IB, EL, and PW is supported in part by NSF grant PHY-2112699
. The work of IB is also supported in part by the Simons Collaboration on Global Categorical Symmetries.
FB is supported by the European Union’s Horizon 2020 Framework: ERC Consolidator Grant 682608.
FB is supported by STFC Consolidated Grant ST/T000864/1.


\appendix
\addtocontents{toc}{\protect\setcounter{tocdepth}{1}}


\section{\boldmath Resolution of $S^4/\mathbb{Z}_k$ orbifold singularities}\label{orbifold_singularity_resolution_appendix}

Here we explicitly describe the blow-up of the orbifold singularities at the level of the metric. Under a coordinate transformation $\mu = \cos{\eta}$, the orbifold metric \eqref{S4_Zk_metric} becomes
\begin{equation}
	ds^2(S^4/\mathbb{Z}_k) = \frac{d\mu^2}{1 - \mu^2} + (1 - \mu^2) \bigg[\frac{1}{k^2} \, D\varphi^2 + \frac{1}{4} \, \big(d\theta^2 + \sin^2 \theta \, d\psi^2\big)\bigg] \, ,
\end{equation}
Expanding around $\mu = \pm 1$, we obtain a single-center Taub-NUT space with charge $k$,
\begin{align}
	ds^2_\pm(S^4/\mathbb{Z}_k) & = \frac{1}{V_\pm} \, D\varphi^2 + V_\pm \Big[dR_\pm^2 + R_\pm^2 \, \big(d\theta^2 + \sin^2 \theta \, d\psi^2\big)\Big] \, ,\label{Taub_NUT_space}
\end{align}
where $R_\pm = (1 \mp \mu)/k$ and $V_\pm = k/2R_\pm$. The singularities at $\mu = \pm 1$ are resolved by locally replacing the Taub-NUT space with a multi-center Gibbons-Hawking space,
\begin{equation}
	ds^2_\pm = \frac{1}{V_\pm} \, \big(d\varphi + A_\pm\big)^2 + V_\pm \, ds^2(\mathbb{R}^3) \, ,\label{gibbons_hawking_space}
\end{equation}
with the harmonic potential $V_\pm$ having the general form,
\begin{equation}
V_\pm(\vec{R}) = v_\pm + \frac{1}{2} \sum_{i=1}^{i_\pm^\mathrm{max}} \frac{n_{\pm,i}}{R_{\pm,i}} \, ,\label{gibbons_hawking_potential}
\end{equation}
where $R_{\pm,i}$ is the distance between the position $\vec{R}$ in $\mathbb{R}^3$ and the $i$-th Kaluza-Klein monopole at $\mu = \pm 1$. The potential behaves asymptotically the same as that in \eqref{Taub_NUT_space} if and only if
\begin{equation}
	v_\pm = 0 \, , \qquad \sum_{i=1}^{N_\pm} n_{\pm,i} = k \, .
\end{equation}
The former condition states that the size of the $S^1_\varphi$ circle goes to zero at each monopole. In this work, we focus on the case where both orbifold singularities are fully resolved by fixing $i_\pm^\mathrm{max} = k$ and $n_{\pm,i} = 1$ for all $i$, such that the space around each monopole is locally Euclidean, i.e.~$\mathbb{R}^4/\mathbb{Z}_{n_{\pm,i}} = \mathbb{R}^4/\mathbb{Z} \cong \mathbb{R}^4$.\footnote{This can be easily verified by performing the coordinate transformation $R_{\pm,i} = r_{\pm,i}^2/2n_{\pm,i}$, which leads to $ds^2_{\pm,i} = dr_{\pm,i}^2 + r_{\pm,i}^2 \big[(1/n_{\pm,i}^2) D\varphi^2 + (1/4) \big(d\theta^2 + \sin^2 \theta \, d\psi^2\big)\big]$.} Furthermore, the connection one-form $A_\pm$ is related to the potential through $dV_\pm = \star_{\mathbb{R}^3} \, dA_\pm$, and has a straightforward solution,
\begin{equation}
	A_\pm = \frac{1}{2} \sum_{i=1}^k \cos{\theta_{\pm,i}} \, d\psi_{\pm,i} \, ,
\end{equation}
where $\theta_{\pm,i}$ and $\psi_{\pm,i}$ are the standard polar and azimuthal angles in $\mathbb{R}^3$ with respect to the $i$-th monopole at $\mu = \pm 1$. We further impose that all the azimuthal angles are identified with the angle $\psi$ of the unresolved orbifold, i.e.~$\psi_{\pm,i} = \psi$, so as to preserve a $\mathrm{U}(1)_R$ subgroup of the original $\mathrm{SU}(2)_R$ isometry. Geometrically, this means that we have $k-1$ two-cycles, separated by $k$ monopoles, aligned along the axis associated with the $\mathrm{U}(1)_\psi$ symmetry.

\begin{figure}[t!]
	\centering
		\begin{tikzpicture}
			\draw (4,4) -- (0,4);
			\draw[opacity=0] (2,4) circle (0) node[above=5,opacity=1]  {\large $\eta = 0$};
			\fill (4,4) circle (0.075);
			\fill (3,4) circle (0.075);
			\fill (2,4) circle (0.075);
			\fill (1,4) circle (0.075);
			\fill (0,4) circle (0.075);
			\draw[opacity=0] (3.5,4) circle (0) node[below,color=blue,opacity=1] {\large $-\tfrac{3}{2}$};
			\draw[opacity=0] (2.5,4) circle (0) node[below,color=blue,opacity=1] {\large $-\tfrac{1}{2}$};
			\draw[opacity=0] (1.5,4) circle (0) node[below,color=blue,opacity=1] {\large $\tfrac{1}{2}$};
			\draw[opacity=0] (0.5,4) circle (0) node[below,color=blue,opacity=1] {\large $\tfrac{3}{2}$};
			\draw (0,4) -- (0,0);
			\draw[opacity=0] (0,2) circle (0) node[left=5,opacity=1]  {\large $\theta = \pi$};
			\draw[opacity=0] (0,2) circle (0) node[right,color=blue,opacity=1]  {\large $\tfrac{5}{2}$};
			\draw (0,0) -- (4,0);
			\draw[opacity=0] (2,0) circle (0) node[below=5,opacity=1]  {\large $\eta = \pi$};
			\fill (4,0) circle (0.075);
			\fill (3,0) circle (0.075);
			\fill (2,0) circle (0.075);
			\fill (1,0) circle (0.075);
			\fill (0,0) circle (0.075);
			\draw[opacity=0] (0.5,0) circle (0) node[above,color=blue,opacity=1] {\large $\tfrac{3}{2}$};
			\draw[opacity=0] (1.5,0) circle (0) node[above,color=blue,opacity=1] {\large $\tfrac{1}{2}$};
			\draw[opacity=0] (2.5,0) circle (0) node[above,color=blue,opacity=1] {\large $-\tfrac{1}{2}$};
			\draw[opacity=0] (3.5,0) circle (0) node[above,color=blue,opacity=1] {\large $-\tfrac{3}{2}$};
			\draw (4,0) -- (4,4);
			\draw[opacity=0] (4,2) circle (0) node[right=5,opacity=1]  {\large $\theta = 0$};
			\draw[opacity=0] (4,2) circle (0) node[left,color=blue,opacity=1]  {\large $-\tfrac{5}{2}$};
			\node (M4) at (2,2) {\Large $M_2$};
		\end{tikzpicture}
	\caption{Value that the function $L$ takes (written in blue) in each boundary interval of $M_2$ separated by monopoles for the case of $k = 5$.}
	\label{L_behavior}
\end{figure}
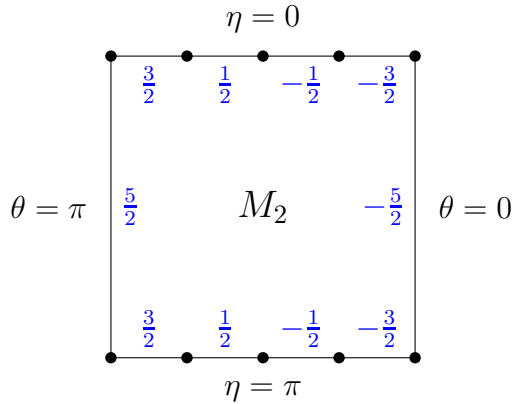

To summarize, the resultant resolved space, $M_4$, has a $\mathrm{U}(1)_\psi \times \mathrm{U}(1)_\varphi$ isometry group, and, omitting the ``$\pm$'' notation for visual clarity, its metric near $\mu = \pm 1$ (or equivalently, $\eta = 0,\pi$) can be written as
\begin{equation}
	ds^2(M_4) = \frac{2}{\sum_{i=1}^k R_i^{-1}} \bigg(d\varphi + \frac{1}{2} \sum_{i=1}^k \cos{\theta_i} \, d\psi\bigg)^{\!2} + \sum_{i=1}^k \frac{1}{2 R_i} \Big[dR_i^2 + R_i^2 \big(d{\theta_i}^2 + \sin^2 \theta_i \, d\psi^2\big)\Big] \, .\label{M4_metric_explicit}
\end{equation}
By construction, $\theta_i$ takes a value of $0$ or $\pi$ at $\mu = \pm 1$, so the size of the circle $S^1_\psi$, which is given by $\sum_{i=1}^{k} R_i \sin^2 \theta_i / 2$ at $\mu = \pm 1$ and $(1 - \mu^2) \sin^2 \theta / 4$ at $|\mu| < 1$, vanishes along the boundary $\partial M_2$, where $M_2$ is defined as the 2d space spanned by $(\mu,\theta)$. It is useful to define the function,
\begin{equation}
	L(\mu,\theta) \equiv \begin{cases} \displaystyle -\frac{1}{2} \sum_{i=1}^k \cos{\theta_i} & \mathrm{if} \ \mu = \pm 1 \, ,\\[3ex] \displaystyle -\frac{k}{2} \cos{\theta} & \mathrm{if} \ |\mu| < 1 \, ,\end{cases}
\end{equation}
such that the global angular form associated with $S^1_\varphi$ can be written as $D\varphi = d\varphi - L \, d\psi$ everywhere in $M_4$. We observe that along $\partial M_2$, the function $L$ is piecewise constant and periodic. Specifically, if we trace the value of $L$ anticlockwise along $\partial M_2$, it starts with $1-k/2$ right next to $(\eta,\theta) = (0,0)$, increases by $1$ whenever we cross a monopole till reaching $k/2$ right after $(\eta,\theta) = (0,\pi)$, then it decreases by 1 whenever we cross a monopole till reaching $-k/2$ right after $(\eta,\theta) = (\pi,0)$, and finally returns to $1-k/2$ after crossing the first monopole at $(\eta,\theta) = (0,0)$. As an example, we illustrate the behavior of $L$ along $\partial M_2$ for $k=5$ in figure \ref{L_behavior}.

\section{\boldmath Cohomology class representatives of $M_6$}\label{cohomology_class_representatives_appendix}

The various homological relations in $M_6$ can be derived by working with its dual de Rham cohomology groups. We start with the relatively trivial case of constructing cohomology class representatives of $H^1(M_6)$. The most general one-form that is covariant under the action of the $\mathrm{U}(1)_\psi \times \mathrm{U}(1)_\varphi$ isometry group of $M_6$ can be written as
\begin{equation}
	\lambda_1 = X_0^u \lambda_{1,u} + X_0^\psi \, \frac{D\psi}{2\pi} + X_0^\varphi \, \frac{D\varphi}{2\pi} + X_1^{\eta\theta} \, ,\label{lambda1_parameterization}
\end{equation}
where $X_0^u$, $X_0^\psi$, $X_0^\varphi$ are zero-forms and $X_1^{\eta\theta}$ is a one-form, all defined on $M_2$, and the index $u$ is summed over $1,\dots,b_1(M_6)$. We want $\lambda_1$ to be closed and invariant under the isometries of $M_6$. The former condition amounts to requiring
\begin{equation}
	dX_0^u = 0 \, , \quad \chi (X_0^\psi - L X_0^\varphi) + \zeta X_0^\varphi = 0 \, , \quad d(X_0^\psi - L X_0^\varphi) = -L \, dX_0^\varphi = 0 \, , \quad dX_1^{\eta\theta} = 0 \, ,
\end{equation}
which immediately tells us that $X_0^u$ are constants. We can also combine the constraints above to deduce
\begin{equation}
	d(X_0^\psi - L X_0^\varphi) = -\frac{\zeta}{\chi} \, dX_0^\varphi = -L \, dX_0^\varphi = 0 \, ,
\end{equation}
which together with \eqref{lambda1_parameterization} implies $X_0^\psi = X_0^\varphi = 0$, since $L$ is a position-dependent function (with respect to $M_2$). This automatically makes $\lambda_1$ gauge-invariant. In addition, we conclude that $X_1^{\eta\theta}$ has to be exact because by construction,
\begin{equation}
	\int_{\mathcal{C}_1^{\Sigma,u}} X_1^{\eta\theta} = 0
\end{equation}
for all $u$. Therefore, without loss of generality, we can simply pick $\lambda_{1,u}$ for $u = 1,\dots,2g$ to be the cohomology class representatives of $H^1(M_6)$.

Let us now write down the most general two-form that is covariant under the isometries of $M_6$,
\begin{equation}
	\omega_2 = W_0^{\psi\varphi} \, \frac{D\psi}{2\pi} \wedge \frac{D\varphi}{2\pi} + W_1^\psi \wedge \frac{D\psi}{2\pi} + W_1^\varphi \wedge \frac{D\varphi}{2\pi} + W_0^\Sigma V_2^\Sigma + W_2^{\eta\theta} \, ,
\end{equation}
where $W_0^{\psi\varphi},W_0^\Sigma$ are zero-forms, $W_1^\psi,W_1^\varphi$ are one-forms, and $W_2^{\eta\theta}$ is a two-form, all of which are defined on $M_2$.\footnote{Strictly speaking, we can also add terms linear in $\lambda_{1,u}$ to $\omega_2$, but they can only be trivial if we want $\omega_2$ to be in $H^2(M_6)$.} Imposing closure on $\omega_2$ yields the following conditions,
\begin{equation}
	W_0^{\psi\varphi} = 0 \, , \quad d(W_1^\psi - L W_1^\varphi) = 0 \, , \quad dW_0^\Sigma = -\chi(W_1^\psi - L W_1^\varphi) - \zeta W_1^\varphi \, , \quad dW_2^{\eta\theta} = 0 \, .
\end{equation}
In addition, note that the interior products act on the connection forms as $\iota_\psi D\psi = 1$, $\iota_\psi D\varphi = -L$, $\iota_\varphi D\psi = 0$, $\iota_\varphi D\varphi = 1$. If we take the interior product of $\omega_2$ with respect to the isometries of $M_6$, we get
\begin{equation}
	\iota_\psi \omega_2 = -(W_1^\psi - L W_1^\varphi) \, , \qquad \iota_\varphi \omega_2 = -W_1^\varphi \, .
\end{equation}
Both of these one-forms vanish when integrated over any of the one-cycles in $M_6$, i.e.~$\mathcal{C}_1^{\Sigma,u}$, because they do not contain any factor of $\lambda_{1,u}$ by construction, thus meaning they are exact,
\begin{equation}
	W_1^\psi - L W_1^\varphi = dW_0^\psi \, , \qquad W_1^\varphi = dW_0^\varphi \, ,
\end{equation}
where $W_0^\psi$ and $W_0^\varphi$ are single-valued functions defined on $M_2$. A similar argument implies $W_2^{\eta\theta}$ is also exact. Hence, a generic member of $H^2(M_6)$ can be parameterized as
\begin{equation}
	\omega_2 = \big(dW_0^\psi + L \, dW_0^\varphi\big) \wedge \frac{D\psi}{2\pi} + dW_0^\varphi \wedge \frac{D\varphi}{2\pi} - (\chi W_0^\psi + \zeta W_0^\varphi) V_2^\Sigma \, .\label{omega2_parameterization}
\end{equation}
We check that $\omega_2$ as constructed above is invariant under the isometries of $M_6$,\footnote{This statement can be equivalently phrased in terms of equivariant cohomology. The operator analogous to the ordinary exterior derivative, $d$, is the equivariant exterior derivative, $d_\mathfrak{g} = d - \iota_I$, where $I$ labels the generator of the isometry group. Defining $\hat{\omega}_2 = \big(dW_0^\psi + L \, dW_0^\varphi\big) \wedge (D\psi/2\pi) + dW_0^\varphi \wedge (D\varphi/2\pi)$, it can be shown that $\hat{\omega}_2$ is equivariantly closed, i.e.~$d_\mathfrak{g} \hat{\omega}_2 = 0$. A review of equivariant cohomology can be found, for example, in \cite{2007arXiv0709.3615L}.}
\begin{equation}
	\begin{gathered}
		\mathcal{L}_\psi \omega_2 = d \iota_\psi \omega_2 + \iota_\psi d \omega_2 = -d^2 W_0^\psi = 0 \, , \quad \mathcal{L}_\varphi \omega_2 = d \iota_\varphi \omega_2 + \iota_\varphi d \omega_2 = -d^2 W_0^\varphi = 0 \, .
	\end{gathered}
\end{equation}
The inner products between a generic cohomology class representative \eqref{omega2_parameterization} and the two-cycles \eqref{Sigma_two_cycles}, \eqref{resolution_two_cycles} are given by
\begin{equation}
	\int_{\mathcal{C}_2^{\Sigma,i}} \omega_2 = -\chi W_0^\psi|_{t=t_i} - \zeta W_0^\varphi|_{t=t_i} \, , \quad \int_{\mathcal{C}_2^i} \omega_2 = W_0^\varphi|^{t_{i+1}}_{t_i} \, .\label{omega2_integration_over_two_cycles}
\end{equation}
Recall that $\partial M_2$ is the preimage of the zero section of the $\mathrm{U}(1)_\psi$ bundle, so $\omega_2$ can only be globally defined if terms explicitly containing $D\psi$ vanish on $\partial M_2$. We therefore have to impose the following regularity constraint on $\omega_2$,
\begin{equation}
	\big[dW_0^\psi + L \, dW_0^\varphi\big]_{\partial M_2} = 0 \, ,\label{omega2_regularity_constraint}
\end{equation}
and when combined with \eqref{omega2_integration_over_two_cycles}, it implies
\begin{equation}
	\int_{\mathcal{C}_2^{\Sigma,i+1}} \omega_2 - \int_{\mathcal{C}_2^{\Sigma,i}} \omega_2 = (\chi \ell_i - \zeta) \int_{\mathcal{C}_2^i} \omega_2 \, .
\end{equation}
Since $\omega_2$ is an arbitrary element of $H^2(M_6)$, the relation above holds if and only if
\begin{equation}
	\mathcal{C}_2^{\Sigma,i+1} - \mathcal{C}_2^{\Sigma,i} = (\chi \ell_i - \zeta) \, \mathcal{C}_2^i \, .\label{appendix_two_cycles_recurrence_relation}
\end{equation}
Making use of the single-valuedness of $W_0^\varphi$, the sums of the second relation in \eqref{omega2_integration_over_two_cycles} as well as \eqref{appendix_two_cycles_recurrence_relation} over $i=1,\dots,2k$ respectively lead to two sum rules,
\begin{equation}
	\sum_{i=1}^{2k} \mathcal{C}_2^i = 0 \, , \qquad \sum_{i=1}^{2k} \ell_i \, \mathcal{C}_2^i = 0 \, .
\end{equation}

If we follow the same lines of arguments for the three-forms, we find that a generic cohomology class representative in $H^3(M_6)$ can be parameterized as
\begin{equation}
	\Lambda_3 = \bigg[\big(dS_0^{\psi,u} + L \, dS_0^{\varphi,u}\big) \wedge \frac{D\psi}{2\pi} + dS_0^{\varphi,u} \wedge \frac{D\varphi}{2\pi}\bigg] \wedge \lambda_{1,u} \, ,\label{Lambda3_parameterization}
\end{equation}
where $S_0^{\psi,u}$ and $S_0^{\varphi,u}$ are zero-forms defined on $M_2$, while a generic cohomology class representative in $H^4(M_6)$ can be parameterized as
\begin{equation}
	\Omega_4 = dT_1 \wedge \frac{D\psi}{2\pi} \wedge \frac{D\varphi}{2\pi} + \big(dU_0 + L \chi T_1 - \zeta T_1\big) \wedge \frac{D\psi}{2\pi} \wedge V_2^\Sigma + \chi T_1 \wedge \frac{D\varphi}{2\pi} \wedge V_2^\Sigma \, ,\label{Omega4_parameteriztion}
\end{equation}
where $U_0$ and $T_1$ are respectively zero-forms and one-forms defined on $M_2$. The two sum rules \eqref{three_cycles_sum_rules} for the three cycles can be readily derived based on the previous discussion for the two-cycles. For the four-cycles, on the other hand, we note that
\begin{equation}
	\int_{\mathcal{C}_{4,\mathrm{C}}} \Omega_4 = \int_{M_2} dT_1 = \int_{\partial M_2} T_1 \, , \quad \int_{\mathcal{C}_{4,i}} \Omega_4 = \chi \int_{t_i}^{t_{i+1}} T_1 \, .\label{four_form_integrals}
\end{equation}
Meanwhile, the regularity constraint analogous to \eqref{omega2_regularity_constraint} is
\begin{equation}
	\big[dU_0 + L \chi T_1 - \zeta T_1\big]_{\partial M_2} = 0 \, .\label{Omega4_regularity_constraint}
\end{equation}
It then follows that
\begin{equation}
	\sum_{i=1}^{2k} \mathcal{C}_{4,i} = \chi \mathcal{C}_{4,\mathrm{C}} \, , \qquad \sum_{i=1}^{2k} \ell_i \, \mathcal{C}_{4,i} = \zeta \mathcal{C}_{4,\mathrm{C}} \, .
\end{equation}

The aforementioned parameterizations of $\lambda_1$, $\omega_2$, $\Lambda_3$, $\Omega_4$ are only well-defined on $M_6$. In order for the Kaluza-Klein expansion of $G_4$ to make sense in $M_{11}$, we need to construct the appropriate cohomology class representatives that are also invariant under gauge transformations in $\mathcal{M}_5$. The first step to achieving this goal is to introduce $\mathrm{U}(1)$ connections over $\mathcal{M}_5$ to the global angular forms $D\psi$ and $D\varphi$ respectively by promoting
\begin{equation}
	D\psi \to (D\psi)^\mathrm{g} = D\psi + A_1^\psi \, , \qquad D\varphi \to (D\varphi)^\mathrm{g} = D\varphi + A_1^\varphi \, ,\label{global_angular_form_promotions}
\end{equation}
where $A_1^\psi$ and $A_1^\varphi$ are 5d one-form gauge fields. For completeness, if the Riemann surface is a sphere, i.e.~$g=0$, which we do not consider in this paper unless otherwise specified, then there is an enhanced $\mathrm{SO}(3)_\Sigma$ isometry. In this case, the volume form $V_2^\Sigma$ has to be promoted to the (normalized) global angular form, 
\begin{equation}
	V_2^\Sigma \to \frac{e_2^\Sigma}{2} = \frac{1}{8\pi} \, \epsilon_{abc} (Dy^a \wedge Dy^b y^c - F_2^{ab} y^c) \, ,
\end{equation}
where $y^a$ with $a = 1,3$ are coordinates of $\Sigma_g$ embedded in $\mathbb{R}^3$, and $Dy^a = dy^a - A_1^{ab} y_b$ with $A_1^{ab}$ being an external $\mathrm{SO}(3)_\Sigma$ connection and $F_2^{ab}$ is its field strength. We hereafter use the notation $\lambda_1^\mathrm{g}$, $\omega_2^\mathrm{g}$, $\Lambda_3^\mathrm{g}$, $\Omega_4^\mathrm{g}$ to denote respectively the expressions \eqref{lambda1_parameterization}, \eqref{omega2_parameterization}, \eqref{Lambda3_parameterization}, \eqref{Omega4_parameteriztion} with the replacements \eqref{global_angular_form_promotions} implemented. Except for $\lambda_1^\mathrm{g}$, which is trivial, these gauged forms are not suitable candidates for the cohomology class representatives because they are no longer closed and gauge-invariant. The proper candidates are of the following forms,
\begin{equation}
	\begin{aligned}
		\lambda_1^\mathrm{eq} & = \lambda_1^\mathrm{g} \, ,\\
		\omega_2^\mathrm{eq} & = \omega_2^\mathrm{g} + \frac{F_2^I}{2\pi} \, \omega_{0,I} \, ,\\
		\Lambda_3^\mathrm{eq} & = \Lambda_3^\mathrm{g} + \frac{F_2^I}{2\pi} \, \Lambda_{1,I}^\mathrm{g} \, ,\\
		\Omega_4^\mathrm{eq} & = \Omega_4^\mathrm{g} + \frac{F_2^I}{2\pi} \, \Omega_{2,I}^\mathrm{g} + \frac{F_2^I}{2\pi} \frac{F_2^J}{2\pi} \, \Omega_{0,IJ} \, ,
	\end{aligned}
\end{equation}
for a zero-forms $\omega_{0,I}$, $\Omega_{0,IJ}$, one-form $\Lambda_{1,I}$, two-form $\Omega_{2,I}$, where $\{I,J\} = \{\psi,\varphi\}$ label the isometries. If we impose closure and gauge-invariance on $\lambda_1^\mathrm{eq}$, $\omega_2^\mathrm{eq}$, $\Lambda_3^\mathrm{eq}$, $\Omega_4^\mathrm{eq}$, we find that they satisfy
\begin{equation}
	\begin{gathered}
		2\pi \iota_I \lambda_1 = 0 \, ,\\
		2\pi \iota_I \omega_2 + d\omega_{0,I} = 0 \, ,\\
		2\pi \iota_I \Lambda_3 + d\Lambda_{1,I} = 0 \, , \quad 2\pi \iota_{(I} \Lambda_{1,J)} = 0 \, ,\\
		2\pi \iota_I \Omega_4 + d\Omega_{2,I} = 0 \, , \quad 2\pi \iota_{(I} \Omega_{2,J)} + d\Omega_{0,IJ} = 0 \, ,
	\end{gathered}
\end{equation}
where the following identity,
\begin{equation}
	d\alpha_p^\mathrm{g} + A_1^I (\mathcal{L}_I \alpha_p)^\mathrm{g} = (d\alpha_p)^\mathrm{g} + F_2^I (\iota_I \alpha_p)^\mathrm{g}
\end{equation}
for some generic $p$-form $\alpha_p$ defined on $M_6$, has been used \cite{Bah:2020uev,Bah:2019rgq}. Note that $\omega_{0,I}$, $\Lambda_{1,I}$, $\Omega_{2,I}$, $\Omega_{0,IJ}$ are all defined only up to addition of closed forms.

We introduce below the explicit parameterizations for such forms in the specific context of this work. First of all, we have
\begin{equation}
	\begin{aligned}
		d\Omega_{2,\psi} & = -2\pi \iota_\psi \Omega_4 = -L\, dT_1 \wedge \frac{D\psi}{2\pi} -  dT_1 \wedge \frac{D\varphi}{2\pi} + dU_0 \wedge V_2^\Sigma \, ,\\
		d\Omega_{2,\varphi} & = -2\pi \iota_\varphi \Omega_4 = dT_1 \wedge \frac{D\psi}{2\pi} + \chi T_1 \wedge V_2^\Sigma \, ,
	\end{aligned}
\end{equation}
which admit the globally defined solutions (up to addition of closed two-forms),
\begin{equation}
	\begin{aligned}
		\Omega_{2,\psi} & = -\frac{1}{\chi} \big(dU_0 + L \chi T_1\big) \wedge \frac{D\psi}{2\pi} - T_1 \wedge \frac{D\varphi}{2\pi} + 2 U_0 V_2^\Sigma \, ,\\
		\Omega_{2,\varphi} & = \big(dY_{0,\psi} + L \, dY_{0,\varphi} + T_1\big) \wedge \frac{D\psi}{2\pi} + dY_{0,\varphi} \wedge \frac{D\varphi}{2\pi} - \chi Y_{0,\psi} V_2^\Sigma \, ,
	\end{aligned}
\end{equation}
where $Y_{0,\psi},Y_{0,\varphi}$ are zero-forms defined on $M_2$. Note that the regularity of $\Omega_{2,\psi}$ and $\Omega_{2,\varphi}$ along $\partial M_2$ (where the circle $S^1_\psi$ vanishes) requires $\big[dU_0 + L \chi T_1\big]_{\partial M_2} = \big[dY_{0,\psi} + L \, dY_{0,\varphi} + T_1\big]_{\partial M_2} = 0$. Consequently we can derive (up to additive constants)
\begin{align}
	\Omega_{0,\psi\psi} & = -\frac{1}{\chi} \, U_0 \, , \quad \Omega_{0,\psi\varphi} = \frac{1}{2} \, Y_{0,\psi} \, , \quad \Omega_{0,\varphi\varphi} = Y_{0,\varphi} \, .
\end{align}
The rest of the forms needed to complete the gauge-invariant cohomology class representatives can be similarly solved for to obtain (up to addition of closed one-forms)
\begin{equation}
	\Lambda_{1,x\psi} = S_{0,x}^{\psi,u} \lambda_{1,u} \, , \qquad \Lambda_{1,x\varphi} = S_{0,x}^{\varphi,u} \lambda_{1,u} \, ,
\end{equation}
and (up to additive constants)
\begin{equation}
	\omega_{0,\psi} = W_0^\psi \, , \qquad \omega_{0,\varphi} = W_0^\varphi \, .
\end{equation}

Now we are in a position to explicitly evaluate various intersection numbers associated with cycles of different degrees in $M_6$, as well as miscellaneous integrals involved in the equations of motion \eqref{c3_equation_of_motion} and \eqref{B2_equation_of_motion}. Before doing so, it is instructive to first record an expansion of $E_4^3$ as follows,
\begin{align}
	& \frac{1}{6} \, E_4^3 \supset\nonumber\\
	& N_\alpha N_\beta N_\gamma \, \frac{F_2^I}{2\pi} \frac{F_2^J}{2\pi} \frac{F_2^K}{2\pi} \, \Omega_{0,IJ}^\alpha \big(\Omega_{2,K}^\beta\big)^\mathrm{g} \big(\Omega_4^\gamma\big)^\mathrm{g} + \frac{1}{6} \, N_\alpha N_\beta N_\gamma \, \frac{F_2^I}{2\pi} \frac{F_2^J}{2\pi} \frac{F_2^K}{2\pi} \big(\Omega_{2,I}^\alpha\big)^\mathrm{g} \big(\Omega_{2,J}^\beta\big)^\mathrm{g} \big(\Omega_{2,K}^\gamma\big)^\mathrm{g}\nonumber\\
	& + N N_\beta N_\gamma \, \frac{F_2^I}{2\pi} \frac{F_2^J}{2\pi} \frac{F_2^\alpha}{2\pi} \, \omega_{0,\alpha I} \big(\Omega_{2,J}^\beta\big)^\mathrm{g} \big(\Omega_4^\gamma\big)^\mathrm{g} + N N_\alpha N_\gamma \, \frac{F_2^I}{2\pi} \frac{F_2^J}{2\pi} \frac{F_2^\beta}{2\pi} \, \Omega_{0,IJ}^\alpha (\omega_{2,\beta})^\mathrm{g} \big(\Omega_4^\gamma\big)^\mathrm{g}\nonumber\\
	& + \frac{1}{2} \, N N_\alpha N_\beta \, \frac{F_2^I}{2\pi} \frac{F_2^J}{2\pi} \frac{F_2^\gamma}{2\pi} \big(\Omega_{2,I}^\alpha\big)^\mathrm{g} \big(\Omega_{2,J}^\beta\big)^\mathrm{g} (\omega_{2,\gamma})^\mathrm{g} + N^2 N_\gamma \, \frac{F_2^I}{2\pi} \frac{F_2^\alpha}{2\pi} \frac{F_2^\beta}{2\pi} \, \omega_{0,\alpha I} (\omega_{2,\beta})^\mathrm{g} \big(\Omega_4^\gamma\big)^\mathrm{g}\nonumber\\
	& + \frac{1}{2} \, N^2 N_\alpha \, \frac{F_2^I}{2\pi} \frac{F_2^\beta}{2\pi} \frac{F_2^\gamma}{2\pi} \big(\Omega_{2,I}^\alpha\big)^\mathrm{g} (\omega_{2,\beta})^\mathrm{g} (\omega_{2,\gamma})^\mathrm{g} + \frac{1}{6} \, N^3 \, \frac{F_2^\alpha}{2\pi} \frac{F_2^\beta}{2\pi} \frac{F_2^\gamma}{2\pi} (\omega_{2,\alpha})^\mathrm{g} (\omega_{2,\beta})^\mathrm{g} (\omega_{2,\gamma})^\mathrm{g}\nonumber\\
	& - \frac{1}{2} \, N^2 N_\alpha \, \frac{f_1^x}{2\pi} \frac{f_1^y}{2\pi} \frac{F_2^I}{2\pi} \frac{F_2^J}{2\pi} \, \Omega_{0,IJ}^\alpha (\Lambda_{3,x})^\mathrm{g} (\Lambda_{3,y})^\mathrm{g} - N^2 N_\alpha \, \frac{f_1^x}{2\pi} \frac{f_1^y}{2\pi} \frac{F_2^I}{2\pi} \frac{F_2^J}{2\pi} \, \Lambda_{1,xI} \big(\Omega_{2,J}^\alpha\big)^\mathrm{g} (\Lambda_{3,y})^\mathrm{g}\nonumber\\
	& - \frac{1}{2} \, N^2 N_\alpha \, \frac{f_1^x}{2\pi} \frac{f_1^y}{2\pi} \frac{F_2^I}{2\pi} \frac{F_2^J}{2\pi} \, \Lambda_{1,xI} \, \Lambda_{1,yJ} \big(\Omega_4^\alpha\big)^\mathrm{g} - \frac{1}{2} \, N^3 \, \frac{f_1^x}{2\pi} \frac{f_1^y}{2\pi} \frac{F_2^I}{2\pi} \frac{F_2^\alpha}{2\pi} \, \omega_{0,\alpha I} (\Lambda_{3,x})^\mathrm{g} (\Lambda_{3,y})^\mathrm{g}\nonumber\\
	& - N^3 \, \frac{f_1^x}{2\pi} \frac{f_1^y}{2\pi} \frac{F_2^I}{2\pi} \frac{F_2^\alpha}{2\pi} \, \Lambda_{1,xI} (\omega_{2,\alpha})^\mathrm{g} (\Lambda_{3,y})^\mathrm{g} - N^2 N_\alpha \, \frac{f_1^x}{2\pi} \frac{F_2^I}{2\pi} \frac{H_3^u}{2\pi} \, \Lambda_{1,xI} \, \lambda_{1,u} \big(\Omega_4^\alpha\big)^\mathrm{g}\nonumber\\
	& + N^2 N_\alpha \, \frac{f_1^x}{2\pi} \frac{F_2^I}{2\pi} \frac{H_3^u}{2\pi} \, \lambda_{1,u} \big(\Omega_{2,I}^\alpha\big)^\mathrm{g} (\Lambda_{3,x})^\mathrm{g} + N^3 \, \frac{f_1^x}{2\pi} \frac{F_2^\alpha}{2\pi} \frac{H_3^u}{2\pi} \, \lambda_{1,u} (\omega_{2,\alpha})^\mathrm{g} (\Lambda_{3,x})^\mathrm{g}\nonumber\\
	& - \frac{1}{2} \, N^3 \, \frac{f_1^x}{2\pi} \frac{f_1^y}{2\pi} \frac{\gamma_4}{2\pi} (\Lambda_{3,x})^\mathrm{g} (\Lambda_{3,y})^\mathrm{g} - \frac{1}{2} \, N^2 N_\alpha \, \frac{H_3^u}{2\pi} \frac{H_3^v}{2\pi} \, \lambda_{1,u} \, \lambda_{1,v} \big(\Omega_4^\alpha\big)^\mathrm{g}\nonumber\\
	& + N N_\alpha N_\beta \, \frac{F_2^I}{2\pi} \frac{\gamma_4}{2\pi} \big(\Omega_{2,I}^\alpha\big)^\mathrm{g} \big(\Omega_4^\beta\big)^\mathrm{g} + N^2 N_\beta \, \frac{F_2^\alpha}{2\pi} \frac{\gamma_4}{2\pi} (\omega_{2,\alpha})^\mathrm{g} \big(\Omega_4^\beta\big)^\mathrm{g} \, ,\label{E4_cubed_expansion}
\end{align}
where we kept only terms with six internal legs because they survive under a fiber integration over $M_6$.\footnote{If we include the $g=0$ case in our consideration, then we will have terms cubic in $e_2^\Sigma$ which yield a nonvanishing integral as determined by the Bott-Cattaneo formula \cite{1997dg.ga....10001B}.} Let us also define the choice of basis of (co)homology classes via the expansions,
\begin{equation}
	\mathcal{C}_{4,\mathrm{C}} = a_\mathrm{C}^\alpha \, \mathcal{C}_{4,\alpha} \, , \quad \mathcal{C}_{4,i} = a_i^\alpha \, \mathcal{C}_{4,\alpha} \, , \quad \mathcal{C}_2^{\Sigma,i} = b^{\Sigma,i}_\alpha \, \mathcal{C}_2^\alpha \, , \quad \mathcal{C}_2^i = b^i_\alpha \, \mathcal{C}_2^\alpha \, , \quad	\mathcal{C}_3^{i,u} = c^{i,u}_x \, \mathcal{C}_3^x \, ,\label{cycles_expansions}
\end{equation}
for some real coefficients $a_\mathrm{C}^\alpha$, $a_i^\alpha$, $b^{\Sigma,i}_\alpha$, $b^i_\alpha$, $c^{i,u}_x$, and the cycles $\mathcal{C}_{4,\mathrm{C}}$, $\mathcal{C}_{4,i}$, $\mathcal{C}_2^{\Sigma,i}$, $\mathcal{C}_2^i$, $\mathcal{C}_3^{i,u}$ are defined in section \ref{geometric_setup_section}.\footnote{The expansion of the one-cycles is trivial given that we use the standard $\mathcal{A}$ and $\mathcal{B}$ cycles of the Riemann surface as our basis one-cycles.} In terms of the differential forms introduced earlier to parameterize the cohomology class representatives, the orthonormality between cycles and their dual cohomology class representatives can be rewritten as expressions for these expansion coefficients,
\begin{equation}
	\begin{gathered}
		a_\mathrm{C}^\alpha = \int_{\partial M_2} T_1^\alpha \, , \qquad a_i^\alpha = \chi \int_{t_i}^{t_{i+1}} T_1^\alpha \, ,\\
		b^{\Sigma,i}_\alpha = -\chi W_{0,\alpha}^\psi(t_i) \, , \quad b^i_\alpha = W_{0,\alpha}^\varphi(t_{i+1}) - W_{0,\alpha}^\varphi(t_i) \, , \quad c^{i,u}_x = S_{0,x}^{\varphi,u}(t_{i+1}) - S_{0,x}^{\varphi,u}(t_i) \, .
	\end{gathered}
\end{equation}

The intersection numbers that are relevant in the computation of $I_6^\mathrm{inflow}$ are
\begin{equation}
	\begin{gathered}
		\mathcal{K}_{uv} \equiv \int_{\Sigma_g} \lambda_{1,u} \wedge \lambda_{1,v} \, , \quad \mathcal{K}_\alpha^\beta \equiv \int_{M_6} \omega_{2,\alpha} \wedge \Omega_4^\beta \, , \quad \mathcal{K}_{xy} \equiv \int_{M_6} \Lambda_{3,x} \wedge \Lambda_{3,y} \, ,\\
		\mathcal{K}_{uv}^\alpha \equiv \int_{M_6} \lambda_{1,u} \wedge \lambda_{1,v} \wedge \Omega_4^\alpha \, , \quad \mathcal{K}_{u \alpha x} \equiv \int_{M_6} \lambda_{1,u} \wedge \omega_{2,\alpha} \wedge \Lambda_{3,x} \, ,
	\end{gathered}
\end{equation}
with $u,v = 1,\dots,2g$ and $\alpha,\beta = 1,\dots,2k-1$ and $x,y = 1,\dots,4g(k-1)$. As discussed in section \ref{geometric_setup_section}, by choosing $\lambda_{1,u}$ to be orthonormal to the standard $\mathcal{A}$ and $\mathcal{B}$ cycles of the Riemann surface $\Sigma_g$, the intersection number $\mathcal{K}_{uv}$ can be compactly written in the following form,
\begin{equation}
	\mathcal{K}_{uv} = \begin{pmatrix} 0 & \delta_{pq} \\ -\delta_{pq} & 0 \end{pmatrix} \, .\label{Kuv_intersection_matrix}
\end{equation}
On the other hand, recall that the cohomology class representatives $\omega_{2,\alpha}$, $\Lambda_{3,x}$, $\Omega_4^\alpha$ can be respectively parameterized as
\begin{align}
	\omega_{2,\alpha} & = \big(dW_{0,\alpha}^\psi + L \, dW_{0,\alpha}^\varphi\big) \wedge \frac{D\psi}{2\pi} + dW_{0,\alpha}^\varphi \wedge \frac{D\varphi}{2\pi} - \chi W_{0,\alpha}^\psi V_2^\Sigma \, ,\label{omega2_parametriztion}\\
	\Lambda_{3,x} & = \big(dS_{0,x}^{\psi,u} + L \, dS_{0,x}^{\varphi,u}\big) \wedge \frac{D\psi}{2\pi} \wedge \lambda_{1,u} + dS_{0,x}^{\varphi,u} \wedge \frac{D\varphi}{2\pi} \wedge \lambda_{1,u} \, ,\\
	\Omega_4^\alpha & = dT_1^\alpha \wedge \frac{D\psi}{2\pi} \wedge \frac{D\varphi}{2\pi} + \big(dU_0^\alpha + L \chi T_1^\alpha\big) \wedge \frac{D\psi}{2\pi} \wedge V_2^\Sigma + \chi T_1^\alpha \wedge \frac{D\varphi}{2\pi} \wedge V_2^\Sigma \, .
\end{align}
Let us first focus on \eqref{omega2_parametriztion}, the regularity constraint \eqref{omega2_regularity_constraint} implies that within each open interval $(t_i,t_{i+1})$ on $\partial M_2$, the functions $W_{0,\alpha}^\psi$ and $W_{0,\alpha}^\varphi$ are locally related by
\begin{equation} \label{b39}
	W_{0,\alpha}^\psi(t_i < t < t_{i+1}) = w_{\alpha,i} - \ell_i W_{0,\alpha}^\varphi(t) \, ,
\end{equation}
with $w_{\alpha,i}$ being a real constant, which can be expressed in terms of the expansion coefficients in $\mathcal{C}_2^{\Sigma,i} = b^{\Sigma,i}_\alpha \, \mathcal{C}_2^\alpha$ and $\mathcal{C}_2^i = b^i_\alpha \, \mathcal{C}_2^\alpha$ for a generic basis of two-cycles $\mathcal{C}_2^\alpha$ as
\begin{equation}
	w_{\alpha,i} = -\frac{1}{\chi} \, b^{\Sigma,i}_\alpha + \ell_i \Bigg[W_{0,\alpha}^\varphi(t_1) +  \sum_{j=1}^{i-1} b^j_\alpha\Bigg] \, ,\label{w_alpha_i_expression}
\end{equation}
for some reference value $W_{0,\alpha}^\varphi(t_1)$. Here and elsewhere in this paper sums from $j=1$ to $0$ are understood to be zero. 

Following an analogous approach we can also derive
\begin{gather}
	s_{x,i}^u = S_{0,x}^{\psi,u}(t_1) + \ell_i S_{0,x}^{\varphi,u}(t_1) +  \sum_{j=1}^{i-1} (\ell_i - \ell_j) \, c^{j,u}_x \, ,\\
	U_0^\alpha(t_i) = U_0^\alpha(t_1) -  \sum_{j=1}^{i-1} \ell_j a_j^\alpha \, ,
\end{gather}
for some reference values $S_{0,x}^{\psi,u}(t_1)$, $S_{0,x}^{\varphi,u}(t_1)$, $U_0^\alpha(t_1)$. As an aside, the exterior derivatives (on $\partial M_2$) of regularity constraints like \eqref{omega2_regularity_constraint} restrict
\begin{equation}
	dW_{0,\alpha}^\psi\big|_{t=t_i} = dW_{0,\alpha}^\varphi\big|_{t=t_i} = dS_{0,x}^{\psi,u}\big|_{t=t_i} = dS_{0,x}^{\varphi,u}\big|_{t=t_i} = dU_0^\alpha\big|_{t=t_i} = T_1^\alpha\big|_{t=t_i} = 0\label{regulairty_constraint_at_monopoles}
\end{equation}
for all $i=1,\dots,2k$.

To evaluate a given intersection number, one can convert the integral over $M_6$ into an integral over the boundary $\partial M_2$ via Stokes' theorem, which can be further broken into a sum of integrals over all the open intervals $(t_i,t_{i+1})$. Note that \eqref{regulairty_constraint_at_monopoles} guarantees that such integrals do not receive any singular contribution potentially induced by discontinuities of the function $L$ at the positions of the KK monopoles. For instance, we can deduce that
\begin{equation}
	\mathcal{K}_\alpha^\beta = \int_{\partial M_2} \! \Big(W_{0,\alpha}^\varphi \, dU_0^\beta - \chi W_{0,\alpha}^\psi \, T_1^\beta\Big) = -\sum_{i=1}^{2k} w_{\alpha,i} a_i^\beta \, .
\end{equation}
At first sight the expression above may na\"{i}vely seem to depend explicitly on the reference value $W_{0,\alpha}^\varphi(t_1)$. However, if we shift this reference value by $\delta W_{0,\alpha}^\varphi(t_1)$, then
\begin{equation}
	\delta\mathcal{K}_\alpha^\beta = -\delta W_{0,\alpha}^\varphi(t_1) \sum_{i=1}^{2k} \ell_i a_i^\beta = 0 \, ,
\end{equation}
where we made use of the second sum rule in \eqref{four_cycles_sum_rules} (with $\zeta = 0$) in the second equality. This shows that the intersection number $\mathcal{K}_\alpha^\beta$ is in fact independent of the choice of the reference value, $W_{0,\alpha}^\varphi(t_1)$. Similarly, we find that
\begin{equation}
	\mathcal{K}_{xy} = -\mathcal{K}_{uv} \sum_{i=1}^{2k} \Big[c^{i,u}_x S_{0,y}^{\psi,v}(t_i) + c^{i,v}_y S_{0,x}^{\psi,u}(t_i) - \ell_i c^{i,u}_x c^{i,v}_y\Big] \, ,
\end{equation}
where the value of $S_{0,x}^{\psi,u}(t_i)$ is constrained by the regularity of $\Lambda_{3,x}$ to be
\begin{equation}
	S_{0,x}^{\psi,u}(t_i) = S_{0,x}^{\psi,u}(t_1) -  \sum_{j=1}^{i-1} \ell_j c^{j,u}_x \, .
\end{equation}
The intersection number $\mathcal{K}_{xy}$ is invariant under shifts in the reference values $S_{0,x}^{\psi,u}(t_1)$ and $S_{0,y}^{\psi,v}(t_1)$, i.e.
\begin{equation}
	\delta\mathcal{K}_{xy} = -\mathcal{K}_{uv} \Bigg[\delta S_{0,y}^{\psi,v}(t_1) \sum_{i=1}^{2k} c_x^{i,u} + \delta S_{0,x}^{\psi,u}(t_1) \sum_{i=1}^{2k} c_y^{i,v}\Bigg] = 0
\end{equation}
by virtue of the first sum rule in \eqref{three_cycles_sum_rules}. In addition, $\mathcal{K}_{uv}^\alpha$ is trivially given by
\begin{equation}
	\mathcal{K}_{uv}^\alpha = \mathcal{K}_{uv} a_\mathrm{C}^\alpha \, ,
\end{equation}
which obviously does not depend on the reference value of any auxiliary function, while the remaining intersection number,
\begin{equation}
	\mathcal{K}_{u \alpha x}  = -\mathcal{K}_{uv} \sum_{i=1}^{2k} \bigg[b^i_\alpha S_{0,x}^{\psi,v}(t_i) - \frac{1}{\chi} \, b^{\Sigma,i}_\alpha c^{i,u}_x - \ell_i b^i_\alpha c^{i,u}_x\bigg] \, ,
\end{equation}
again does not change under a shift in the reference value, $S_{0,x}^{\psi,v}(t_1)$, i.e.
\begin{equation}
	\delta \mathcal{K}_{u \alpha x} = -\mathcal{K}_{uv} \, \delta S_{0,x}^{\psi,v}(t_1) \sum_{i=1}^{2k} b^i_\alpha = 0 \, ,
\end{equation}
which follows from the first sum rule in \eqref{two_cycles_sum_rules}.

Next we proceed to study the following integrals which appear in the equations of motion \eqref{c3_equation_of_motion} and \eqref{B2_equation_of_motion},
\begin{equation}
	\begin{aligned}
		\mathcal{J}_I^{\alpha\beta} & \equiv \frac{1}{2} \int_{M_6} \! \Big(\Omega_{2,I}^\alpha \wedge \Omega_4^\beta + \Omega_{2,I}^\beta \wedge \Omega_4^\alpha\Big) \, ,\\
		\mathcal{J}_{Iux}^\alpha & \equiv \int_{M_6} \! \Big(\Lambda_{1,xI} \wedge \lambda_{1,u} \wedge \Omega_4^\alpha - \lambda_{1,u} \wedge \Omega_{2,I}^\alpha \wedge \Lambda_{3,x}\Big) \, ,
	\end{aligned}
\end{equation}
with $I = \psi,\varphi$. We can follow essentially the same procedure as before to obtain
\begin{align}
	\mathcal{J}_\psi^{\alpha\beta} & = \frac{1}{\chi} \sum_{i=1}^{2k} \Big[a_i^\alpha \, U_0^\beta(t_i) + a_i^\beta \, U_0^\alpha(t_i) - \ell_i a_i^\alpha a_i^\beta\Big] \, ,\\
	\mathcal{J}_\varphi^{\alpha\beta} & = \frac{\chi}{2} \sum_{i=1}^{2k} \bigg(\bigg[\frac{1}{\chi} \, a_i^\alpha + \ell_i \big(Y_{0,\varphi}^\alpha(t_{i+1}) - Y_{0,\varphi}^\alpha(t_i)\big)\bigg] \bigg[\frac{1}{\chi} \, a_i^\beta + \ell_i \big(Y_{0,\varphi}^\beta(t_{i+1}) - Y_{0,\varphi}^\beta(t_i)\big)\bigg]\nonumber\\
	& \phantom{=\ } - \bigg[\frac{1}{\chi} \, a_i^\alpha + \ell_i \big(Y_{0,\varphi}^\alpha(t_{i+1}) - Y_{0,\varphi}^\alpha(t_i)\big)\bigg] Y_{0,\psi}^\beta(t_i)\nonumber\\
	& \phantom{=\ } - \bigg[\frac{1}{\chi} \, a_i^\beta + \ell_i \big(Y_{0,\varphi}^\beta(t_{i+1}) - Y_{0,\varphi}^\beta(t_i)\big)\bigg] Y_{0,\psi}^\alpha(t_i)\nonumber\\
	& \phantom{=\ } + \ell_i Y_{0,\psi}^\alpha(t_i) \big(Y_{0,\varphi}^\beta(t_{i+1}) - Y_{0,\varphi}^\beta(t_i)\big) - \ell_i \bigg[\frac{1}{\chi} \, a_i^\alpha + \ell_i \big(Y_{0,\varphi}^\alpha(t_{i+1}) - Y_{0,\varphi}^\alpha(t_i)\big)\bigg] Y_{0,\varphi}^\beta(t_{i+1})\nonumber\\
	& \phantom{=\ } + \ell_i Y_{0,\psi}^\beta(t_i) \big(Y_{0,\varphi}^\alpha(t_{i+1}) - Y_{0,\varphi}^\alpha(t_i)\big) - \ell_i \bigg[\frac{1}{\chi} \, a_i^\beta + \ell_i \big(Y_{0,\varphi}^\beta(t_{i+1}) - Y_{0,\varphi}^\beta(t_i)\big)\bigg] Y_{0,\varphi}^\alpha(t_{i+1})\nonumber\\
	& \phantom{=\ } + \ell_i^2 \big(Y_{0,\varphi}^\alpha(t_{i+1}) Y_{0,\varphi}^\beta(t_{i+1}) - Y_{0,\varphi}^\alpha(t_i) Y_{0,\varphi}^\beta(t_i)\big)\bigg) \, ,\\
	\mathcal{J}_{\psi ux}^\alpha & = -\frac{1}{\chi} \, \mathcal{K}_{uv} \sum_{i=1}^{2k} s_{x,i}^v \, a_i^\alpha \, ,\\
	\mathcal{J}_{\varphi ux}^\alpha & = \mathcal{K}_{uv} \sum_{i=1}^{2k} s_{x,i}^v \big(Y_{0,\varphi}^\alpha(t_{i+1}) - Y_{0,\varphi}^\alpha(t_i)\big) \, .
\end{align}
As usual, the regularity constraint that has to be imposed on $\Omega_{2,\varphi}^\alpha$ is given by
\begin{equation}
	\big[dY_{0,\psi}^\alpha + L \, dY_{0,\varphi}^\alpha + T_1^\alpha\big]_{\partial M_2} = 0 \, .\label{omega2varphi_regularity_constraint}
\end{equation}
Integrating both sides of the relation above over $\partial M_2$ yields a continuity condition on $Y_{0,\psi}^\alpha(t)$,
\begin{equation}
	a_\mathrm{C}^\alpha = -\sum_{i=1}^{2k} \ell_i \big(Y_{0,\varphi}^\alpha(t_{i+1}) - Y_{0,\varphi}^\alpha(t_i)\big) = \sum_{i=1}^k Y_{0,\varphi}^\alpha(t_i) - \sum_{i=k+1}^{2k} Y_{0,\varphi}^\alpha(t_i) \, ,
\end{equation}
which constrains the otherwise arbitrary values of $Y_{0,\varphi}^\alpha(t_i)$ for $i = 1,\dots,2k$. For example, one convenient choice of these values is
\begin{equation}
	Y_{0,\varphi}^\alpha(t_i) = \begin{cases} \displaystyle \frac{a_\mathrm{C}^\alpha}{2k} & \mathrm{if} \ 1 \leq i \leq k \, ,\\[2ex] \displaystyle -\frac{a_\mathrm{C}^\alpha}{2k} & \mathrm{if} \ k + 1 \leq i \leq 2k \, .\end{cases}\label{Omega_varphi_choice}
\end{equation}
Regardless of the choice of $Y_{0,\varphi}^\alpha(t_i)$, the regularity constraint \eqref{omega2varphi_regularity_constraint} further determines
\begin{equation}
	Y_{0,\psi}^\alpha(t_i) = Y_{0,\psi}^\alpha(t_1) - (1 - \delta_{i,1}) \sum_{j=1}^{i-1} \bigg[\frac{1}{\chi} \, a_j^\alpha + \ell_j \big(Y_{0,\varphi}^\alpha(t_{j+1}) - Y_{0,\varphi}^\alpha(t_j)\big)\bigg]
\end{equation}
for some reference value $Y_{0,\psi}^\alpha(t_1)$.

Unlike the intersection numbers, the integrals $\mathcal{J}_\psi^{\alpha\beta}$, $\mathcal{J}_\varphi^{\alpha\beta}$, $\mathcal{J}_{\psi ux}^\alpha$, $\mathcal{J}_{\varphi ux}^\alpha$ are sensitive to the reference values $U_0^\alpha(t_1)$, $Y_{0,\psi}^\alpha(t_1)$, $S_{0,x}^{\psi,u}(t_1)$, $S_{0,x}^{\varphi,u}(t_1)$. Nevertheless, if we impose the convention
\begin{equation}
	N_\alpha N_\beta \, \mathcal{J}_I^{\alpha\beta} = 0
\end{equation}
for each $I = \psi,\varphi$, then we can uniquely fix
\begin{align}
	N_\alpha U_0^\alpha(t_1) & = \frac{1}{2 \chi N} \Bigg[\sum_{i=1}^{2k} \ell_i \big(N_\beta a_i^\beta\big)^2 + 2 \sum_{i=2}^{2k} N_\beta a_i^\beta \sum_{j=1}^{i-1} \ell_j N_\gamma a_j^\gamma\Bigg] \, ,\\
	N_\alpha Y_{0,\psi}^\alpha(t_1) & = \frac{1}{2 \chi^2 N} \Bigg[\sum_{i=1}^{2k} \Big[\big(N_\beta a_i^\beta\big)^2 - 2 \chi N_\beta U_0^\beta(t_{i+1}) N_\gamma \big(Y_{0,\varphi}^\gamma(t_{i+1}) - Y_{0,\varphi}^\gamma(t_i)\big)\Big]\nonumber\\
	& \phantom{=\ } + 2 \sum_{i=2}^{2k} N_\beta a_i^\beta \sum_{j=1}^{i-1} \Big[N_\gamma a_j^\gamma + \chi \ell_j N_\gamma \big(Y_{0,\varphi}^\gamma(t_{j+1}) - Y_{0,\varphi}^\gamma(t_j)\big)\Big]\Bigg] \, .
\end{align}
Similarly, if we impose the convention
\begin{equation}
	N_\alpha \mathcal{J}_{Iux}^\alpha = 0
\end{equation}
for each $I = \psi,\varphi$, each $u = 1,\dots,2g$, and each $x = 1,\dots,4g(k-1)$, then we can uniquely fix
\begin{align}
	S_{0,x}^{\psi,u}(t_1) & = -\frac{1}{\chi N} \sum_{i=2}^{2k} N_\alpha a_i^\alpha \sum_{j=1}^{i-1} (\ell_i - \ell_j) \, c_x^{j,u} \, ,\\
	S_{0,x}^{\varphi,u}(t_1) & = \frac{1}{N} \sum_{i=2}^{2k} N_\alpha \big[Y_{0,\varphi}^\alpha(t_{i+1}) - Y_{0,\varphi}^\alpha(t_i)\big] \sum_{j=1}^{i-1} (\ell_i - \ell_j) \, c_x^{j,u} \, .
\end{align}


\section{The full inflow anomaly polynomial}\label{full_anomaly_polynomial_appendix}

With some algebra, it can be shown that in a given (co)homology basis parameterized by the expansion coefficients $a_i^\alpha$, $b^{\Sigma,i}_\alpha$, $b^i_\alpha$, $c^{i,u}_x$ as defined in \eqref{cycles_expansions}, the full large-$N$ inflow anomaly polynomial for arbitrary flux configurations is
\begin{align}
	& I_6^{\textrm{inflow,large-}N} =\nonumber\\
	& \frac{2}{3 \chi^2} \sum_{i=1}^{2k} \bigg[\ell_i^2 N_i^3 + 3 N_i \, \tilde{U}_{0,i} \, \tilde{U}_{0,i+1}\bigg] \bigg(\frac{F_2^\psi}{2\pi}\bigg)^{\!3}\nonumber\\
	& - \frac{3}{2\chi} \sum_{i=1}^{2k} \bigg[\frac{2}{3 \chi} \, \ell_i N_i^3 + N_i \, \tilde{U}_{0,i} \, \tilde{Y}_{0,\psi,i+1} + N_i \, \tilde{Y}_{0,\psi,i} \, \tilde{U}_{0,i+1}\nonumber\\
	& \phantom{- \frac{3}{2\chi} \sum_{i=1}^{2k} \bigg[\ } + \big(\tilde{U}_{0,i+1}^2 + \ell_i N_i \tilde{U}_{0,i}\big) \big(\tilde{Y}_{0,\varphi,i+1} - \tilde{Y}_{0,\varphi,i}\big)\bigg] \bigg(\frac{F_2^\psi}{2\pi}\bigg)^{\!2} \, \frac{F_2^\varphi}{2\pi}\nonumber\\
	& + \sum_{i=1}^{2k} \bigg[\frac{1}{3 \chi^2} \, N_i^3 + N_i \, \tilde{Y}_{0,\psi,i} \, \tilde{Y}_{0,\psi,i+1} - \frac{1}{\chi} \, \ell_i N_i^2 \tilde{Y}_{0,\varphi,i} - \frac{2}{\chi} \, N_i \tilde{U}_{0,i+1} \tilde{Y}_{0,\varphi,i+1}\nonumber\\
	& \phantom{+ \sum_{i=1}^{2k} \bigg[\ } + \bigg(\frac{1}{\chi} \, N_i \tilde{U}_{0,i+1} + \tilde{U}_{0,i} \tilde{Y}_{0,\psi,i} + \tilde{U}_{0,i+1} \tilde{Y}_{0,\psi,i+1}\bigg) \big(\tilde{Y}_{0,\varphi,i+1} - \tilde{Y}_{0,\varphi,i}\big)\bigg] \frac{F_2^\psi}{2\pi} \bigg(\frac{F_2^\varphi}{2\pi}\bigg)^{\!2}\nonumber\\
	& + \frac{1}{2} \sum_{i=1}^{2k} \bigg[\frac{1}{\chi} \, N_i^2 \tilde{Y}_{0,\varphi,i+1} + 2 N_i \tilde{Y}_{0,\psi,i+1} \tilde{Y}_{0,\varphi,i+1} - \bigg(N_i \tilde{Y}_{0,\psi,i} + \chi \tilde{Y}_{0,\psi,i} \tilde{Y}_{0,\psi,i+1}\nonumber\\
	& \phantom{+ \frac{1}{2} \sum_{i=1}^{2k} \bigg[\ } - \tilde{U}_{0,i} \big(\tilde{Y}_{0,\varphi,i} + \tilde{Y}_{0,\varphi,i+1}\big) + \frac{\chi}{3} \, \ell_i^2 \big(\tilde{Y}_{0,\varphi,i+1} - \tilde{Y}_{0,\varphi,i}\big)^2\bigg) \big(\tilde{Y}_{0,\varphi,i+1} - \tilde{Y}_{0,\varphi,i}\big)\bigg] \bigg(\frac{F_2^\varphi}{2\pi}\bigg)^{\!3}\nonumber\\
	& - \frac{3 N}{\chi} \sum_{i=1}^{2k} \bigg[\frac{1}{2} \, w_{\alpha,i} \ell_i N_i^2 + w_{\alpha,i} N_i \, \tilde{U}_{0,i+1}\bigg] \bigg(\frac{F_2^\psi}{2\pi}\bigg)^{\!2} \, \frac{F_2^\alpha}{2\pi}\nonumber\\
	& + \frac{2 N}{\chi} \sum_{i=1}^{2k} \bigg[\frac{1}{2} \, w_{\alpha,i} N_i^2 + \chi w_{\alpha,i} N_i \, \tilde{Y}_{0,\psi,i+1} + \chi w_{\alpha,i} \tilde{U}_{0,i} \big(\tilde{Y}_{0,\varphi,i+1} - \tilde{Y}_{0,\varphi,i}\big)\bigg] \frac{F_2^\psi}{2\pi} \frac{F_2^\varphi}{2\pi} \frac{F_2^\alpha}{2\pi}\nonumber\\
	& - \chi N \sum_{i=1}^{2k} \bigg[w_{\alpha,i} \tilde{Y}_{0,\psi,i+1} \tilde{Y}_{0,\varphi,i+1} - w_{\alpha,i} \tilde{Y}_{0,\psi,i} \tilde{Y}_{0,\varphi,i} + \frac{1}{2} \, w_{\alpha,i} \ell_i \big(\tilde{Y}_{0,\varphi,i+1}^2 - \tilde{Y}_{0,\varphi,i}^2\big)\bigg] \bigg(\frac{F_2^\varphi}{2\pi}\bigg)^{\!2} \frac{F_2^\alpha}{2\pi}\nonumber\\
	& + N^2 \sum_{i=1}^{2k} \bigg[w_{\alpha,i} w_{\beta,i} N_i - (w_{\alpha,i+1} - w_{\alpha,i}) (w_{\beta,i+1} - w_{\beta,i}) (\ell_{i+1} - \ell_i) \, \tilde{U}_{0,i+1}\bigg] \frac{F_2^\psi}{2\pi} \frac{F_2^\alpha}{2\pi} \frac{F_2^\beta}{2\pi}\nonumber\\
	& + \frac{\chi N^2}{2} \sum_{i=1}^{2k} \bigg[(w_{\alpha,i+1} - w_{\alpha,i}) (w_{\beta,i+1} - w_{\beta,i}) (\ell_{i+1} - \ell_i) \, \tilde{Y}_{0,\psi,i+1}\nonumber\\
	& \phantom{+ \frac{\chi N^2}{2} \sum_{i=1}^{2k} \bigg[\ } - w_{\alpha,i} w_{\beta,i} \big(\tilde{Y}_{0,\varphi,i+1} - \tilde{Y}_{0,\varphi,i}\big)\bigg] \frac{F_2^\varphi}{2\pi} \frac{F_2^\alpha}{2\pi} \frac{F_2^\beta}{2\pi}\nonumber\\
	& + \frac{\chi N^3}{6} \sum_{i=1}^{2k} \bigg[(w_{\alpha,i+1} - w_{\alpha,i}) (w_{\beta,i+1} - w_{\beta,i}) (w_{\gamma,i+1} - w_{\gamma,i}) (\ell_i + \ell_{i+1})\nonumber\\
	& \phantom{+ \frac{\chi N^3}{6} \sum_{i=1}^{2k} \bigg[\ } - 3 (w_{\alpha,i+1} - w_{\alpha,i}) (w_{\beta,i+1} - w_{\beta,i}) (w_{\gamma,i+1} \ell_{i+1} - w_{\gamma,i} \ell_i)\nonumber\\
	& \phantom{+ \frac{\chi N^3}{6} \sum_{i=1}^{2k} \bigg[\ } + 3 (w_{\alpha,i+1} - w_{\alpha,i}) (w_{\beta,i+1} w_{\gamma,i+1} - w_{\beta,i} w_{\gamma,i}) (\ell_{i+1} - \ell_i)\bigg] \frac{F_2^\alpha}{2\pi} \frac{F_2^\beta}{2\pi} \frac{F_2^\gamma}{2\pi}\nonumber\\
	& + \frac{N^2}{2\chi} \, \mathcal{K}_{uv} \sum_{i=1}^{2k} \bigg[s_{x,i}^u s_{y,i}^v N_i + c_x^{i,u} \tilde{U}_{0,i+1} \, S_{0,y,i}^{\psi,v} + c_y^{i,v} \tilde{U}_{0,i+1} \, S_{0,x,i}^{\psi,u} - s_{x,i}^u \ell_i N_i \, S_{0,y,i}^{\varphi,v}\nonumber\\
	& \phantom{+ \frac{N^2}{2\chi} \, \mathcal{K}_{uv} \sum_{i=1}^{2k} \bigg[\ } - s_{y,i}^v \ell_i N_i \, S_{0,x,i}^{\varphi,u} - \ell_i c_x^{i,u} c_y^{i,v} \tilde{U}_{0,i+1} + \ell_i^2 N_i \, S_{0,x,i}^{\varphi,u} \, S_{0,y,i}^{\varphi,v}\bigg] \frac{f_1^x}{2\pi} \frac{f_1^y}{2\pi} \bigg(\frac{F_2^\psi}{2\pi}\bigg)^{\!2}\nonumber\\
	& + \frac{N^2}{2\chi} \, \mathcal{K}_{uv} \sum_{i=1}^{2k} \bigg[c_x^{i,u} c_y^{i,v} \tilde{U}_{0,i} + c_x^{i,u} \tilde{U}_{0,i} \, S_{0,y,i}^{\varphi,v} + c_y^{i,v} \tilde{U}_{0,i} \, S_{0,x,i}^{\varphi,u} - \chi c_x^{i,u} \tilde{Y}_{0,\psi,i} \, S_{0,y,i}^{\psi,v}\nonumber\\
	& \phantom{+ \frac{N^2}{2\chi} \, \mathcal{K}_{uv} \sum_{i=1}^{2k} \bigg[\ } - \chi c_y^{i,v} \tilde{Y}_{0,\psi,i} \, S_{0,x,i}^{\psi,u} + N_i \, S_{0,x,i+1}^{\psi,u} \, S_{0,y,i+1}^{\varphi,v} + N_i \, S_{0,y,i+1}^{\psi,v} \, S_{0,x,i+1}^{\varphi,u}\nonumber\\
	& \phantom{+ \frac{N^2}{2\chi} \, \mathcal{K}_{uv} \sum_{i=1}^{2k} \bigg[\ } - \chi s_{x,i}^u s_{y,i}^v \big(\tilde{Y}_{0,\varphi,i+1} - \tilde{Y}_{0,\varphi,i}\big)\bigg] \frac{f_1^x}{2\pi} \frac{f_1^y}{2\pi} \frac{F_2^\psi}{2\pi} \frac{F_2^\varphi}{2\pi}\nonumber\\
	& - \frac{N^3}{2} \, \mathcal{K}_{uv} \sum_{i=1}^{2k} \bigg[s_{x,i}^u \big(\tilde{Y}_{0,\varphi,i+1} S_{0,y,i+1}^{\varphi,v} - \tilde{Y}_{0,\varphi,i} S_{0,y,i}^{\varphi,v}\big) + s_{y,i}^v \big(\tilde{Y}_{0,\varphi,i+1} S_{0,x,i+1}^{\varphi,u} - \tilde{Y}_{0,\varphi,i} S_{0,x,i}^{\varphi,u}\big)\nonumber\\
	& \phantom{- \frac{N^3}{2} \, \mathcal{K}_{uv} \sum_{i=1}^{2k} \bigg[\ } - \ell_i \big(\tilde{Y}_{0,\varphi,i+1} S_{0,x,i+1}^{\varphi,u} S_{0,y,i+1}^{\varphi,v} - \tilde{Y}_{0,\varphi,i} S_{0,x,i}^{\varphi,u} S_{0,y,i}^{\varphi,v}\big)\bigg] \frac{f_1^x}{2\pi} \frac{f_1^y}{2\pi} \bigg(\frac{F_2^\varphi}{2\pi}\bigg)^{\!2}\nonumber\\
	& - \frac{N^3}{2\chi} \, \mathcal{K}_{uv} \sum_{i=1}^{2k} \bigg[\ell_i b^{\Sigma,i+1}_\alpha c_x^{i,u} c_y^{i,v} - b^{\Sigma,i+1}_\alpha c_x^{i,u} S_{0,y,i}^{\psi,v} - b^{\Sigma,i+1}_\alpha c_y^{i,v} S_{0,x,i}^{\psi,u}\nonumber\\
	& \phantom{- \frac{N^3}{2\chi} \, \mathcal{K}_{uv} \sum_{i=1}^{2k} \bigg[\ } + \chi b^i_\alpha S_{0,x,i}^{\psi,u} \, S_{0,y,i}^{\psi,v} + 2 \chi \ell_i^2 b^i_\alpha S_{0,x,i}^{\varphi,u} \, S_{0,y,i}^{\varphi,v}\bigg] \frac{f_1^x}{2\pi} \frac{f_1^y}{2\pi} \frac{F_2^\psi}{2\pi} \frac{F_2^\alpha}{2\pi}\nonumber\\
	& - \frac{N^3}{2} \, \mathcal{K}_{uv} \sum_{i=1}^{2k} \bigg[w_{\alpha,i} c_x^{i,u} S_{0,y,i}^{\varphi,v} + w_{\alpha,i} c_y^{i,v} S_{0,x,i}^{\varphi,u} + w_{\alpha,i} c_x^{i,u} c_y^{i,v}\bigg] \frac{f_1^x}{2\pi} \frac{f_1^y}{2\pi} \frac{F_2^\varphi}{2\pi} \frac{F_2^\alpha}{2\pi} \ ,\label{inflow_anomaly_polynomial}
\end{align}
where the shorthand notation,
\begin{equation}
	N_i \equiv N_\alpha a_i^\alpha \, , \quad \tilde{U}_{0,i} \equiv N_\alpha U_0^\alpha(t_i) \, , \quad \tilde{Y}_{0,\psi,i} \equiv N_\alpha Y_{0,\psi}^\alpha(t_i) \, , \quad \tilde{Y}_{0,\varphi,i} \equiv N_\alpha Y_{0,\varphi}^\alpha(t_i) \, ,
\end{equation}
is employed. We introduced the constants $\ell_i$ in \eqref{l_i_expression}, and the intersection matrix $\mathcal{K}_{uv}$ is explicitly written in \eqref{Kuv_intersection_matrix}, whereas the definitions of the other auxiliary functions are recorded in appendix \ref{cohomology_class_representatives_appendix}.

For concreteness, let us we focus on the natural basis of homology classes that we described towards the end of section \ref{geometric_setup_section}. It is straightforward to deduce that the choice of basis four-cycles,
\begin{equation}
	\mathcal{C}_{4,1 \leq \alpha \leq k-1} = \mathcal{C}_{4,1 \leq i \leq k-1} \, , \quad \mathcal{C}_{4,\alpha=k} = \mathcal{C}_{4,\mathrm{C}} \, , \quad \mathcal{C}_{4,k+1 \leq \alpha \leq 2k-1} = \mathcal{C}_{4,k+1 \leq i \leq 2k-1} \, ,
\end{equation}
can be equivalently written in terms of the coefficients,
\begin{equation}
	\begin{gathered}
		a_\mathrm{C}^\alpha = \delta_k^\alpha \, , \qquad a_{i \neq k,2k}^\alpha = \delta_i^\alpha \, ,\\
		a_{i=k}^{1 \leq \alpha \leq k-1} = -\frac{\alpha}{k} \, , \quad a_{i=k}^{\alpha=k} = \frac{\chi}{2} \, , \quad a_{i=k}^{k+1 \leq \alpha \leq 2k-1} = -\frac{2k - \alpha}{k} \, ,\\
		a_{i=2k}^{1 \leq \alpha \leq k-1} = -\frac{k - \alpha}{k} \, , \quad a_{i=2k}^{\alpha=k} =   \frac{\chi}{2} \, , \quad a_{i=2k}^{k+1 \leq \alpha \leq 2k-1} = -\frac{\alpha - k}{k} \, .
	\end{gathered}
\end{equation}
Note that we used the two four-cycle sum rules \eqref{four_cycles_sum_rules} to determine the coefficients in the latter two lines. The two-cycles that are Poincar\'{e}-dual to the basis four-cycles satisfy
\begin{equation}
	\mathcal{K}^\alpha_\beta = \int_{M_6} \Omega_4^\alpha \wedge \omega_{2,\beta} = -\sum_{i=1}^{2k} a_i^\alpha w_{\beta,i} = \delta^\alpha_\beta \, .\label{Poincare_dual_basis_convention}
\end{equation}
Applying the condition above to the ``natural'' four-cycles yields the constraints,
\begin{equation}
	\begin{aligned}
		-\delta_\beta^{1 \leq \alpha \leq k-1} & = w_{\beta,\alpha} - \frac{\alpha}{k} \, w_{\beta,k} - \frac{k - \alpha}{k} \, w_{\beta,2k} \, ,\\
		-\delta_\beta^{\alpha=k} & = \frac{\chi}{2} (w_{\beta,k} + w_{\beta,2k}) \, ,\\
		-\delta_\beta^{k+1 \leq \alpha \leq 2k-1} & = w_{\beta,\alpha} - \frac{2k - \alpha}{k} \, w_{\beta,k} - \frac{\alpha - k}{k} \, w_{\beta,2k} \, ,
	\end{aligned}
\end{equation}
which have the solution,
\begin{equation}
	w_{\beta,i \neq k,2k} = -\delta_\beta^i - \frac{\delta_\beta^k}{\chi} \, , \qquad w_{\beta,k} = w_{\beta,2k} = -\frac{\delta_\beta^k}{\chi} \, .\label{w_solution}
\end{equation}
The $b_\beta^{\Sigma,i}$ and $b_\beta^i$ coefficients are related to the constants $w_{\beta,i}$ through
\begin{equation}
	b_\beta^{\Sigma,i} = \chi \, \frac{\ell_{i-1} w_{\beta,i} - \ell_i w_{\beta,i-1}}{\ell_i - \ell_{i-1}} \, , \quad b_\beta^i = \frac{w_{\beta,i+1} - w_{\beta,i}}{\ell_{i+1} - \ell_i} - \frac{w_{\beta,i} - w_{\beta,i-1}}{\ell_i - \ell_{i-1}} \, ,\label{b_coefficients_expression}
\end{equation}
as can be derived utilizing \eqref{b39} and the single-valuedness of $W_{0,\beta}^\varphi(t)$.\footnote{Recall that the quantity $n_i = \ell_i - \ell_{i-1}$ is the charge of the $i$-th Kaluza-Klein monopole along $\partial M_2$. For a fully resolved setup, we have $n_i = +1$ for $1 \leq i \leq k$ and $n_i = -1$ for $k+1 \leq i \leq 2k$.} 
We also defined the ``natural'' three-cycles in section \ref{geometric_setup_section} to be $\mathcal{C}_3^{x=(\beta \neq k,v)} = \mathcal{C}_2^{\alpha \neq k} \times \mathcal{C}_1^{\Sigma,v}$, with the index $x$ parameterized here as a tuple $(\beta \neq k,v)$. It follows from \eqref{cycles_expansions} that
\begin{equation}
	c_{x=(\beta \neq k,v)}^{i,u} = b_\beta^i \, \delta^u_v \, .
\end{equation}
The collection of expansion coefficients presented above fully specifies our basis of homology classes that is relevant to the anomaly polynomial.

As explained in detail in appendix \ref{cohomology_class_representatives_appendix}, the reference values of the auxiliary functions used in \eqref{inflow_anomaly_polynomial} are uniquely fixed under the convention \eqref{isometry_cohomology_decoupled_convention}, apart from $Y_{0,\varphi}^\alpha$ for which we choose to parameterize as in \eqref{Omega_varphi_choice}. To summarize, we have
\begin{align}
	\tilde{U}_{0,i} & = \frac{1}{2 \chi N} \Bigg(\sum_{j=1}^{2k} \ell_j N_j^2 + 2 \sum_{j=2}^{2k} N_j \sum_{m=1}^{j-1} \ell_m N_m\Bigg) - \sum_{j=1}^{i-1} \ell_j N_j \, ,\\
	\tilde{Y}_{0,\psi,i} & = \frac{1}{2 \chi^2 N} \Bigg(\sum_{j=1}^{2k} N_j^2 - \frac{2 \chi N}{k} \sum_{j=1}^k \ell_j N_j + 2 \sum_{j=2}^{2k} N_j \sum_{m=1}^{j-1} \Big[N_m + \frac{\chi N}{k} \, \ell_m (\delta_{m,2k} - \delta_{m,k})\Big]\Bigg)\nonumber\\
	& \phantom{=\ } - \sum_{j=1}^{i-1} \bigg[\frac{1}{\chi} \, N_j + \frac{N}{k} \, \ell_j (\delta_{j,2k} - \delta_{j,k})\bigg] \ ,\\
	\tilde{Y}_{0,\varphi,i} & = \begin{cases} \displaystyle \frac{N}{2k} & \mathrm{if} \ 1 \leq i \leq k \, ,\\[2ex] \displaystyle -\frac{N}{2k} & \mathrm{if} \ k + 1 \leq i \leq 2k \, ,\end{cases}\\
	S_{0,x,i}^{\psi,u} & = -\frac{1}{\chi N} \sum_{j=2}^{2k} N_j \sum_{m=1}^{j-1} (\ell_j - \ell_m) \, c_x^{m,u} - \sum_{j=1}^{i-1} \ell_j c^{j,u}_x \, ,\\
	S_{0,x,i}^{\varphi,u} & = -\frac{1}{k} \sum_{j=1}^{2k-1} \bigg(\frac{k}{2} + \ell_j\bigg) \, c_x^{j,u} - \frac{1}{k} \sum_{j=1}^{k-1} \bigg(\frac{k}{2} - \ell_j\bigg) \, c_x^{j,u} + \sum_{j=1}^{i-1} c_x^{j,u} \, ,\\
	s_{x,i}^u & = -\frac{1}{\chi N} \sum_{j=2}^{2k} N_j \sum_{m=1}^{j-1} (\ell_j - \ell_m) \, c_x^{m,u} -\frac{1}{k} \, \ell_i \sum_{j=1}^{2k-1} \bigg(\frac{k}{2} + \ell_j\bigg) \, c_x^{j,u} - \frac{1}{k} \, \ell_i \sum_{j=1}^{k-1} \bigg(\frac{k}{2} - \ell_j\bigg) \, c_x^{j,u}\nonumber\\
	& \phantom{=\ } + \sum_{j=1}^{i-1} (\ell_i - \ell_j) \, c^{j,u}_x \, ,
\end{align}
where the sums over $j$ from 1 to $i-1$ are to be understood to not contribute if $i=1$. Lastly, it is worth emphasizing again that all the auxiliary functions listed above are (rational) functions of only $\chi$, $N$, $N_\alpha$, $a_i^\alpha$, $b^{\Sigma,i}_\alpha$, $b^i_\alpha$, $c^{i,u}_x$, so there is no extra data required to compute $I_6^{\textrm{inflow,large-}N}$.


\section{Basis-(in)dependence of the anomaly polynomial}\label{basis_independence_appendix}

The explicit expression of $I_6^{\textrm{inflow,large-}N}$ depends on the specific choices of
\begin{enumerate}
	\item the basis of cohomology classes, i.e.~$\Omega_4^\alpha$, $\Lambda_{3,x}$, $\omega_{2,\alpha}$, $\lambda_{1,u}$;
	\item the non-closed forms associated with isometries, i.e.~$\Omega_{2,I}^\alpha$, $\Lambda_{1,xI}$, $\omega_{0,\alpha I}$;
\end{enumerate}
used in constructing $E_4$, both of which can be further shifted with exact forms. We argue that the invariance of $I_6^{\textrm{inflow,large-}N}$ can be restored with appropriate redefinitions of either or both the flux parameters and the external field strengths.\footnote{We also expect the $\mathcal{O}(N)$ contribution to $I_6^\mathrm{inflow}$ from $-E_4 X_8 \subset \mathcal{I}_{12}$ to be invariant, but we refrain from discussing it in detail given that we do not explicitly construct $X_8$ in this paper.}

Let us first consider a generic change of basis of cohomology classes (related to the basis of homology classes via orthonormality), plus a shift in the choice of representative within each cohomology class,
\begin{equation}
	\begin{gathered}
		\big(\Omega_4^\alpha\big)' = (\mathcal{R}_4)^\alpha_\beta \, \Omega_4^\beta + d\Omega_3^\alpha \, , \quad \Lambda_{3,x}' = (\mathcal{R}_3)_x^y \, \Lambda_{3,y} + d\Lambda_{2,x} \, ,\\
		\omega_{2,\alpha}' = (\mathcal{R}_2)_\alpha^\beta \, \omega_{2,\beta} + d\omega_{1,\alpha} \, , \quad \lambda_{1,u}' = (\mathcal{R}_1)_u^v \, \lambda_{1,v} + d\lambda_{0,u} \, ,
	\end{gathered}
\end{equation}
for some constant matrices $\mathcal{R}_p \in \mathrm{GL}\big(b_p(M_6),\mathbb{R}\big)$ and globally defined, gauge-invariant forms $\Omega_3^\alpha$, $\Lambda_{2,x}$, $\omega_{1,\alpha}$, $\lambda_{0,u}$. It results in the shifts below by solving the closure constraints on $E_4'$,
\begin{equation}
	\begin{gathered}
		\big(\Omega_{2,I}^\alpha\big)' = (\mathcal{R}_4)^\alpha_\beta \, \Omega_{2,I}^\beta + 2\pi \iota_I \Omega_3^\alpha \, , \ \big(\Omega_{0,IJ}^\alpha\big)' = (\mathcal{R}_4)^\alpha_\beta \, \Omega_{0,IJ}^\beta \, ,\\
		\Lambda_{1,xI}' = (\mathcal{R}_3)_x^y \, \Lambda_{1,yI} + 2\pi \iota_I \Lambda_{2,x} \, , \ \omega_{0,\alpha I}' = (\mathcal{R}_2)_\alpha^\beta \, \omega_{0,\beta I} + 2\pi \iota_I \omega_{1,\alpha} \, .
	\end{gathered}
\end{equation}
We can check that under the redefinitions,\footnote{Strictly speaking, if we were to preserve the integral quantization condition (\ref{N_alpha_quantization}), then we should limit $\mathcal{R}_4 \in \mathrm{GL}(2k-1,\mathbb{Z})$.}
\begin{equation}
	N_\alpha' = (\mathcal{R}_4^{-1})_\alpha^\beta \, N_\beta \, , \quad (f_1^x)' = (\mathcal{R}_3^{-1})^x_y \, f_1^y \, , \quad (F_2^\alpha)' = (\mathcal{R}_2^{-1})^\alpha_\beta \, F_2^\beta \, , \quad (H_3^u)' = (\mathcal{R}_1^{-1})^u_v \, H_3^v \, ,
\end{equation}
the four-form flux $E_4$ merely acquires an additional globally defined exact piece,\footnote{Here we implicitly used the identity $d(\omega_p)^\mathrm{g} + A^I \big(\mathcal{L}_I \omega_p\big)^\mathrm{g} = \big(d\omega_p\big)^\mathrm{g} + F^I \big(\iota_I\big)^\mathrm{g}$ with $\mathcal{L}_I \omega_p = 0$ for some gauge-invariant $p$-form $\omega_p$ \cite{Bah:2019rgq}.}
\begin{align}
	E_4' & = E_4 + d\bigg[N_\alpha' \big(\Omega_3^\alpha\big)^\mathrm{g} + N \, \frac{(f_1^x)'}{2\pi} (\Lambda_{2,x})^\mathrm{g} + N \, \frac{(F_2^\alpha)'}{2\pi} (\omega_{1,\alpha})^\mathrm{g} + N \, \frac{(H_3^u)'}{2\pi} (\lambda_{0,u})^\mathrm{g}\bigg] \, ,
\end{align}
implying that $I_6^{\textrm{inflow,large-}N}$ is invariant. It guarantees as well that the equations of motion (\ref{c3_equation_of_motion}) and (\ref{B2_equation_of_motion}) are automatically preserved.

Alternatively, we may consider shifts of the non-closed forms in $E_4$ associated with isometries of $M_6$, by linear combinations of harmonic forms plus exact forms,
\begin{equation}
	\begin{gathered}
		\big(\Omega_{2,I}^\alpha\big)' = \Omega_{2,I}^\alpha + (\mathcal{T}_{2,I})^{\alpha\beta} \, \omega_{2,\beta} + d\Omega_{1,I}^\alpha \, , \quad \Lambda_{1,xI}' = \Lambda_{1,xI} + (\mathcal{T}_{1,I})^u_x \, \lambda_{1,u} + d\Lambda_{0,xI} \, ,\label{non_closed_forms_shifts}\\
		\omega_{0,\alpha I}' = \omega_{0,\alpha I} + (\mathcal{T}_{0,I})_\alpha \, ,
	\end{gathered}
\end{equation}
for some constant real matrices $\mathcal{T}_{2,I}$, $\mathcal{T}_{1,I}$, constant real vectors $\mathcal{T}_{0,I}$, and some globally defined, gauge-invariant forms $\Omega_{1,I}^\alpha$ and $\Lambda_{0,xI}$. The matrices $\mathcal{T}_{2,I}$ and $\mathcal{T}_{1,I}$ are not totally unconstrained as we will soon see. Demanding closure of $E_4'$ requires that
\begin{equation}
	\big(\Omega_{0,IJ}^\alpha\big)' = \Omega_{0,IJ}^\alpha + (\mathcal{T}_2)^{\alpha\beta}_{(I} \omega_{0,\beta |J)} + 2\pi \iota_{(I} \Omega_{1,J)}^\alpha + (\mathcal{T}_{0,IJ})^\alpha \, ,
\end{equation}
for some constant real vectors $\mathcal{T}_{0,IJ}$. It is again straightforward to check that under the redefinitions,\footnote{If we were to impose that these redefined field strengths are also appropriately quantized, then it would require $N_\beta (\mathcal{T}_{2,I})^{\alpha\beta} / N, (\mathcal{T}_{1,I})^u_x, N_\alpha (\mathcal{T}_{0,IJ})^\alpha / N, (\mathcal{T}_{0,I})_\alpha \in \mathbb{Z}$.}
\begin{equation}
	\begin{gathered}
		\frac{(F_2^\alpha)'}{2\pi} = \frac{F_2^\alpha}{2\pi} - \frac{N_\beta}{N} (\mathcal{T}_{2,I})^{\alpha\beta} \, \frac{F_2^I}{2\pi} \, , \quad \frac{(H_3^u)'}{2\pi} = \frac{H_3^u}{2\pi} - (\mathcal{T}_{1,I})^u_x \, \frac{f_1^x}{2\pi} \frac{F_2^I}{2\pi} \, ,\\
		\frac{\gamma_4'}{2\pi} = \frac{\gamma_4}{2\pi} - \frac{N_\alpha}{N} (\mathcal{T}_{0,IJ})^\alpha \, \frac{F_2^I}{2\pi} \frac{F_2^J}{2\pi} - (\mathcal{T}_{0,I})_\alpha \, \frac{F_2^\alpha}{2\pi} \frac{F_2^I}{2\pi} \, ,
	\end{gathered}
\end{equation}
the four-form flux $E_4$ is shifted with an exact piece,
\begin{equation}
	E_4' = E_4 + d\bigg[N_\alpha \, \frac{F_2^I}{2\pi} \big(\Omega_{1,I}^\alpha\big)^\mathrm{g} + N \, \frac{f_1^x}{2\pi} \frac{F_2^I}{2\pi} (\Lambda_{0,xI})^\mathrm{g}\bigg] \, .
\end{equation}
Note that, however, this is not sufficient to keep $I_6^{\textrm{inflow,large-}N}$ invariant.\footnote{The paradox can be resolved by understanding the fact that $f_1^x$, $F_2^I$, $F_2^\alpha$, $H_3^u$ are not mutually independent, so (\ref{E4_local_expression}) can be interpreted as an expansion in an overcomplete basis of external field strengths.} We also need to enforce the equivalence between the shifted equations of motion,
\begin{equation}
	N_\alpha N_\beta \bigg\{\mathcal{J}_I^{\alpha\beta} + \mathcal{K}_\gamma^\alpha \big[(\mathcal{T}_{2,I})^{\beta\gamma} - (\mathcal{T}_{2,I})^{\gamma\beta}\big]\bigg\} \frac{F_2^I}{2\pi} + N N_\beta \, \mathcal{K}_\alpha^\beta \, \frac{F_2^\alpha}{2\pi} - \frac{1}{2} \, N^2 \, \mathcal{K}_{xy} \, \frac{f_1^x}{2\pi} \frac{f_1^y}{2\pi} = 0 \, ,\label{modified_c3_equation_of_motion}\\
\end{equation}
\begin{multline}
	N_\alpha \bigg\{\mathcal{J}_{Iux}^\alpha - \frac{3}{2} \, \mathcal{K}_{uv}^\alpha (\mathcal{T}_{1,I})^v_x - \mathcal{K}_{u \beta x} \big[(\mathcal{T}_{2,I})^{\alpha\beta} - (\mathcal{T}_{2,I})^{\beta\alpha}\big]\bigg\} \frac{f_1^x}{2\pi}\frac{F_2^I}{2\pi} + \frac{1}{2} \, N_\alpha \, \mathcal{K}_{uv}^\alpha \, \frac{H_3^v}{2\pi}\\
	- N \, \mathcal{K}_{u \alpha x} \, \frac{f_1^x}{2\pi} \frac{F_2^\alpha}{2\pi} = 0 \, ,\label{modified_B2_equation_of_motion}
\end{multline}
and (\ref{c3_equation_of_motion}) and (\ref{B2_equation_of_motion}) respectively, which amounts to imposing the following constraints on $\mathcal{T}_{2,I}$ and $\mathcal{T}_{1,I}$,
\begin{equation}
	N_\alpha N_\beta \mathcal{K}_\gamma^\alpha \big[(\mathcal{T}_{2,I})^{\beta\gamma} - (\mathcal{T}_{2,I})^{\gamma\beta}\big] = 0 \, , \quad \frac{3}{2} \, N_\alpha \, \mathcal{K}_{uv}^\alpha (\mathcal{T}_{1,I})^v_x + N_\alpha \, \mathcal{K}_{u \beta x} \big[(\mathcal{T}_{2,I})^{\alpha\beta} - (\mathcal{T}_{2,I})^{\beta\alpha}\big] = 0 \, ,
\end{equation}
so as to maintain the invariance of $I_6^{\textrm{inflow,large-}N}$. We verified as a consistency check that after the aforementioned redefinitions of the field strengths, the expression we obtain using \eqref{inflow_anomaly_polynomial} for $I_6^{\textrm{inflow,large-}N}$ at $k=2$ correctly reproduces the independently derived result of \cite{Bah:2019vmq}.

\bibliographystyle{./JHEP}
\bibliography{./references}


\end{document}